\documentclass[referee]{aa} 
\newcommand\etal{{\it et al.}}
\newcommand\lae{\mathrel{<\kern-1.0em\lower0.9ex\hbox{$\sim$}}}
\newcommand\gae{\mathrel{\kern-1.0em\lower0.9ex\hbox{$\sim$}}}

\newcommand\mone{$^{-1}$}
\newcommand\mtwo{$^{-2}$}
\newcommand\mthree{$^{-3}$}
\newcommand{\Msun}{$M_{\sun}$}

\usepackage{lscape}
\usepackage{graphicx}
\usepackage{longtable}

\begin{document}

\title{Physical Properties of Very Powerful FRII Radio Galaxies}
\author{Christopher P. O'Dea\inst{1}, Ruth A. Daly\inst{2},
Preeti Kharb\inst{3}, Kenneth  A. Freeman\inst{2}, and
  Stefi A. Baum\inst{3}}

\institute{
{Department of Physics,
Rochester Institute of Technology, 54 Lomb Memorial Drive,
Rochester, NY 14623}
\and
Department of Physics, Penn State University,
Berks Campus, P. O.
Box 7009, Reading, PA 19610
\and
Center for Imaging Science,
Rochester Institute of Technology, 54 Lomb Memorial Drive,
Rochester, NY 14623}

\date{ }


\abstract
{} {We estimate ages and physical properties of powerful radio galaxies.}
{An analysis of new multi-wavelength VLA observations of
eleven very powerful classical double (FRIIb) radio galaxies with redshifts
between 0.4 and 1.3 is presented.
We estimate ages and velocities for each side
of each source. The eleven new sources are combined with previously
studied samples and the characteristics of the full
sample of 31 sources are studied; the full sample includes sources with
redshifts between 0.056 and 1.79, and
core-hot spot sizes of about 30 to 400 kpc.}
{The velocities are independent
of core-hotspot separation, suggesting the rate of growth of a
given source is roughly constant over the source lifetime. We combine
the rate of growth, width, and pressure of a source
to study the beam power, lifetime, energy, and ambient
gas density using standard methods previously applied to smaller samples.
Typical beam powers are in the range from $10^{44}$
to $10^{46}$ erg/s; we show that this quantity is insensitive to
assumptions regarding minimum energy conditions. The beam powers
are independent of core-hotspot separation suggesting that
the beam power of a given source is roughly constant over
the source lifetime.
Typical total source lifetimes are found to be about a few $\times (10^6 - 10^7) $
years, and typical total outflow
energies ($E/c^2$) are found to be about $5 \times (10^5 - 10^6)$ M$_{\odot}$.
Ambient gas densities are found to
decrease with increasing core-hotspot distance,
but have no redshift dependence.
Overall, the results obtained with the sample of 31 sources studied here
are consistent with those obtained earlier with smaller samples.}
{}


\titlerunning{Powerful Radio Galaxies}
\authorrunning{O'Dea et al.}

\maketitle

\section{INTRODUCTION}

Radio galaxies are beacons that pinpoint sites of active nuclei of galaxies
containing the most massive black holes.
They provide important information on the properties of the active nuclei,
the large-scale environments of these distant galaxies, and the interaction
between outflows from active nuclei and the source environment.
In this paper, we focus
on the properties of very powerful classical double radio galaxies, referred
to as FRIIb sources.

Classical double radio sources, also known as FRII sources (Fanaroff \&
Riley 1974), exhibit a variety of structures.  Leahy \& Williams (1984)
identified five types of FRII sources including sources with no
bridge distortion (type 1) and four types of bridge
distortion (types 2-5).  Leahy, Muxlow, \& Stephens (1989) found
a correlation between radio bridge (also called lobe)
structure and radio power:
the most powerful classical double radio galaxies, those with
178 MHz radio
powers greater than about $3 h^{-2} \times 10^{26}$ W/Hz/sr, have
quite regular bridge structure, that is, they have
type 1 bridge structure. The regular radio bridge structure of
these very high power classical double radio galaxies
was taken as an indication that back flow was likely to
be negligible in these sources. Given that FRII sources
can be divided into those with regular radio bridge structure,
type 1 of Leahy \& Williams (1984), and those with distorted
radio bridge structure, types 2 - 5 of Leahy \& Williams (1984),
Daly (2002) divided the sources into types FRIIa and FRIIb based
on the power dependence found by Leahy, Muxlow, \& Stephens (1989).
Thus, Daly (2002) defined FRII sources
with 178 MHz
radio powers greater than about $3 h^{-2} \times 10^{26}$ W/Hz/sr
as FRIIb sources, and all FRII sources with lower power to be
FRIIa sources.  It is the FRIIb sources that have high radio power and
regular radio bridge structure, and thus are likely to have
negligible back flow in the radio bridge region.

Only FRIIb sources are included in the present study. Thus, most of
the sources are found at relatively high redshift;
only a few local sources belong to this population.
An estimate of source age is presented for 11 new radio sources, as discussed
in section \ref{spectral}. The new sources are combined with
previously studied samples of 14 and 6 radio galaxies to form a
sample of 31 sources.
The physical properties of the sample of 31 radio
sources such as the source velocities, pressures, widths, beam powers,
ambient gas densities, lifetimes, and energies are discussed in section
\ref{properties}, and the results are summarized in section \ref{summary}.
Throughout the paper, we assume a Lambda Cold Dark Matter cosmology with
the standard cosmological parameters
$\Omega_m$=0.3 and $\Omega_{\Lambda}$=0.7, and a value of
Hubble's constant of $H_0$=70 km s$^{-1}$ Mpc$^{-1}$.

\section{Estimates of Source Age for the New Sample of Eleven Radio Galaxies}
\label{spectral}

In the current paradigm (the ``standard model") for classical double radio
galaxies supersonic jets propagate outwards through the ambient medium
(e.g., Blandford \& Rees 1974; Scheuer 1974; Leahy, Muxlow, \& Stephens 1989;
Begelman \& Cioffi 1989; Daly 1990; Leahy 1991). The energy of the jets is
thermalized in a terminal shock (called the hotspot). X-ray observations
show that particle acceleration occurs in the hot spots (e.g., Hardcastle,
Croston \& Kraft 2007). The shocked  jet material
flows sideways from the hot spots to inflate a cocoon (called the
radio lobes or bridges).
This cocoon material is left behind as the jet continues to push forwards.
Radiative losses will steepen the radio spectrum (e.g., Kardashev 1962;
Pacholczyk 1970; Alexander 1987) with increasing distance from the 
acceleration site 
(e.g., Burch 1977). Thus, multi frequency radio imaging can allow
the spectral steepening to be measured, and in the context of the standard
model, the source age and overall advance speed to be estimated.
Note that it is the radio spectral index, and not the radio flux density at
one given frequency, that enters into the spectral aging approach.  
The sources are considered to have a region in which the electrons are
accelerated to very high energies (identified with the hot spot and its 
immediate vicinity), and regions in which these electrons age 
significantly due to synchrotron and inverse Compton losses (identified
as regions distinct from the acceleration site, that is, regions in
the radio lobe).

Many multi-wavelength studies of radio sources have shown that this approach
gives reasonable estimates of source ages (e.g., Myers \& Spangler 1985;
Alexander 1987; Alexander \& Leahy 1987; Leahy, Muxlow \& Stephens 1989;
Carilli et al. 1991; Liu, Pooley, \& Riley 1992;
Mack \etal\ 1998; Guerra, Daly, \& Wan 2000; Jamrozy et al. 2008).
Observations at 90 GHz of some radio sources show spectral steepening 
consistent
with the expectations of spectral aging models (Hardcastle \& Looney 2008).
Studies of large radio sources show that spectral aging analyses give source
ages that are in good agreement with estimates based on dynamical models
(e.g., Parma \etal\ 1998; Murgia \etal\ 1999; Kaiser 2000; Machalski \etal\
2007).
VLBI studies of hot spot proper motions in compact radio sources give direct
measurements
of dynamical ages which are in good agreement with estimates from spectral 
aging
(e.g., Polatidis and Conway 2003; Murgia 2003; Nagai et al. 2006;
Orienti, Dallacasa \& Stanghellini 2007). These studies indicate that spectral
aging analysis provides a useful estimate of the source age.

\subsection{The Injection Spectral Index}
\label{sec:inj}

The injection spectral index is the spectral index the electrons have
in and around the acceleration site (the hot spot and its immediate vicinity).  
We combined snapshot data from the A and C configurations at 20 cm, 
and from the B and D configurations at 6 cm to give us sensitivity 
to structure on a range of size scales with a nominal resolution 
of about 1.2 arcsec (Kharb et al. 2008). In order to optimize the 
quality of the maps and our sensitivity to spectral index gradients
in the low surface brightness lobe emission we tapered the uv data to achieve
a resolution of between 2-2.5 arcsec. 
The spectral index maps used are shown in Figures 51.1, and 51.3 through 
51.12 of the electronic version of Kharb et al. (2008), and each figure 
includes an insert showing the beam size. 

As expected for hot spots of powerful radio galaxies (e.g., Black et al. 1992;
Hardcastle, Croston, and Kraft 2007),
our sources show sub-arcsecond scale structure in the hot spots (see the 8.4
GHz images at $\sim 0.25$ arcsec resolution in Kharb et al. 2008). Thus, the
2-2.5 arcsec resolution images used here will average over substructure
in the hot spots. In order to indicate how much substructure
is included in our estimates of the injection spectral index,  we show
the  $\sim 0.25$ arcsec resolution 8.4 GHz images of the hot spots
in Figures 1-21  with two circles centered on the peak brightness of the hot spot.
One circle has a diameter equal to our CLEAN beam FHWM (between 2 and 2.5 arcsec) 
and  one has a diameter equal to 1 arcsec. 

One potential concern is that the spectral index we determine for the 
hot spot and vicinity (and thus the injection index) is affected by the substructure 
which we include, or perhaps by lobe emission where the hot spot is 
seen projected against the lobe.  
To check this, we remade the images with a CLEAN restoring beam of 1.0
arcsec, calculated the 1.4 to 4.8 GHz spectral index image, and determined the 
spectral index for the hot spot.  
The 1 arcsec beam is close to the nominal 1.2 arcsec and so is not a 
large extrapolation of our data.  The 1 arcsec CLEAN beam has an area 
which is at least 4 times less than that of the 2-2.5 arcsec resolution 
data and thus the contribution to each beam from extended emission would be 
lower by that factor of 4.
Table ~\ref{tab:comp} shows that the hot spot spectral indices determined at
the two resolutions (1 arcsec and 2-2.5 arcsec) are consistent with each other -
i.e., there is no significant offset, but a scatter of about 0.1.
Thus, we conclude that the spectral index measurements of the  hot spots are 
not significantly affected by  including hot spot sub-structure or faint 
emission from the lobes on scales between $\sim 1-2$ arcsec.
In addition, we note that the spectral index of the lobe
emission near the hot spot is close to that of the hot spot itself
(as indicated in the analysis described above, and shown in 
Figures 51.1, and 51.3 through 51.12 of the electronic version
of Kharb et al. 2008).  This is presumably because the lobe material 
near the hot spot has not experienced substantial radiative aging.

Leahy, Muxlow and Stephens (1989) used the 151 MHz to 1.5 GHz hot
spot spectral index obtained at resolutions of 3 to 4 arcsec 
as the injection index, and Liu, Pooley and Riley (1992) used 
the 38 MHz to 1 GHz spectral index integrated over the source
as the injection index. The data of Liu, Pooley, \& Riley (1992)
were analyzed by Wellman, Daly, \& Wan (1997; hereafter WDW)
the 1.4 to 5 GHz hot spot spectral index obtained at about 
1.2 arcsec resolution as the injection index who showed 
that the spectral aging results were in very good agreement with
those obtained by Liu, Pooley, \& Riley (1992).  Thus, the two
approaches are consistent, suggesting that at low frequencies, the
electrons have not experienced significant radiative losses and
the spectral index is that of the original injection spectrum.

In Table~\ref{tab:comp} we compare our 20-6 cm hot spot spectral indices with
the 38 MHz to 1.4 GHz spectral indices based on {\it integrated\/} flux
densities, which is similar to the approach of Liu, Pooley, \& Riley (1992).
We see that there are no significant offsets (i.e., one is not systematically
steeper than the other), but there is a scatter of about  0.1.
Thus, our hot spot spectral indices measured between 1.4 and 4.8 GHz are
consistent with those measured (for the whole source) at low frequencies (38
MHz to 1.4 GHz) and are a useful measure of the injection spectrum.
Because both of these comparisons result in scatter of
about 0.1, we adopt this value as a conservative estimate of the possible
contribution of systematic errors to the spectral index measurements.

As an additional consistency check, we note that the hotspot spectral index is
correlated with the lobe spectral index and is flatter than the lobe spectral
index (Fig 2 of Kharb et al. 2008), as expected if the spectra of the radiating
electrons steepen after they leave the hot spot.

\subsection{Estimating Source Age}

To estimate the source ages, we followed the procedure outlined by
Leahy, Muxlow, \& Stephens (1989).
The spectral age is characterized by a break frequency $\nu_T$. The source age
$t$ is given by
\begin{equation}
t =  50.2 {B^{1/2} \over B^2 + B_{\rm MWB}^2 } \nu_T^{-1/2}
\end{equation}
where t is in Myr, B and $B_{\rm MWB}$ are in units of
$10 \mu$G,
$\nu_T$ is in GHz and is simply related to the observed break frequency
$\nu_{0T}$, $\nu_T = \nu_{0T}(1+z)$,
and $B_{\rm MWB} = 3.18(1+z)^2$ $\mu G$ is the equivalent magnetic field
strength of the cosmic microwave background (Jaffe \& Perola 1973).
The minimum energy magnetic field strength
is obtained using the expression of
Miley (1980) assuming no relativistic protons, a filling factor of unity, a
tangled magnetic field , cylindrical symmetry with random orientation to
the line of sight, and a power-law spectrum with cutoffs at 10 MHz and 100 
GHz.
The expected spectral index for a given injection spectral index
as a function of break frequency was computed
by numerically integrating the
equations given by Meyers \& Spangler (1985).  Meyers and Spangler
showed that the spectral index between two frequencies near the break
was sufficient to allow estimation of the break frequency.
We performed the Myers and Spangler calculations for both the
Kardashev-Pacholczyk (KP) and Jaffe-Perola (JP) models. They give
similar results and we quote the results for the  JP model.

Total intensity maps at 20 and 6 cm at matched angular resolutions of
2 to 2.5 arcsec with circular Gaussian restoring beams
were corrected for the primary beam attenuation and were
used to obtain the spectral index maps.
The spectral index between two frequencies $\nu_1$ and $\nu_2$ is given by
\begin{equation}
\alpha = {\log (S_1/S_2) \over \log(\nu_1/\nu_2) }
\end{equation} where $S_1$ and $S_2$ are the respective flux densities.
The error in the spectral index is given by
\begin{equation}\label{spixerr}
\Delta \alpha = { \log e \over log  (\nu_1/\nu_2) } \left( \left({\Delta S_1
\over S_1}\right)^2
+ \left({\Delta S_2 \over S_2}\right)^2 \right)^{1/2}
\end{equation}
where $\Delta S_1$ and $\Delta S_2$ are the one sigma errors in the flux
densities.
The errors in flux density includes a contribution from the thermal noise due
to the sensitivity of the array which is straightforward to determine
(e.g., Crane and Napier 1985) plus additional contributions whose magnitude
is difficult to determine e.g., the missing uv spacings, and problems with
the calibration, self-calibration, and deconvolution processes (e.g.,
Perley 1985; Leahy, Muxlow and Stephens 1989). To account
for the additional flux density errors, we adopted values for $\Delta S_1$ 
and
$\Delta S_2$ that were factors of three larger than the rms noise in the image
at locations far from the source. However, the hot spots are so bright,
that even with the conservative values of $\Delta S$,
the calculated values of $\Delta \alpha$
are very small. So as discussed in \S~\ref{sec:inj}
we adopted an additional conservative upper limit for the uncertainty
in the spectral index of $\Delta \alpha = 0.1$ We also blanked the images at
values less than 3 times the rms noise determined at distances far from the
source.

Following the work on Leahy, Muxlow, \& Stephens (1989), Liu, Pooley,
\& Riley (1992), and WDW, the spectral index gradient is measured between
the hot spot and one lobe position for
each side of each source; in the Appendix, it is shown that
very similar results are obtained by measuring the spectral gradient
between many points on each side of each source.
The lobe position was chosen to be
in a region of high signal to noise, and far from the hot spot so
that a significant spectral index difference existed between the hot spot
and the lobe.  The break frequency
was determined by comparing the observed spectral
index difference
with the calculations of the JP model using the Myers and Spangler method. The
one sigma range of
the spectral index difference was used to obtain the uncertainty of the break
frequency.
The age of the electrons was obtained using equation (1), with
the magnetic field strengths described in section \ref{v}; the best
fit parameters are listed in Table \ref{tabaging}.
Using the age and the distance from the hot spot, the rate of growth of
each side of each source
(the expansion velocity) is obtained, and is listed
in Table \ref{tabprop}.
In the Appendix, velocities are determined at multiple locations
on each side of each source. It turns out that these velocities are
in very good agreement with those obtained here and listed in
Table \ref{tabprop}.
on the

\section{Physical Parameters of the Combined Sample of 31 Radio Galaxies}
\label{properties}

The physical parameters of the new sample of 11 radio galaxies
of Kharb et al. (2008)
are combined with those published earlier for samples of 6
(Guerra, Daly, \& Wan 2000) and
14 radio galaxies (Leahy, Muxlow, \& Stephens 1989; Liu,
Pooley, \& Riley 1992; WDW)
to yield a sample of 31 radio galaxies. Parameters are determined
for each side of each source, though two sources (\object{3C 239} and
\object{3C 324})
only had radio bridge data on one side of the source, so
we study a total of 60 radio bridges.

The lobe propagation velocity for each side of each source is
presented and discussed in section \ref{v} and
listed in Table \ref{tabprop}.  Converting the velocity from the
cosmological model adopted for the earlier studied samples
of 6 and 14 radio galaxies to that adopted here is non-trivial,
so these are listed in Table \ref{tabprop} for the cosmology adopted here.
Beam powers, ambient gas densities, total source
lifetimes, and total source energies are
obtained for the new sample of 11 sources and
combined with those published earlier for the samples
of 6 and 14 sources. The results obtained for the full
sample are
presented in sections \ref{Lj}, \ref{na}, \ref{T}, and \ref{E}
respectively.

\subsection{Projection Effects}

In this study, radio galaxies are considered so as to minimize
projection effects. If the sources
are not in the plane of the sky, our estimated velocities will be lower
limits to the true velocity, e.g., a source which is oriented 30 degrees
from the plane of the sky will have a velocity 15\% larger than our estimate.
Projection effects were studied in
detail by Wan \& Daly (1998), and are found to be quite small for the
quantities studied here except when the outflow axis of the source is
quite close to the line of sight of the observer.

In Kharb et al (2008), we showed that several diagnostics of alignment
and Doppler boosting were consistent with the eleven galaxies studied here
lying close to the plane of the sky. The  UV/optical spectra show
that these are all narrow-line radio galaxies, which suggests that they
are oriented at angles $ >45$ degrees to the line of sight,
under the standard Unified scheme (e.g., Barthel 1989; Urry \& Padovani 1995).

We noted that the ratio of the angular extent of the larger radio lobe
(distance between hotspot and core) to the smaller one, known as the
arm-length ratio ($Q$), was $\sim 1.1$ for the majority of the radio galaxies,
with
only three galaxies exhibiting ratios$ >1.5$ (3C44 (1.6), 3C54 (1.8) and
3C441 (2.4)). Assuming that the jets are intrinsically symmetric and any
apparent asymmetry is due to the difference in light travel times between
the two sides (e.g., Longair \& Riley 1979; Kapahi \& Saikia 1982)
and using the relation
$Q=(1+\beta \cos \theta)/(1-\beta \cos \theta)$, an arm-length ratio of 1.1
would give a $\beta \cos \theta$ of 0.047 or $\theta$ = 62 degrees
relative to the line of sight for $\beta$ = 0.1 (e.g., Bridle \& Perley
1984).

Another statistical indicator of beaming/orientation is the radio core
prominence parameter (R$_{\rm c}$, Orr \& Browne 1982), which is the ratio of
the
(beamed) core to the (unbeamed) extended flux density (see Appendix C,
Urry \& Padovani 1995). For these radio galaxies, $-4.3 < \log R_{\rm c} <
-2.5$ (Kharb et al. 2008).
Kharb \& Shastri (2004) estimated the R$_{\rm c}$ values (which are linearly
correlated with
$\theta$) for a large sample of radio galaxies and their respective beamed
counterparts (see Fig. 3 of Kharb \& Shastri).
Assuming that these sources are randomly oriented
in the sky and span the entire $\theta$ range (0 to 90 degrees), a
comparison with Fig. 3 indicates that our eleven galaxies lie
at angles larger than 60 degrees to the line of sight, or within
30 degrees of the plane of the sky.

For sources lying at small angles to line of sight, intrinsic misalignment 
between the jet and counter-jet appears to get amplified (e.g., Kapahi \& Saikia 1982). 
The observed misalignment ($\zeta$) is related to the intrinsic misalignment 
($\zeta_{int}$), by $sin(\theta) = tan(\zeta_{int})/tan(\zeta)$, for the 
simplified case in which the azimuth of the jet bending is $90\degr$ 
(see Appl, S. et al. 1996). We found that the jet and counter-jet were 
misaligned by less than 5 degrees for the majority of our radio galaxies, 
with only one galaxy showing a misalignment $>10$ degrees (3C169.1 (15 degrees)). 
If we adopt an intrinsic misalignment angle of $\sim4\degr$ (e.g., Appl, S. et al.
1996), this would imply a jet orientation angle $\theta > 50\degr$ relative
to the line of sight for the majority of the radio galaxies.

As seen in Fig. 5 of Kharb et al (2008), the lobe-to-lobe differences in
depolarization are not correlated with either the radio core prominence
or the misalignment angle. In other words, we do not find evidence for the
`Laing-Garrington effect' (Garrington et al. 1988; Laing 1988)
in our radio galaxies. The lack of a correlation is consistent
with the picture of these radio galaxies lying close to the plane of the sky.
This is also consistent with the fact that all these galaxies lack
bright one-sided jets, typically observed in quasars.

The fact that all of the sources are narrow-line radio galaxies, and have
arm-length ratios close to unity, low radio core prominence parameters,
small misalignments between the jet and counter-jet, and do not exhibit the
Laing-Garrington effect all suggest that the sources are within
about 30 degrees of the plane of the sky.

\subsection{Velocities}
\label{v}

The velocities are obtained using the spectral age given by equation (1) and
the separation $\theta$ between the hotspot and the location where
$\alpha_{\rm final}$
was measured. The average minimum energy
magnetic field strength in the radio lobe region was obtained in
a manner identical to that used earlier so the new results can be
combined with those obtained previously; thus,
$B_{min}$ in the radio lobe region
was taken to be $(B_{10}B_{25})^{0.5}$, where
$B_{10}$ and $B_{25}$ are the minimum energy magnetic field strengths at
distances of
10 and 25 $h^{-1}$ kpc behind the hotspot (toward the core).
In addition to considering a magnetic field strength given by
the minimum energy
value
$B_{min}$, we also consider the possibility that the magnetic field 
strength is
less
than the minimum energy value by a factor of 4, so $B = 0.25 B_{min}$,
parameterized by $b = B/B_{min}$. This is the same offset that was
studied previously, and thus provides a good comparison to results
obtained previously, and there are good reasons for this choice,
as described below.

An offset from minimum energy conditions was suggested by
Carilli et al. (1991) for Cygnus A and Perley \& Taylor (1991) for
\object{3C 295}
to obtain agreement between the
ambient gas density indicated by X-ray observations and that indicated
by the ram pressure confinement of the forward region of the
radio lobe (see also De Young 2002).
WDW showed that the dispersion
in the offset is likely to be less than about 15 \% for very powerful radio
galaxies of the type studied here.

Note that only FRIIb radio galaxies are included in the
study.  These sources have very regular bridge structure suggesting
that back flow is not important in these sources (Alexander and Leahy 1987;
Leahy, Muxlow, and Stephens 1989; WDW; Kaiser 2000; Daly 2002), and that the
sources are growing at a rate that
is well into the supersonic regime.  As described by
Alexander and Leahy (1987) and Leahy, Muxlow, and Stephens (1989),
significant back flow will collide with the relativistic plasma in the
bridge causing side flows and bridge distortions.
As shown by Machalski et al. (2007), a spectral aging analysis of
radio sources provides a reliable parametric fit to the source age.

The overall rate of growth or lobe propagation velocity is listed for each
source for $b=0.25$ and $b=1$ in Table \ref{tabprop} and shown in
Figures \ref{figvda} and \ref{figvdb} as functions of core-hotspot distance
and in Figures \ref{figvza} and \ref{figvzb} as functions of redshift.
When a value of $b=0.25$ is taken, a fractional error in $b$ of 15 \% is
included in the uncertainties of the velocity, pressure, and beam power.
When a value of $b = 1$ is taken, no uncertainty in $b$ is
propagated through other quantities.
Typically, the difference in the total error is not strongly affected by
the uncertainty in $b$. Several previous studies of spectral
aging
have obtained velocities assuming $b=1$ with no uncertainty in this quantity.
Thus, the values and uncertainties of $v$ we obtain with $b=1$ can be
directly compared with those obtained by other groups, and, to illustrate the
effect of including the uncertainty of $b$ on the uncertainty of $v$, this
is included for the case $b=0.25$.

The velocities obtained do not correlate with core-hotspot separation,
as indicated in Tables 5 and 6; this result is also obtained in
the Appendix.  This suggests that the velocity of a particular
source is roughly constant over the lifetime of that source.
The velocities do not correlate with redshift, suggesting that
no redshift dependent selection effects or systematic errors are
skewing the synchrotron ages.

Thus, we find that propagation speeds
of powerful radio galaxies on scales of tens to hundreds of kpc
are independent of source size or redshift.  This is consistent
with the results of observational, analytical, and numerical simulation 
studies
of radio galaxy evolution which suggest that large radio sources
propagate at roughly constant velocity (e.g., Fanti \etal\ 1995; Neeser et al.
1995;  Begelman 1996; Kaiser \& Alexander 1997; Kaiser, Dennett-Thorpe, \&
Alexander 1997;
De Young 1997; O'Dea and Baum 1997; Blundell, Rawlings \& Willott 1999; 
Snellen
\etal\ 2000; Alexander 2000; De Young 1997; Carvalho \& O'Dea 2002a,b).

\subsection{Beam Powers}
\label{Lj}

The beam power of a source may  be estimated based on the
rate of growth of the source, $v$, the pressure in the radio
bridge, $P$, and the cross-sectional width of the bridge, $a$,
as described, for example, by
Daly (1990) and Rawlings \& Saunders (1991).
Here, we use eqs. (2) and (3) of Wan, Daly, and Guerra (2000),
who show that
$L_j = 3.6 \times 10^{44} (a/kpc)^2 (v/c)(P/(10^{-10}
\hbox{dynes cm}^{-2})) \hbox{ergs s}^{-1}$ and
$P=[(4/3)b^{-1.5}+b^2](B_{min}/24\pi)$, so $P(b=1.0)
\approx 0.218 P(b=.25)$.
Widths and pressure for the 11 new sources were obtained in a manner
identical to the way they were obtained for the
previously studied samples of 6 and 14 radio galaxies (see WDW and
Wan, Daly, \& Guerra 2000). The lobe pressure as a function of
core - hot spot separation and redshift are shown in Figures \ref{figpda}
and \ref{figpdb}, respectively.
Values of $L_j$ obtained for the sample are listed in Table \ref{tabprop}
for $b=0.25$,
and shown in Figures \ref{figljda}, \ref{figljdb}, \ref{figljza}
and \ref{figljzb}  for
$b=0.25$ and $b=1$.

The quantity $L_j$ is rather insensitive to the value
of $b \equiv B/B_{min}$  for the current sample, and thus is
insensitive to offsets from minimum energy conditions.
The reason is that
the dependence of $v$ on $b$ just about cancels the dependence of $P$ on
$b$. When $B \equiv bB_{min} \geq B_{MB}$, as is the case here, equation (1)
indicates that $v \propto b^{1.5}$, and for $b \leq 1$, which is expected
to be the case, $P \propto b^{-1.5}$, so $L_j \propto vP$
is quite insensitive to
the value of $b$.  Thus, the values of $L_j$ obtained here
remain valid even if the relativistic plasma in the sources is not
close to minimum energy.

The values of $L_j$ range from about $10^{44}$
to $10^{46}$ erg/s or so.  There is no dependence of $L_j$ on core-hotspot
separation, as indicated in Tables 4 and 5, suggesting that the beam
power of a particular source is roughly constant over the lifetime of that
source. There is a correlation between
beam power and redshift, as is expected from the fact that the parent
population is a subset of the flux limited 3CR sample.

\subsection{Ambient Gas Densities}
\label{na}

The equation of ram pressure confinement (e.g., De Young 2002)
indicates that the ambient gas density is given by
$n_a \propto P/v^2$, with the constant of proportionality equal to
$1/(1.4 \mu m_p)$, where $\mu$ is the mean molecular weight of the gas
taken to be 0.63 for solar abundances, and $m_p$ is the proton rest mass.
It is well known that the ambient gas density obtained with
this equation for $b=1$ is well below that known to be present from
X-ray observations.  This is the primary reason to consider values
of $b$ less than one.  The studies of Carilli et al. (1991) of
Cygnus A and Perley and Taylor (1991) of \object{3C 295}
indicate that b is likely to be less than one.
The study of WDW suggests that the dispersion of
b for very powerful classical double radio galaxies
like those included in this study is less than about 15 \%.

The ambient gas density in the vicinity of the hotspot was
determined for each side of each source  for the 11 new radio galaxies
for b = 0.25 and for b = 1, following the procedure of WDW,
and combined with the previously determined values of the ambient
gas density for the other 20 sources.  The results are shown in
Figures \ref{fignda}, \ref{figndb}, \ref{fignza} and
\ref{fignzb}
as a function of core-hotspot distance and redshift, and listed
in Table \ref{tabprop} for b = 0.25.

The ambient gas density falls with distance from the source center,
as indicated in Table 5 and 6.  For b = 0.25, $n_a \propto r^{-1.9 \pm 0.6}$;
the uncertainty of $0.6$ is estimated by correcting the best fit uncertainty
of $0.2$ by $\sqrt(\chi^2/58)$, since there are 60 data points or
58 degrees of freedom included in the fit. For b = 1,
$n_a \propto r^{-1.38 \pm 0.28}$ with the uncertainty corrected from
the raw value of 0.13 by  $\sqrt(\chi^2/58)$.
The magnitude of the ambient gas density is similar to that in
cluster-type gaseous environments for b = 0.25, and well below
that for b=1.
The values of the ambient gas density and offsets from minimum energy
conditions obtained and
studied here are consistent with recent X-ray studies (Belsole et al. 2007;
Croston et al. 2007).

\subsection{Total Outflow Lifetimes}
\label{T}

The total outflow lifetime $T_T$ is an estimate of the
total lifetime of the jet (past and future)
that is powering that side of the source in its current cycle of
radio activity.
It is obtained from the beam power using the relationship,
$T_T \propto L_j^{-\beta/3}$
for the best fit value of $\beta = 1.5$
obtained by Daly et al. (2007), and it is normalized using Cygnus A.
$T_T$ is not the current synchrotron or dynamical age of the source, but
is the total time that the outflow will be
supplying beam power to the source hot spots.
The empirical basis for the relationship and its derivation
are described in detail by Daly et al. (2007).
The total lifetime is interesting because the total energy that will
be channeled through the jet over its lifetime can be obtained from using
the relation $E_* = L_jT_T$.

The total lifetime of the jet on each side of each source is shown in
Figures \ref{figtda}, \ref{figtdb}, \ref{figtza}, and \ref{figtzb} for
$b=1$ and $b=0.25$; although the beam power is nearly independent of
$b$, the normalization does depend upon this parameter.
We see that the
total lifetime determined in this way
does not depend on the distance of the hotspot from the
core of the source, indicating that we are randomly sampling the source
during its lifetime, and the estimated total lifetime does not depend
on the age of the source at the instant it is observed.  The source
lifetime does depend on redshift, as expected from the dependence
of beam power on redshift.  This follows from the fact that the sample
is flux limited, so we only observe the most powerful sources at high
redshift; these are the sources with large beam power and correspondingly
small total lifetimes.

The estimated total lifetimes are roughly a few $\times 10^{6}$ yr for
$b=1$ and are about a factor of 5 higher for $b=0.25$.
This is the lifetime of the radio luminous outflow. However, there
is a correlation between optical emission
line luminosity and radio luminosity in powerful radio galaxies (e.g., Baum \&
Heckman 1989; Rawlings \& Saunders 1991; Xu, Livio \& Baum 1999;
Willott et al 1999). This requires the radio outflow phase to be accompanied
by an optically luminous QSO phase which can ionize the emission line
nebulae.
Thus, in these powerful radio galaxies, the QSO phase must persist for
at least the lifetime of the radio source giving an independent lower limit
to the QSO lifetime of a few $\times 10^{6}$ to a few $\times 10^{7}$ yr
(depending on the value of $b$). This has implications for the question of
how much mass is accreted by the black hole during a bright phase (e.g., 
Martini
2004).

\subsection{Total Outflow Energy}
\label{E}

The total outflow energy $E_T$ is an estimate of the total energy
that the jet will deposit into the radio source over the entire
lifetime
of the jet $T_T$, and is obtained using the expression
$E_T = L_j T_T$.
The total outflow energies ($E/c^2$) are a few $\times 10^5$
M$_\odot$ for $b=1$ and about a factor of 5 higher for $b=0.25$.

The total energy that will be processed by the central engine and expelled
through the jet during the entire lifetime of the jet is shown in
Figures \ref{figeda}, \ref{figedb}, \ref{figeza} and \ref{figezb} for two
values of b.
The total energy is independent of core-hotspot separation (see Tables
5 and 6), indicating that we are observing the
source at a random epoch during the lifetime of the source and our estimate
of total outflow energy does not depend
on the actual age of the source when we observe it.
The total energy does depend upon redshift as expected from the
dependence of beam power on redshift and the fact that
$E_T \propto L_j^{0.5}$; this is an artifact of the fact
that we are working with a flux limited sample.

\section{SUMMARY}
\label{summary}

We present an analysis of new multi-wavelength VLA observations of
eleven very powerful classical double radio galaxies with redshifts 
between 0.4
and 1.3. We estimate spectral ages and velocities for each radio bridge.
Pressures and widths for each side of each source are also
determined.  These quantities are used to solve for the beam power,
ambient gas density, total source lifetime, and total energy
channeled through the jets over the source lifetime both assuming
minimum energy conditions, and allowing for an offset from these
conditions.

Results obtained with the eleven new sources were combined with a
sample of 20 radio galaxies, and the characteristics of the full
sample of 31 sources were studied. The velocities are independent
of core-hotspot separation, suggesting that the velocity of a
given source is roughly constant over that source lifetime.
The velocities are independent of redshift, suggesting that there are
no redshift-dependent systematic effects/errors in the computation of the
velocity.  Our result that source expansion speeds are roughly constant
is consistent with independent studies based on the statistics of
powerful radio galaxies, theoretical modeling, and numerical simulations.

Typical beam powers are in the range from $10^{44}$
to $10^{46}$ erg/s, and this quantity is insensitive to offsets
from minimum energy conditions due to the fact that
$L_j \propto v~P$, and $v \propto b^{1.5}$ and
$P \propto b^{-1.5}$ for $b \leq 1$, and $B \geq B_{MB}$,
which is the range of magnetic field strengths relevant for
these sources.
We find that the beam powers are independent of core-hotspot separation,
suggesting that the beam power of a given source is roughly constant over
that source lifetime.

The estimated total lifetimes are roughly a few $\times 10^{6}$ yr for $b = 1$
(equipartition), and about 5 times higher for $b=0.25$.
The observed correlation between  radio luminosity and
optical emission line luminosity requires the radio source to be accompanied
by an optically bright QSO phase. Thus, the total source lifetimes
derived here are lower limits on the QSO phase of the activity.

The total outflow energies ($E/c^2$) are a few $\times 10^5$
M$_\odot$ for $b=1$ (equipartition) and about a factor of 5 higher for
$b=0.25$.

Typical ambient gas densities are similar to
those in clusters of galaxies at the current epoch for an offset
from minimum energy conditions by a factor of about 0.25.  The
ambient gas densities (obtained from the ram pressure confinement of
the forward region of the source) decrease  as the  core-hotspot distance
increases, as expected for source sizes of about 30 to 400 kpc,
but have no redshift dependence.

\begin{acknowledgements}

We would like to thank the referee for helpful comments and suggestions.
We would like to acknowledge helpful conversations
with Roger Blandford,
Greg Taylor,  Paul Wiita, Dan Harris, and Alan Marscher.
We are grateful to the Penn State University Computer Center for
the use of their IMSL libraries. Support for this work was provided in part
by US National Science Foundation grant number AST-0507465 (R. A. D.).
This research made use of (1) the NASA/IPAC Extragalactic Database
(NED) which is operated by the Jet Propulsion Laboratory, California
Institute of Technology, under contract with the National Aeronautics and
Space Administration; and (2)  NASA's Astrophysics Data System Abstract
Service.

\end{acknowledgements}


\begin{table*}
\caption{Comparison of Hot Spot Spectral Indices}   
\label{tab:comp}        
\begin{tabular}{lrrrrr}   
\hline\hline                    
Source &  $\alpha$ (2", N) & $\alpha$ (1", N) & $\alpha$ (2", S) & $\alpha$
(1", S) & $\alpha_{38-1400}$ (int) \\
(1)    &    (2)         &    (3)          &   (4)           &  (5) 
      &       (6)             \\
\hline                          
3C 6.1  &    0.84& 0.84   & 0.82  &  0.82 & 0.63 \\
3C 34   &    0.81& 0.78    & 0.84  & 0.88 & 1.07 \\
3C 41   &    0.71& 0.70    & 0.52  & 0.46 & 0.69 \\
3C 44  &     0.89&1.00     & 0.90  & 0.96 & 0.94 \\
3C 54  &     0.74&0.73     & 0.78  & 0.93 & 0.81 \\
3C 114 &     0.82&0.98     & 0.84  & 0.98 & 0.90 \\
3C 142.1 &   0.89&0.94     & 0.81  & 0.77 & 0.87 \\
3C 169.1 &   0.79&...      & 0.77  & 0.91 & 0.88 \\
3C 172  &    0.88&1.03     & 0.91  & 1.19& 0.78 \\
3C 441  &    0.65&0.57     & 0.95  & 0.88& 0.82 \\
3C469.1 &    1.11&1.36     & 0.95  & 1.11 & 0.86 \\
\hline
\end{tabular}
\\
Col 1: Source Name.
Col 2-5: The spectral index of the hotspot between 20 and 6 cm.
Col 2: 2-2.5 arcsec resolution, North side.
Col 3: 1.0   arcsec resolution, North side.
Col 4: 2-2.5 arcsec resolution, South side.
Col 5: 1.0  arcsec resolution, South side.
Col 6: Spectral index from integrated flux densities at 38 and 1400 MHz.
(Except for 3C 34 North and South are really  East and West, respectively.)
\end{table*}

\begin{table*}
\caption{Estimates of Ages of the Eleven Radio Galaxies}   
\label{tabaging}      
\begin{tabular}{lrrrrrrrr}        
\hline\hline                 
Source & Side & $\alpha_{hs} $ & $\alpha_{\rm final} $ & $\theta$ & 
$\nu_{0T}$ &
$B_{min}$ & $t_{(b=1)}$ & $t_{(b=0.25)}$ \\
      &  &  &  & arcsec & GHz & $\mu$ G & Myr  & Myr \\
\hline                        
3C 
6.1  &       N       &       0.84    &       1.37    &       9.5     &$ 
   10.6    \pm     2.9     $&$     50.8    \pm     0.6     $&$     1.0 
\pm     0.1
$&$     4       \pm     1$      \\
3C 
6.1  &       S       &       0.82    &       1.34    &       9       &$ 
   12.6    \pm     3.7     $&$     39.2    \pm     0.6     $&$     1.2 
\pm     0.2
$&$     5       \pm     1       $\\
3C 
34   &       E       &       0.81    &       1.72    &       20      &$ 
   5.6     \pm     0.6     $&$     28      \pm     0.4     $&$     3.2 
\pm     0.2     $&$
10      \pm     1       $\\
3C 
34   &       W       &       0.84    &       1.95    &       18.5    &$ 
   4.5     \pm     0.4     $&$     27.6    \pm     0.3     $&$     3.5 
\pm     0.2
$&$     12      \pm     1       $\\
3C 
41   &       N       &       0.71    &       1.42    &       10      &$ 
   6.6     \pm     1.4     $&$     30.2    \pm     0.9     $&$     2.5 
\pm     0.3     $&$
8       \pm     1       $\\
3C 
41   &       S       &       0.52    &       1.57    &       11      &$ 
   3.5     \pm     0.3     $&$     45      \pm     0.4     $&$     2.0 
\pm     0.1     $&$     9
\pm     1       $\\
3C 
44   &       N       &       0.89    &       1.97    &       30.5    &$ 
   4.1     \pm     0.6     $&$     40      \pm     1       $&$     2.3 
\pm     0.2     $&$
11      \pm     2       $\\
3C 
44   &       S       &       0.90    &       2.03    &       19      &$ 
   3.7     \pm     0.4     $&$     39.5    \pm     0.7     $&$     2.5 
\pm     0.2     $&$
12      \pm     2       $\\
3C 
54   &       N       &       0.74    &       1.5     &       21      &$ 
   6.4     \pm     1.1     $&$     32      \pm     1       $&$     2.4 
\pm     0.2     $&$     7
\pm     1       $\\
3C 
54   &       S       &       0.78    &       1.98    &       15.5    &$ 
   3.3     \pm     0.4     $&$     48.4    \pm     0.2     $&$     1.8 
\pm     0.1
$&$     9       \pm     1       $\\
3C 
114  &       N       &       0.82    &       2.07    &       19.5    &$ 
   3.3     \pm     0.3     $&$     28.2    \pm     0.1     $&$     3.8 
\pm     0.1
$&$     11      \pm     1       $\\
3C 
114  &       S       &       0.84    &       1.82    &       22.5    &$ 
   4.1     \pm     0.6     $&$     27.4    \pm     0.3     $&$     3.5 
\pm     0.2
$&$     10      \pm     1       $\\
3C 
142.1        &       N       &       0.89    &       1.75    &       13.5 
   &$      5.5     \pm     0.9     $&$     39.6    \pm     0.1     $&$ 
2.2     \pm
0.2
$&$     13      \pm     2       $\\
3C 
142.1        &       S       &       0.81    &       1.2     &       30 
   &$      13      \pm     4       $&$     25.7    \pm     0.5     $&$ 
2.7     \pm     0.4     $&$
12      \pm     2       $\\
3C 
169.1        &       N       &       0.79    &       1.33    &       16 
   &$      9.3     \pm     2.5     $&$     19.8    \pm     0.1     $&$ 
3.9     \pm     0.5
$&$     9       \pm     1       $\\
3C 
169.1        &       S       &       0.77    &       1.5     &       23.5 
   &$      6.3     \pm     1.4     $&$     25.8    \pm     0.1     $&$ 
3.4     \pm     0.4
$&$     11      \pm     2       $\\
3C 
172  &       N       &       0.91    &       1.22    &       20      &$ 
   21      \pm     9       $&$     28.6    \pm     0.6     $&$     1.7 
\pm     0.3     $&$
7       \pm     2       $\\
3C 
172  &       S       &       0.88    &       1.01    &       21.5    &$ 
   86      \pm     83      $&$     25      \pm     1       $&$     1.1 
\pm     0.5     $&$
4       \pm     1.5$\\
3C 
441  &       N       &       0.65    &       1.1     &       5       &$ 
   9.6     \pm     2       $&$     27.6    \pm     0.2     $&$     2.4 
\pm     0.2     $&$
8       \pm     1       $\\
3C 
441  &       S       &       0.95    &       1.35    &       16.5    &$ 
   17      \pm     6       $&$     37.8    \pm     0.7     $&$     1.2 
\pm     0.2     $&$
5       \pm     1       $\\
3C 
469.1        &       N       &       1.11    &       1.76    &       10.5 
   &$      9.8     \pm     2.1     $&$     65      \pm     4       $&$ 
0.6     \pm     0.1
$&$     2.4     \pm     0.3     $\\
3C 
469.1        &       S       &       0.95    &       1.96    &       14.5 
   &$      4.6     \pm     0.7     $&$     57.3    \pm     0.5     $&$ 
1.0     \pm
0.1
$&$     3.6     \pm     0.3     $\\
\hline
\end{tabular}
\\
Col 1: Source Name. Col 2: Side of the source on which the spectral aging
was determined. Col 3: The spectral index of the hotspot between 20 and 6 cm.
This is adopted as the injection spectral index for the electrons in the lobe.
Col 4: The spectral index in the lobe furthest from the hot spot along the
brightness ridge line.
The errors in hotspot and lobe spectral index are likely dominated by
systematic errors rather than by thermal noise and are estimated to be $\sim
0.1$. Note that any
constant multiplicative error on the flux density (e.g., due to an absolute
flux density
calibration error) will subtract out when the difference between the hotspot
and lobe spectral
index is determined.  Col 5: The distance from the hotspot to the location of
the maximum spectral
index. The estimated error is of order 1 arcsec.  Col 6: The estimated break
frequency.
Col 7: The equipartition magnetic field in the lobe obtained from $B_{min} =
(B_{10}B_{25})^{0.5}$
using the minimum energy fields obtained at distances of 10 and 25 kpc 
from the
hotspot along the
ridge line towards the nucleus.  Col 8: The radiative loss age at the distance
$\theta$ from the hotspot,
  assuming the magnetic field is equal to the minimum energy value. Col
9: The radiative
loss age at the distance $\theta$ from the hotspot, assuming the magnetic 
field
is equal to
1/4 of the minimum energy value.
\end{table*}

\clearpage

\begin{longtable}{lrrrrrrrr}
\caption{\label{tabprop} Summary of Source Properties}\\
\hline\hline
Source &  z & r &$(v/c)$ & $(v/c)$ & $a_{L}$ & $P_{10}$ &  $L_j$ & $n_{a}$ \\
    &  & kpc & (b=0.25) & (b=1) & kpc & $10^{-10}$ dynes cm\mtwo & $10^{44}$
ergs s\mone     &
$10^{-3}$ cm\mthree \\
\hline
\endfirsthead
\caption{continued.}\\
\hline\hline
Source &  z & r &$(v/c)$ & $(v/c)$ & $a_{L}$ & $P_{10}$ &  $L_j$ & $n_{a}$ \\
    &  & kpc & (b=0.25) & (b=1) & kpc & $10^{-10}$ dynes cm\mtwo & $10^{44}$
ergs s\mone     &
$10^{-3}$ cm\mthree \\
\hline
\endhead
\hline
\endfoot
3C 
239  &       1.790   &       69.5    &$      0.044   \pm     0.008   $&$ 
   0.29    \pm     0.05    $&$     10.0    \pm     0.4     $&$
15.3    \pm     1.8     $&$     242     \pm     86      $&$     0.6 
\pm     0.3     $\\
3C 
322  &       1.681   &       139.4   &$      0.036   \pm     0.006   $&$ 
   0.21    \pm     0.03    $&$     19.6    \pm     1.0
$&$     5.0     \pm     0.6     $&$     250     \pm     89      $&$ 
0.3     \pm     0.1     $\\
3C 
322  &       1.681   &       185.3   &$      0.050   \pm     0.015   $&$ 
   0.31    \pm     0.10    $&$     11.5    \pm     1.2
$&$     11.3    \pm     1.6     $&$     264     \pm     121     $&$ 
0.3     \pm     0.2     $\\
3C 68.2 
&       1.575   &       104.7   &$      0.055   \pm     0.015   $&$ 
0.37    \pm     0.11    $&$     8.2     \pm     1.9
$&$     6.1     \pm     1.2     $&$     80      \pm     43      $&$ 
0.1     \pm     0.1     $\\
3C 68.2 
&       1.575   &       115.8   &$      0.053   \pm     0.009   $&$ 
0.37    \pm     0.07    $&$     9.3     \pm     1.1
$&$     8.9     \pm     1.3     $&$     144     \pm     55      $&$ 
0.2     \pm     0.1     $\\
3C 
437  &       1.480   &       168.4   &$      0.049   \pm     0.004   $&$ 
   0.28    \pm     0.03    $&$     18.1    \pm     1.3
$&$     5.2     \pm     0.6     $&$     294     \pm     98      $&$ 
0.2     \pm     0.1     $\\
3C 
437  &       1.480   &       160.5   &$      0.080   \pm     0.007   $&$ 
   0.54    \pm     0.06    $&$     16.7    \pm     1.3
$&$     9.6     \pm     1.2     $&$     760     \pm     249     $&$ 
0.11    \pm     0.04    $\\
3C 
324  &       1.210   &       35.2    &$      0.038   \pm     0.010   $&$ 
   0.26    \pm     0.07    $&$     7.1     \pm     1.4     $&$
11.5    \pm     2.0     $&$     77      \pm     39      $&$     0.6 
\pm     0.4     $\\
3C 
194  &       1.190   &       50.3    &$      0.023   \pm     0.002   $&$ 
   0.15    \pm     0.02    $&$     7.2     \pm     1.6     $&$
5.5     \pm     1.0     $&$     23      \pm     11      $&$     0.8 
\pm     0.3     $\\
3C 
194  &       1.190   &       79.6    &$      0.025   \pm     0.002   $&$ 
   0.17    \pm     0.02    $&$     8.9     \pm     1.3     $&$
4.8     \pm     0.7     $&$     35      \pm     13      $&$     0.6 
\pm     0.2     $\\
3C 
267  &       1.144   &       175.1   &$      0.032   \pm     0.007   $&$ 
   0.21    \pm     0.05    $&$     8.8     \pm     1.2     $&$
11.2    \pm     1.7     $&$     97      \pm     41      $&$     0.8 
\pm     0.4     $\\
3C 
267  &       1.144   &       165.6   &$      0.029   \pm     0.005   $&$ 
   0.19    \pm     0.03    $&$     10.7    \pm     0.9
$&$     7.9     \pm     1.0     $&$     95      \pm     34      $&$ 
0.7     \pm     0.3     $\\
3C 
356  &       1.079   &       377.1   &$      0.048   \pm     0.012   $&$ 
   0.26    \pm     0.07    $&$     13.6    \pm     2.2
$&$     3.4     \pm     0.5     $&$     105     \pm     49      $&$ 
0.11    \pm     0.07    $\\
3C 
356  &       1.079   &       251.4   &$      0.045   \pm     0.010   $&$ 
   0.28    \pm     0.06    $&$     13.6    \pm     2.2
$&$     6.5     \pm     1.0     $&$     190     \pm     84      $&$ 
0.24    \pm     0.14    $\\
3C 
280  &       0.996   &       49.3    &$      0.027   \pm     0.004   $&$ 
   0.18    \pm     0.03    $&$     7.7     \pm     0.5     $&$
12.0    \pm     1.4     $&$     69      \pm     24      $&$     1.2 
\pm     0.5     $\\
3C 
280  &       0.996   &       60.4    &$      0.014   \pm     0.002   $&$ 
   0.08    \pm     0.01    $&$     7.7     \pm     0.3     $&$
7.7     \pm     0.9     $&$     22      \pm     8       $&$     3.1 
\pm     1.3     $\\
3C 
268.1        &       0.974   &       201.8   &$      0.025   \pm     0.005 
   $&$     0.15    \pm     0.03    $&$     14.5    \pm     0.7
$&$     2.6     \pm     0.3     $&$     48      \pm     17      $&$ 
0.31    \pm     0.15    $\\
3C 
268.1        &       0.974   &       171.6   &$      0.041   \pm     0.007 
   $&$     0.26    \pm     0.05    $&$     13.8    \pm     0.7
$&$     4.1     \pm     0.5     $&$     114     \pm     40      $&$ 
0.18    \pm     0.08    $\\
3C 
289  &       0.967   &       44.5    &$      0.019   \pm     0.003   $&$ 
   0.13    \pm     0.02    $&$     8.9     \pm     0.3     $&$
7.2     \pm     0.8     $&$     38      \pm     12      $&$     1.6 
\pm     0.6     $\\
3C 
289  &       0.967   &       46.1    &$      0.021   \pm     0.003   $&$ 
   0.14    \pm     0.02    $&$     9.7     \pm     0.4     $&$
7.2     \pm     0.8     $&$     50      \pm     16      $&$     1.3 
\pm     0.5     $\\
3C 
325  &       0.860   &       77.7    &$      0.027   \pm     0.005   $&$ 
   0.18    \pm     0.03    $&$     5.2     \pm     1.8     $&$
5.1     \pm     1.3     $&$     14      \pm     8       $&$     0.52 
\pm     0.27    $\\
3C 
325  &       0.860   &       55.0    &$      0.046   \pm     0.010   $&$ 
   0.33    \pm     0.08    $&$     7.0     \pm     1.3     $&$
17.6    \pm     2.8     $&$     144     \pm     63      $&$     0.62 
\pm     0.34    $\\
3C 
265  &       0.811   &       250.0   &$      0.031   \pm     0.005   $&$ 
   0.21    \pm     0.04    $&$     9.5     \pm     1.3     $&$
4.9     \pm     0.7     $&$     48      \pm     19      $&$     0.38 
\pm     0.18    $\\
3C 
265  &       0.811   &       294.3   &$      0.021   \pm     0.004   $&$ 
   0.11    \pm     0.02    $&$     9.5     \pm     1.3     $&$
3.2     \pm     0.5     $&$     21      \pm     9       $&$     0.56 
\pm     0.28    $\\
3C 
247  &       0.749   &       39.5    &$      0.027   \pm     0.005   $&$ 
   0.17    \pm     0.03    $&$     7.8     \pm     0.8     $&$
5.4     \pm     0.7     $&$     31      \pm     12      $&$     0.57 
\pm     0.27    $\\
3C 
247  &       0.749   &       64.7    &$      0.017   \pm     0.003   $&$ 
   0.12    \pm     0.02    $&$     5.8     \pm     0.4     $&$
7.5     \pm     0.9     $&$     15      \pm     5       $&$     1.9 
\pm     0.8     $\\
3C 
55   &       0.720   &       261.7   &$      0.044   \pm     0.009   $&$ 
   0.29    \pm     0.06    $&$     11.6    \pm     1.0     $&$
4.5     \pm     0.6     $&$     95      \pm     35      $&$     0.18 
\pm     0.09    $\\
3C 
55   &       0.720   &       268.0   &$      0.055   \pm     0.011   $&$ 
   0.37    \pm     0.08    $&$     9.5     \pm     1.3     $&$
4.6     \pm     0.7     $&$     82      \pm     32      $&$     0.12 
\pm     0.06    $\\
3C 
337  &       0.630   &       196.1   &$      0.018   \pm     0.005   $&$ 
   0.09    \pm     0.03    $&$     9.7     \pm     0.7     $&$
1.4     \pm     0.2     $&$     8       \pm     3       $&$     0.32 
\pm     0.20    $\\
3C 
337  &       0.630   &       109.5   &$      0.014   \pm     0.001   $&$ 
   0.08    \pm     0.01    $&$     16.5    \pm     0.6
$&$     1.6     \pm     0.2     $&$     22      \pm     7       $&$ 
0.61    \pm     0.22    $\\
3C 
427.1        &       0.572   &       82.8    &$      0.010   \pm     0.002 
   $&$     0.06    \pm     0.01    $&$     13.7    \pm     0.4
$&$     2.7     \pm     0.3     $&$     18      \pm     6       $&$ 
2.1     \pm     0.8     $\\
3C 
427.1        &       0.572   &       75.0    &$      0.011   \pm     0.002 
   $&$     0.08    \pm     0.01    $&$     8.9     \pm     0.6
$&$     4.4     \pm     0.5     $&$     14      \pm     5       $&$ 
2.5     \pm     1.0     $\\
3C 
330  &       0.549   &       205.9   &$      0.031   \pm     0.005   $&$ 
   0.21    \pm     0.04    $&$     10.3    \pm     1.4
$&$     4.0     \pm     0.6     $&$     47      \pm     18      $&$ 
0.3     \pm     0.1     $\\
3C 
330  &       0.549   &       199.7   &$      0.029   \pm     0.005   $&$ 
   0.19    \pm     0.04    $&$     10.3    \pm     0.9
$&$     3.3     \pm     0.4     $&$     36      \pm     13      $&$ 
0.28    \pm     0.13    $\\
3C 
244.1        &       0.430   &       155.9   &$      0.015   \pm     0.001 
   $&$     0.08    \pm     0.01    $&$     8.5     \pm     0.5
$&$     1.4     \pm     0.2     $&$     5       \pm     2       $&$ 
0.49    \pm     0.16    $\\
3C 
244.1        &       0.430   &       141.4   &$      0.024   \pm     0.002 
   $&$     0.16    \pm     0.02    $&$     7.2     \pm     0.5
$&$     3.1     \pm     0.4     $&$     14      \pm     4       $&$ 
0.40    \pm     0.12    $\\
3C 
405  &       0.056   &       63.8    &$      0.009   \pm     0.001   $&$ 
   0.07    \pm     0.01    $&$     10.7    \pm     0.3     $&$
7.4     \pm     0.9     $&$     26      \pm     6       $&$     7.7 
\pm     2.7     $\\
3C 
405  &       0.056   &       72.5    &$      0.009   \pm     0.001   $&$ 
   0.07    \pm     0.01    $&$     10.7    \pm     0.3     $&$
6.1     \pm     0.7     $&$     22      \pm     5       $&$     6.3 
\pm     2.4     $\\
& & & & & & & & 
\\ 

3C 
6.1  &       0.840   &       106.0   &$      0.052   \pm     0.009   $&$ 
   0.25    \pm     0.04    $&$     4.99    \pm     0.07
$&$     8.09    \pm     0.07    $&$     38      \pm     7       $&$ 
0.22    \pm     0.06    $\\
3C 
6.1  &       0.840   &       92.2    &$      0.047   \pm     0.008   $&$ 
   0.18    \pm     0.03    $&$     11.29   \pm     0.03
$&$     3.04    \pm     0.01    $&$     64      \pm     15      $&$ 
0.10    \pm     0.03    $\\
3C 
34   &       0.690   &       170.4   &$      0.045   \pm     0.004   $&$ 
   0.15    \pm     0.01    $&$     13.91   \pm     0.01
$&$     2.01    \pm     0.01    $&$     63      \pm     13      $&$ 
0.07    \pm     0.01    $\\
3C 
34   &       0.690   &       149.9   &$      0.038   \pm     0.002   $&$ 
   0.12    \pm     0.01    $&$     11.28   \pm     0.04
$&$     1.63    \pm     0.01    $&$     28      \pm     5       $&$ 
0.09    \pm     0.01    $\\
3C 
41   &       0.794   &       93.8    &$      0.032   \pm     0.005   $&$ 
   0.10    \pm     0.01    $&$     7.28    \pm     0.06    $&$
2.86    \pm     0.02    $&$     17      \pm     4       $&$     0.21 
\pm     0.06    $\\
3C 
41   &       0.794   &       84.8    &$      0.030   \pm     0.004   $&$ 
   0.14    \pm     0.01    $&$     7.49    \pm     0.01    $&$
6.71    \pm     0.01    $&$     40      \pm     7       $&$     0.56 
\pm     0.12    $\\
3C 
44   &       0.660   &       280.0   &$      0.065   \pm     0.008   $&$ 
   0.31    \pm     0.02    $&$     5.90    \pm     0.05    $&$
3.44    \pm     0.02    $&$     28      \pm     4       $&$     0.06 
\pm     0.01    $\\
3C 
44   &       0.660   &       172.2   &$      0.038   \pm     0.004   $&$ 
   0.18    \pm     0.01    $&$     8.35    \pm     0.02    $&$
2.94    \pm     0.01    $&$     28      \pm     4       $&$     0.15 
\pm     0.02    $\\
3C 
54   &       0.827   &       201.9   &$      0.071   \pm     0.008   $&$ 
   0.22    \pm     0.02    $&$     9.74    \pm     0.02    $&$
2.92    \pm     0.01    $&$     70      \pm     15      $&$     0.04 
\pm     0.01    $\\
3C 
54   &       0.827   &       187.6   &$      0.045   \pm     0.006   $&$ 
   0.21    \pm     0.02    $&$     7.86    \pm     0.01    $&$
4.86    \pm     0.01    $&$     48      \pm     7       $&$     0.18 
\pm     0.03    $\\
3C 
114  &       0.815   &       192.1   &$      0.045   \pm     0.003   $&$ 
   0.13    \pm     0.01    $&$     18.09   \pm     0.01
$&$     1.37    \pm     0.002   $&$     72      \pm     15      $&$ 
0.05    \pm     0.01    $\\
3C 
114  &       0.815   &       209.5   &$      0.058   \pm     0.004   $&$ 
   0.16    \pm     0.01    $&$     15.19   \pm     0.01
$&$     1.47    \pm     0.002   $&$     70      \pm     15      $&$ 
0.03    \pm     0.00    $\\
3C 
142.1        &       0.406   &       99.2    &$      0.019   \pm     0.003 
   $&$     0.11    \pm     0.01    $&$     6.77    \pm     0.00
$&$     4.39    \pm     0.005   $&$     13      \pm     2       $&$ 
0.95    \pm     0.17    $\\
3C 
142.1        &       0.406   &       185.1   &$      0.047   \pm     0.008 
   $&$     0.20    \pm     0.03    $&$     9.49    \pm     0.01

$&$     1.82    \pm     0.003   $&$     27      \pm     6       $&$ 
0.06    \pm     0.02    $\\
3C 
169.1        &       0.633   &       131.4   &$      0.039   \pm     0.005 
   $&$     0.09    \pm     0.01    $&$     20.17   \pm     0.01

$&$     0.58    \pm     0.002   $&$     33      \pm     8       $&$ 
0.03    \pm     0.01    $\\
3C 
169.1        &       0.633   &       189.7   &$      0.048   \pm     0.006 
   $&$     0.16    \pm     0.02    $&$     12.98   \pm     0.02
$&$     1.16    \pm     0.003   $&$     34      \pm     7       $&$ 
0.04    \pm     0.01    $\\
3C 
172  &       0.519   &       294.7   &$      0.058   \pm     0.013   $&$ 
   0.24    \pm     0.05    $&$     6.01    \pm     0.11
$&$     2.01    \pm     0.03    $&$     15      \pm     4       $&$ 
0.04    \pm     0.02    $\\
3C 
172  &       0.519   &       333.0   &$      0.119   \pm     0.058   $&$ 
   0.43    \pm     0.21    $&$     14.14   \pm     0.01
$&$     1.92    \pm     0.004   $&$     163     \pm     83      $&$ 
0.01    \pm     0.01    $\\
3C 
441  &       0.707   &       69.4    &$      0.015   \pm     0.001   $&$ 
   0.049   \pm     0.004   $&$     19.16   \pm     0.01
$&$     1.40    \pm     0.002   $&$     28      \pm     6       $&$ 
0.45    \pm     0.06    $\\
3C 
441  &       0.707   &       169.7   &$      0.076   \pm     0.016   $&$ 
   0.33    \pm     0.06    $&$     10.14   \pm     0.02
$&$     3.11    \pm     0.01    $&$     87      \pm     21      $&$ 
0.04    \pm     0.02    $\\
3C 
469.1        &       1.336   &       306.1   &$      0.124   \pm     0.020 
   $&$     0.49    \pm     0.07    $&$     6.97    \pm     0.04
$&$     17.17   \pm     0.07    $&$     368     \pm     79      $&$ 
0.08    \pm     0.02    $\\
3C 
469.1        &       1.336   &       327.5   &$      0.112   \pm     0.012 
   $&$     0.40    \pm     0.04    $&$     13.54   \pm     0.02
$&$     6.48    \pm     0.01    $&$     473     \pm     96      $&$ 
0.04    \pm     0.01    $\\
\end{longtable}

All calculations follow WDW and assume b = 0.25 except where noted.
The top part of the table contains the properties of the previous sample of 20
radio sources recalculated for the currently adopted cosmology.  The bottom
part
of the table contains the new sample of 11 radio galaxies.
Col 1. Source Name. Col 2. redshift.
Col 3. The distance between the core and hotspot.
Col 4. The source expansion speed.
Col 5. The source expansion speed for b= 1.
Col 6. The half-width of the lobe at a projected distance of 10 kpc along the
ridge
line from the hotspot towards the core.
Col 7.   The minimum energy pressure in the lobe at a projected distance of 10
kpc
along the ridge line from the hotspot towards the core.
Col 8. The beam power.
Col 9. The ambient density.

\clearpage

\begin{longtable}{lrrrr}
\caption{\label{tabpropII} Summary of Source Properties. II}  \\
\hline\hline                 
Source & $T_{T (b=0.25)}$ & $E_{T (b=0.25)}$ & $T_{T (b=1)}$ & $E_{T 
(b=1)}$ \\
        & $10^{7}$ yr      & $10^{6}$ \Msun & $10^{7}$ yr & $10^{6}$ \Msun \\
\hline                        
\endfirsthead
\caption{continued.}\\
\hline\hline
Source & $T_{T (b=0.25)}$ & $E_{T (b=0.25)}$ & $T_{T (b=1)}$ & $E_{T 
(b=1)}$ \\
           & $10^{7}$ yr   & $10^{6}$ \Msun & $10^{7}$ yr & $10^{6}$ \Msun \\
\hline
\endhead
\hline
\endfoot
\object{3C 239} 
&$      1.2     \pm     0.4     $&$     5.1     \pm     1.5     $&$ 
0.17    \pm     0.04    $&$     1.02    \pm
0.26    $\\
\object{3C 322} 
&$      1.2     \pm     0.3     $&$     5.2     \pm     1.5     $&$ 
0.18    \pm     0.05    $&$     0.97    \pm
0.25    $\\
3C 
322  &$      1.2     \pm     0.4     $&$     5.4     \pm     1.7     $&$ 
   0.16    \pm     0.05    $&$     1.04    \pm     0.31    $\\
\object{3C 
68.2}        &$      2.1     \pm     0.7     $&$     3.0     \pm     1.0 
   $&$     0.29    \pm     0.09    $&$     0.59    \pm
0.19    $\\
3C 68.2 
&$      1.6     \pm     0.5     $&$     4.0     \pm     1.2     $&$ 
0.21    \pm     0.06    $&$     0.81    \pm     0.22    $\\
\object{3C 437} 
&$      1.1     \pm     0.3     $&$     5.7     \pm     1.6     $&$ 
0.16    \pm     0.04    $&$     1.05    \pm
0.26    $\\
3C 
437  &$      0.7     \pm     0.2     $&$     9.1     \pm     2.6     $&$ 
   0.09    \pm     0.02    $&$     1.83    \pm     0.46    $\\
\object{3C 324} 
&$      2.2     \pm     0.7     $&$     2.9     \pm     1.0     $&$ 
0.29    \pm     0.09    $&$     0.59    \pm
0.18    $\\
\object{3C 194} 
&$      3.9     \pm     1.3     $&$     1.6     \pm     0.5     $&$ 
0.55    \pm     0.16    $&$     0.31    \pm
0.09    $\\
3C 
194  &$      3.2     \pm     1.0     $&$     1.9     \pm     0.6     $&$ 
   0.45    \pm     0.12    $&$     0.38    \pm     0.10    $\\
\object{3C 267} 
&$      1.9     \pm     0.6     $&$     3.3     \pm     1.0     $&$ 
0.26    \pm     0.07    $&$     0.65    \pm
0.18    $\\
3C 
267  &$      1.9     \pm     0.6     $&$     3.2     \pm     0.9     $&$ 
   0.27    \pm     0.07    $&$     0.63    \pm     0.17    $\\
\object{3C 356} 
&$      1.8     \pm     0.6     $&$     3.4     \pm     1.1     $&$ 
0.28    \pm     0.08    $&$     0.62    \pm
0.18    $\\
3C 
356  &$      1.4     \pm     0.4     $&$     4.6     \pm     1.5     $&$ 
   0.19    \pm     0.06    $&$     0.88    \pm     0.25    $\\
\object{3C 280} 
&$      2.3     \pm     0.7     $&$     2.7     \pm     0.8     $&$ 
0.31    \pm     0.08    $&$     0.54    \pm
0.14    $\\
3C 
280  &$      4.0     \pm     1.2     $&$     1.6     \pm     0.4     $&$ 
   0.58    \pm     0.15    $&$     0.29    \pm     0.08    $\\
\object{3C 
268.1}       &$      2.7     \pm     0.8     $&$     2.3     \pm     0.7 
   $&$     0.40    \pm     0.11    $&$     0.43    \pm
0.11    $\\
3C 
268.1        &$      1.8     \pm     0.5     $&$     3.5     \pm     1.0 
   $&$     0.25    \pm     0.06    $&$     0.69    \pm     0.18    $\\
\object{3C 289} 
&$      3.1     \pm     0.9     $&$     2.0     \pm     0.6     $&$ 
0.42    \pm     0.11    $&$     0.41    \pm
0.10    $\\
3C 
289  &$      2.7     \pm     0.8     $&$     2.3     \pm     0.7     $&$ 
   0.36    \pm     0.09    $&$     0.47    \pm     0.12    $\\
\object{3C 325} 
&$      5.1     \pm     2.0     $&$     1.2     \pm     0.5     $&$ 
0.72    \pm     0.24    $&$     0.24    \pm
0.08    $\\
3C 
325  &$      1.6     \pm     0.5     $&$     4.0     \pm     1.3     $&$ 
   0.21    \pm     0.06    $&$     0.82    \pm     0.24    $\\
\object{3C 265} 
&$      2.7     \pm     0.8     $&$     2.3     \pm     0.7     $&$ 
0.37    \pm     0.10    $&$     0.46    \pm
0.12    $\\
3C 
265  &$      4.1     \pm     1.3     $&$     1.5     \pm     0.5     $&$ 
   0.62    \pm     0.17    $&$     0.27    \pm     0.08    $\\
\object{3C 247} 
&$      3.4     \pm     1.0     $&$     1.8     \pm     0.5     $&$ 
0.47    \pm     0.13    $&$     0.36    \pm
0.10    $\\
3C 
247  &$      4.8     \pm     1.4     $&$     1.3     \pm     0.4     $&$ 
   0.65    \pm     0.17    $&$     0.26    \pm     0.07    $\\
\object{3C 
55}  &$      1.9     \pm     0.6     $&$     3.2     \pm     0.9     $&$ 
   0.27    \pm     0.07    $&$     0.64    \pm
0.17    $\\
3C 
55   &$      2.1     \pm     0.6     $&$     3.0     \pm     0.9     $&$ 
   0.29    \pm     0.08    $&$     0.60    \pm     0.17    $\\
\object{3C 337} 
&$      6.6     \pm     2.1     $&$     0.9     \pm     0.3     $&$ 
1.05    \pm     0.30    $&$     0.16    \pm
0.05    $\\
3C 
337  &$      4.0     \pm     1.1     $&$     1.6     \pm     0.4     $&$ 
   0.61    \pm     0.15    $&$     0.28    \pm     0.07    $\\
\object{3C 
427.1}       &$      4.5     \pm     1.3     $&$     1.4     \pm     0.4 
   $&$     0.62    \pm     0.16    $&$     0.28    \pm
0.07    $\\
3C 
427.1        &$      5.0     \pm     1.4     $&$     1.2     \pm     0.4 
   $&$     0.68    \pm     0.18    $&$     0.25    \pm     0.07    $\\
\object{3C 330} 
&$      2.8     \pm     0.8     $&$     2.3     \pm     0.7     $&$ 
0.38    \pm     0.10    $&$     0.45    \pm
0.12    $\\
3C 
330  &$      3.1     \pm     0.9     $&$     2.0     \pm     0.6     $&$ 
   0.44    \pm     0.12    $&$     0.39    \pm     0.10    $\\
\object{3C 
244.1}       &$      8.3     \pm     2.3     $&$     0.8     \pm     0.2 
   $&$     1.24    \pm     0.31    $&$     0.14    \pm
0.03    $\\
3C 
244.1        &$      5.0     \pm     1.4     $&$     1.2     \pm     0.3 
   $&$     0.70    \pm     0.17    $&$     0.24    \pm     0.06    $\\
\object{3C 405} 
&$      3.7     \pm     1.0     $&$     1.7     \pm     0.4     $&$ 
0.48    \pm     0.12    $&$     0.36    \pm
0.09    $\\
3C 
405  &$      4.1     \pm     1.1     $&$     1.5     \pm     0.4     $&$ 
   0.52    \pm     0.13    $&$     0.33    \pm     0.08    $\\
    &        &           &
&            \\ 

\object{3C 6.1} 
&$      3.1     \pm     0.8     $&$     2.0     \pm     0.5     $&$ 
0.50    \pm     0.12    $&$     0.34    \pm
0.08    $\\
3C 
6.1  &$      2.4     \pm     0.6     $&$     2.7     \pm     0.7     $&$ 
   0.42    \pm     0.10    $&$     0.40    \pm     0.10    $\\
\object{3C 
34}  &$      2.4     \pm     0.6     $&$     2.6     \pm     0.7     $&$ 
   0.46    \pm     0.11    $&$     0.37    \pm
0.09    $\\
3C 
34   &$      3.6     \pm     0.9     $&$     1.7     \pm     0.4     $&$ 
   0.71    \pm     0.16    $&$     0.24    \pm     0.06    $\\
\object{3C 
41}  &$      4.6     \pm     1.2     $&$     1.4     \pm     0.4     $&$ 
   0.91    \pm     0.22    $&$     0.19    \pm
0.05    $\\
3C 
41   &$      3.0     \pm     0.7     $&$     2.1     \pm     0.5     $&$ 
   0.49    \pm     0.12    $&$     0.35    \pm     0.08    $\\
\object{3C 
44}  &$      3.6     \pm     0.9     $&$     1.7     \pm     0.4     $&$ 
   0.58    \pm     0.14    $&$     0.29    \pm
0.07    $\\
3C 
44   &$      3.6     \pm     0.9     $&$     1.8     \pm     0.4     $&$ 
   0.58    \pm     0.14    $&$     0.29    \pm     0.07    $\\
\object{3C 
54}  &$      2.3     \pm     0.6     $&$     2.8     \pm     0.7     $&$ 
   0.45    \pm     0.11    $&$     0.38    \pm
0.09    $\\
3C 
54   &$      2.7     \pm     0.7     $&$     2.3     \pm     0.6     $&$ 
   0.44    \pm     0.10    $&$     0.38    \pm     0.09    $\\
\object{3C 114} 
&$      2.2     \pm     0.6     $&$     2.8     \pm     0.7     $&$ 
0.47    \pm     0.11    $&$     0.36    \pm
0.08    $\\
3C 
114  &$      2.3     \pm     0.6     $&$     2.8     \pm     0.7     $&$ 
   0.48    \pm     0.11    $&$     0.35    \pm     0.08    $\\
\object{3C 
142.1}       &$      5.2     \pm     1.2     $&$     1.2     \pm     0.3 
   $&$     0.76    \pm     0.18    $&$     0.23    \pm
0.05    $\\
3C 
142.1        &$      3.6     \pm     0.9     $&$     1.7     \pm     0.4 
   $&$     0.62    \pm     0.15    $&$     0.28    \pm     0.07    $\\
\object{3C 
169.1}       &$      3.3     \pm     0.9     $&$     1.9     \pm     0.5 
   $&$     0.75    \pm     0.18    $&$     0.23    \pm
0.05    $\\
3C 
169.1        &$      3.3     \pm     0.8     $&$     1.9     \pm     0.5 
   $&$     0.64    \pm     0.15    $&$     0.27    \pm     0.06    $\\
\object{3C 172} 
&$      4.9     \pm     1.3     $&$     1.3     \pm     0.3     $&$ 
0.85    \pm     0.21    $&$     0.20    \pm
0.05    $\\
3C 
172  &$      1.5     \pm     0.5     $&$     4.2     \pm     1.5     $&$ 
   0.28    \pm     0.09    $&$     0.62    \pm     0.21    $\\
\object{3C 441} 
&$      3.6     \pm     0.9     $&$     1.8     \pm     0.4     $&$ 
0.71    \pm     0.17    $&$     0.24    \pm
0.06    $\\
3C 
441  &$      2.0     \pm     0.5     $&$     3.1     \pm     0.8     $&$ 
   0.35    \pm     0.09    $&$     0.49    \pm     0.12    $\\
\object{3C 
469.1}       &$      1.0     \pm     0.3     $&$     6.3     \pm     1.6 
   $&$     0.17    \pm     0.04    $&$     0.97    \pm
0.24    $\\
3C 
469.1        &$      0.9     \pm     0.2     $&$     7.2     \pm     1.8 
   $&$     0.16    \pm     0.04    $&$     1.04    \pm     0.25    $\\
\end{longtable}

The top part of the table contains the properties of the previous sample of 20
radio sources recalculated for the
currently adopted cosmology.  The bottom part of the table contains the new
sample of 11 radio galaxies.
Col 1. Source Name.
Col 2. The total lifetime of the outflow (b=0.25).
Col 3. The total energy of the outflow (b=0.25).
Col 4. The total lifetime of the outflow (b = 1).
Col 5. The total energy of the outflow (b = 1).

\begin{table*}
\caption{Fits to the Data       (b = 0.25) }             
\label{tabfits}      
\begin{tabular}{lrrrr}        
\hline\hline                 
Relationship & Slope & Y intercept & $\chi^2$ & Corr. ? \\
\hline
$T$ vs 1 + 
z            &$      -2.00   \pm     0.2             $&$     6.1     \pm 
   0.5     $&      90.6    &Y \\
Log $T$ vs Log (1 + 
z)  &$      -1.80   \pm     0.2             $&$     0.9     \pm     0.06 
  $&      100.8   &Y \\
$T$ vs $r$              &$      -4.00   \pm     1.4     \times 10^{-3}$ 
&$      2.5     \pm     0.3     $&      181.6   &N
\\
Log $T$ vs Log 
$r$      &$      -0.25   \pm     0.11            $&$     1       \pm 
0.2     $&      183     &N \\
$E_T$ vs 1 + 
z          &$      1.70    \pm     0.3             $&$     -1.1    \pm 
0.5     $&      92.8    &Y \\
Log $E_T$ vs Log (1 + 
z)        &$      1.80    \pm     0.2             $&$     -0.1    \pm 
0.06    $&      100.8   &Y  \\
$E_T$ vs $r$            &$      1.40    \pm     1.4     \times 10^{-3}$ 
&$      1.5     \pm     0.2     $&      138     &N\\
Log $E_T$ vs Log 
$r$            &$      0.25    \pm     0.11            $&$     -0.2    \pm 
     0.2     $&      183     &N \\
$n_a$ vs. $r$           &$      -1.90   \pm     0.9     \times 10^{-4}$ 
&$      0.09    \pm     0.04    $&
347.3   &N\\
Log $n_a$ vs.  Log 
$r$          &$      -1.90   \pm     0.2             $&$     3.4     \pm 
   0.5     $&      561.9   &Y \\
Log $n_a$ vs.  Log (1 + 
z)      &$      -1.90   \pm     1.1             $&$     -0.3    \pm 
0.3     $&      1213.6  &N \\
v/c vs.        $r$      &$      1.30    \pm     0.3     \times 10^{-4}$ 
&$      0.006   \pm     0.003   $&
741.1   &N \\
Log v/c vs. Log 
$r$             &$      0.70    \pm     0.1             $&$     -3 
\pm     0.2     $&      740.9   &N \\
v/c vs. 1 + 
z           &$      0.02    \pm     0.004           $&$     -0.01   \pm 
  0.006   $&      747.5   &N\\
Log v/c vs. Log (1 + 
z)         &$      1.50    \pm     0.4             $&$     -1.88   \pm 
0.1     $&      966.9   &N \\
$L_j$ vs. 
$r$           &$      0.01    \pm     0.03            $&$     16      \pm 
    5       $&      364.8   &N \\
Log $L_j$ vs. Log 
$r$           &$      0.60    \pm     0.2             $&$     0.3     \pm 
    0.5     $&      710.6   &N \\
$L_j$ vs. 1 + 
z                 &$      36.50   \pm     9               $&$     -38 
\pm     13      $&      280.1   &N \\
Log $L_j$ vs Log (1 + 
z)        &$      3.80    \pm     0.5             $&$     0.72    \pm 
0.11    $&      361.4   &Y \\
\hline
\end{tabular}
\\
The best fits to the radio source relationships for b = 0.25.
Col 1. The relationship. v/c is the lobe expansion velocity, $r$ is the
core-hotspot separation, $L_j$ ($\times 10^{44}$ ergs s\mone) is the
beam power, $n_a$ ($\times 10^{-3}$ cm \mthree) is the ambient density,
$T_T$ is the source lifetime, and $E_T$ is the total energy.
Col 2. The best fit slope. Col 3. The best fit Y intercept. Col 4. The 
$\chi^2$
for 60 points and 58 degrees of freedom.
Col 5. Is there a correlation? The uncertainty of the best fit
parameter is multiplied by $\sqrt{(\chi^2/58)}$ to obtain the
normalized uncertainty in the best fit parameter,and this is
compared with the value of the best fit parameter to determine
whether the correlation is significant.
\end{table*}


\begin{table*}
\caption{Fits to the Data       (b = 1) }             
\label{tabfitsb1}      
\begin{tabular}{lrrrr}        
\hline\hline                 
Relationship & Slope & Y intercept & $\chi^2$ & Corr. ? \\
\hline
$T$ vs 1 + 
z            &$      -3.20   \pm     0.4             $&$     9.5     \pm 
   0.8     $&      113.9   &Y \\
Log $T$ vs Log (1 + 
z)  &$      -1.90   \pm     0.3             $&$     1.12    \pm     0.07 
  $&      131.5   &Y     \\
$T$ vs $r$              &$      -3.40   \pm     2.5     \times 10^{-3}$ 
&$      3.1     \pm     0.5     $&      255     &N \\
Log $T$ vs Log 
$r$      &$      -0.15   \pm     0.12            $&$     1       \pm 
0.3     $&      253.1   &N \\
$E_T$ vs 1 + 
z          &$      3.50    \pm     0.6             $&$     -2.7    \pm 
1       $&      113.7   &Y \\
Log $E_T$ vs Log (1 + 
z)        &$      1.90    \pm     0.3             $&$     0.11    \pm 
0.07    $&      131.5   &Y \\
$E_T$ vs $r$            &$      1.00    \pm     3       \times 10^{-3}$ 
&$      2.7     \pm     0.4     $&      171.3   &N
\\
Log $E_T$ vs Log 
$r$            &$      0.15    \pm     0.12            $&$     0.3     \pm 
     0.3     $&      253.1   &N     \\
$n_a$ vs. $r$           &$      -5.00   \pm     1       \times 10^{-6}$ 
&$      0.002   \pm     0.003   $&
300.9
  &N \\
Log $n_a$ vs.  Log 
$r$          &$      -1.38   \pm     0.14            $&$     0.3     \pm 
   0.3     $&      328.3   &Y \\
Log $n_a$ vs.  Log (1 + 
z)      &$      -1.60   \pm     0.6             $&$     -2.27   \pm 
0.17    $&      760.3   &N \\
v/c vs.        $r$      &$      7.10    \pm     1.1     \times 10^{-4}$ 
&$      0.02    \pm     0.01    $&
572.8
  &N \\
Log v/c vs. Log 
$r$     &$      0.54    \pm     0.12            $&$     -1.9    \pm 
0.3     $&      827.6   &N \\
v/c vs. 1 + 
z           &$      0.10    \pm     0.03            $&$     -0.07   \pm 
  0.05    $&      823.7   &N \\
Log v/c vs. Log (1 + 
z)         &$      1.80    \pm     0.3             $&$     -1.24   \pm 
0.09    $&      771.5   &N
  \\
$L_j$ vs. 
$r$           &$      0.04    \pm     0.03            $&$     15      \pm 
    5       $&      873.8   &N \\
Log $L_j$ vs. Log 
$r$           &$      0.50    \pm     0.2             $&$     0.4     \pm 
    0.5     $&      2072    &N \\
$L_j$ vs. 1 + 
z         &$      44.00   \pm     10              $&$     -50     \pm 
17      $&      690.5   &N \\
Log $L_j$ vs Log (1 + 
z)        &$      4.70    \pm     0.5             $&$     0.44    \pm 
0.13    $&      933.1   &N \\
\hline                                   
\end{tabular}
\\
The best fits to the radio source relationships for b = 1. Col 1
The relationship. v/c is the lobe expansion velocity, $r$ is the core-hotspot
separation, $L_j$ ($\times 10^{44}$ ergs s\mone) is the beam power, $n_a$
($\times 10^{-3}$ cm \mthree) is the ambient density, $T$ is the source
lifetime,
and $E_T$ is the total energy.  Col 2. The best fit slope. Col 3.
The best fit Y
intercept.
Col 4. The $\chi^2$ for 60 points and 58 degrees of freedom.
Col 5. Is there a correlation?
The uncertainty of the best fit
parameter is multiplied by $\sqrt{(\chi^2/58)}$ to obtain the
normalized uncertainty in the best fit parameter,and this is
compared with the value of the best fit parameter to determine
whether the correlation is significant.
\end{table*}

\clearpage


\begin{figure}
    \centering
    \includegraphics[width=\textwidth]{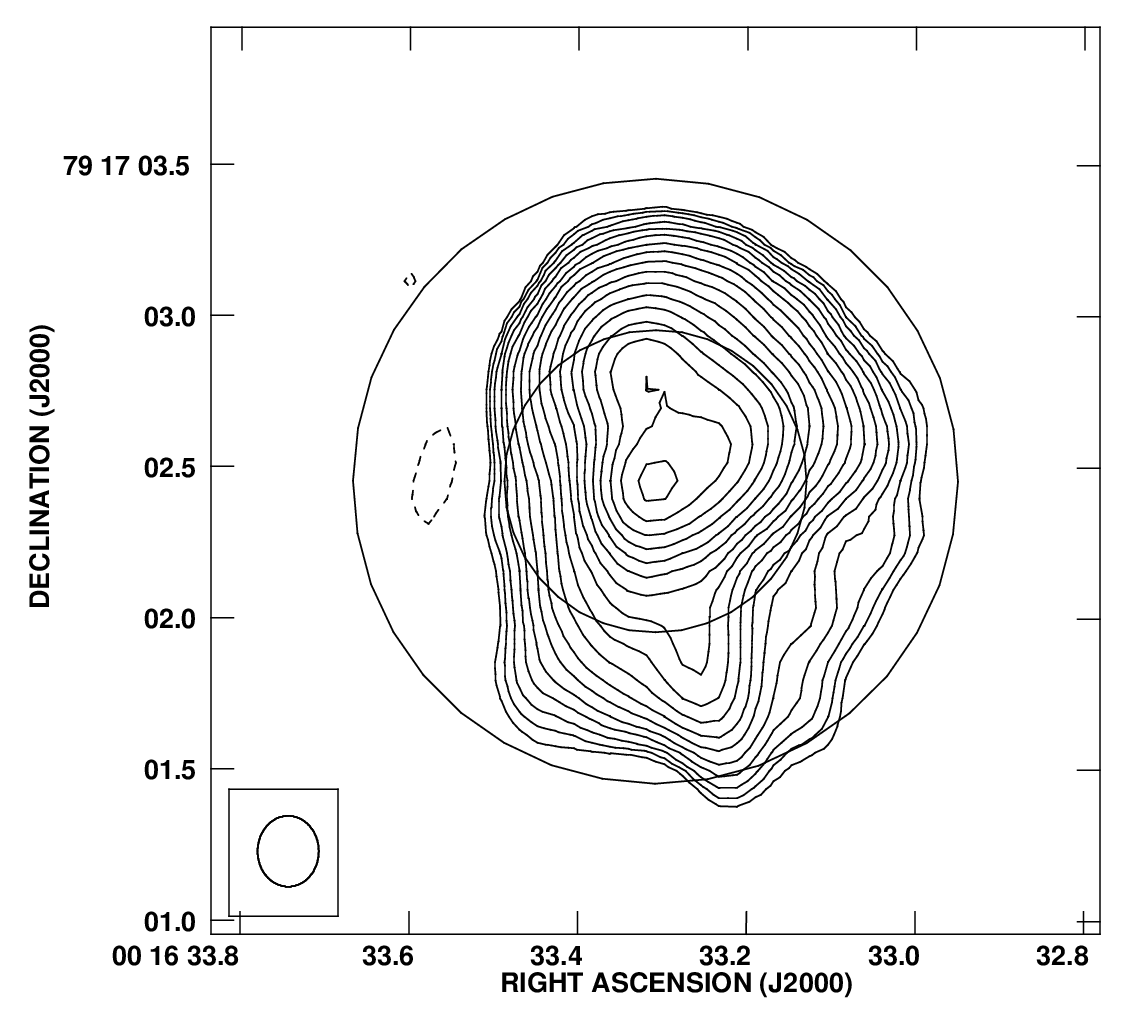}
	\caption{3C6.1 north hot spot. 8.4 GHz VLA image from Kharb et al. (2008) 
with CLEAN restoring beam FWHM $0.23 \times 0.20$ arcsec 
(shown in the inset in the lower left). 
Two circles are shown centered on the brightness peak. The larger circle has a diameter 
of 2.0 arcsec, while the smaller circle has a diameter of 1.0 arcsec. 
}
	  \label{fig3c61_N}
    \end{figure}

\begin{figure}
    \centering
    \includegraphics[width=\textwidth]{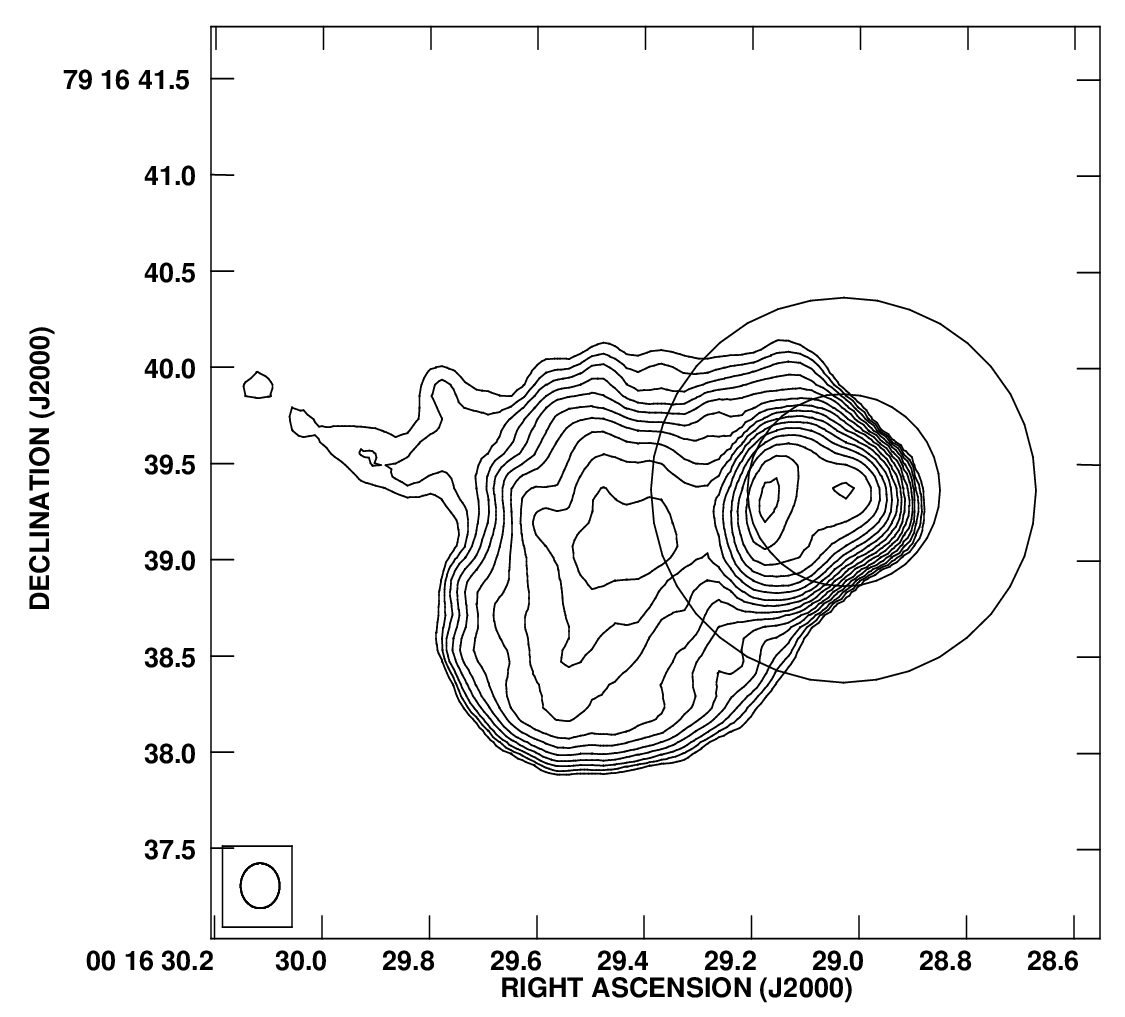} 
		 \caption{3C6.1 south hot spot. 8.4 GHz VLA image from Kharb et al. (2008)
with CLEAN restoring beam FWHM $0.23 \times 0.20$ arcsec 
(shown in the inset in the lower left).
Two circles are shown centered on the brightness peak. The larger circle has a diameter
of 2.0 arcsec, while the smaller circle has a diameter of 1.0 arcsec.
}
	  \label{fig3c61_S}
    \end{figure}

\begin{figure}
    \centering
    \includegraphics[width=\textwidth]{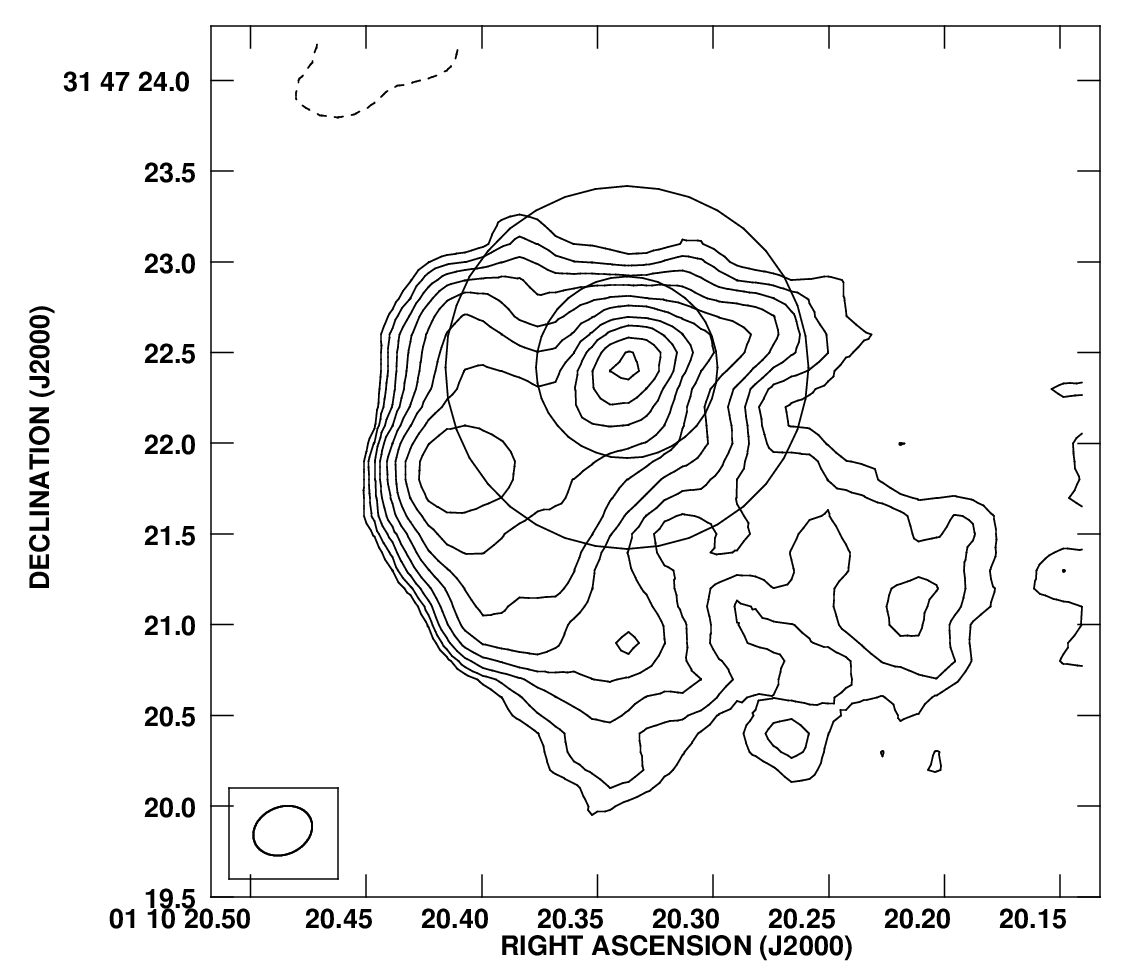}
		\caption{3C34 east  hot spot. 8.4 GHz VLA image from Kharb et al. (2008)
with CLEAN restoring beam FWHM $0.33 \times 0.26$ arcsec
(shown in the inset in the lower left).
Two circles are shown centered on the brightness peak. The larger circle has a diameter
of 2.0 arcsec, while the smaller circle has a diameter of 1.0 arcsec.
}
		  \label{fig3c34_E}
    \end{figure}

\begin{figure}
    \centering
    \includegraphics[width=\textwidth]{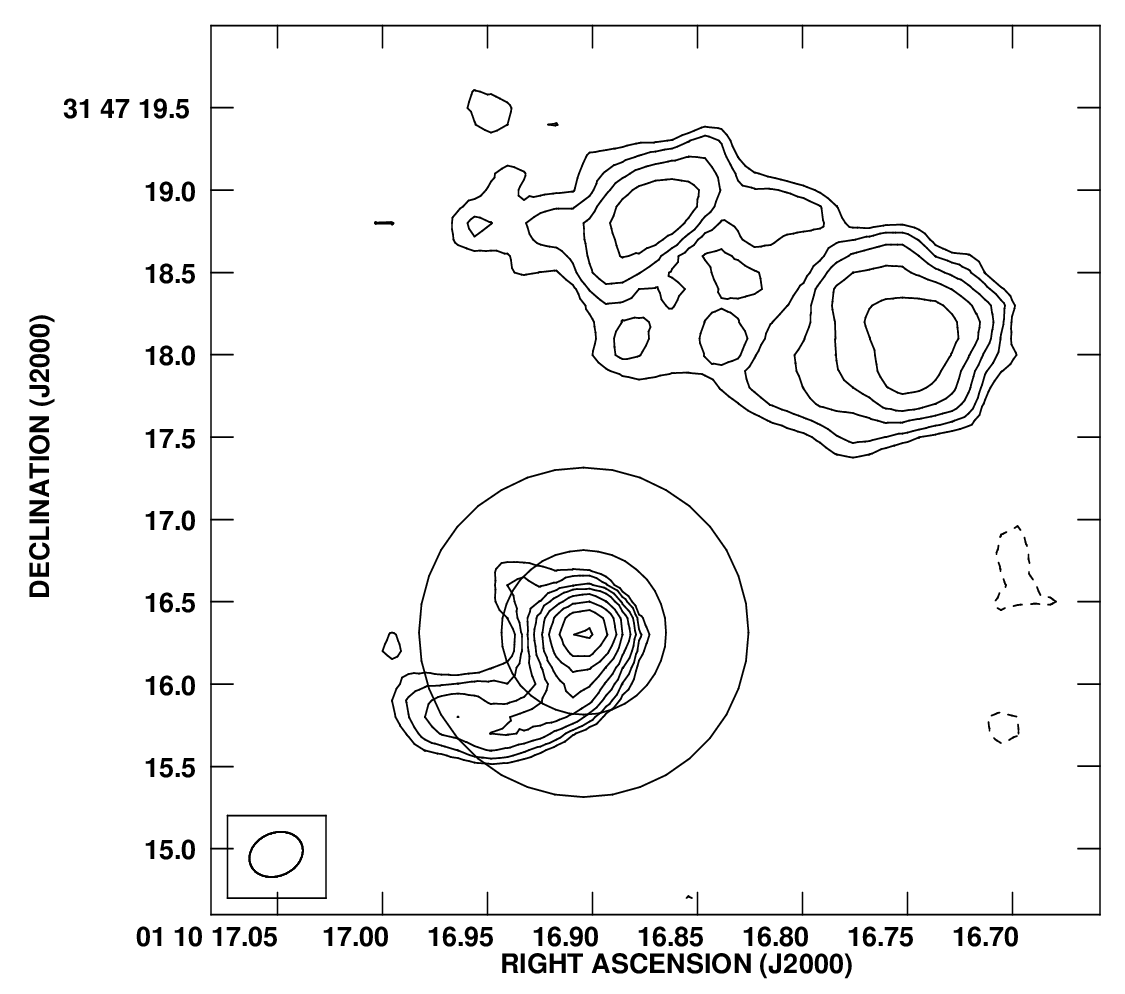}
			 \caption{3C34 west hot spot. 8.4 GHz VLA image from Kharb et al. (2008)
with CLEAN restoring beam FWHM $0.33 \times 0.26$ arcsec
(shown in the inset in the lower left).
Two circles are shown centered on the brightness peak. The larger circle has a diameter
of 2.0 arcsec, while the smaller circle has a diameter of 1.0 arcsec.
}
		  \label{fig3c34_W}
    \end{figure}

\begin{figure}
    \centering
    \includegraphics[width=\textwidth]{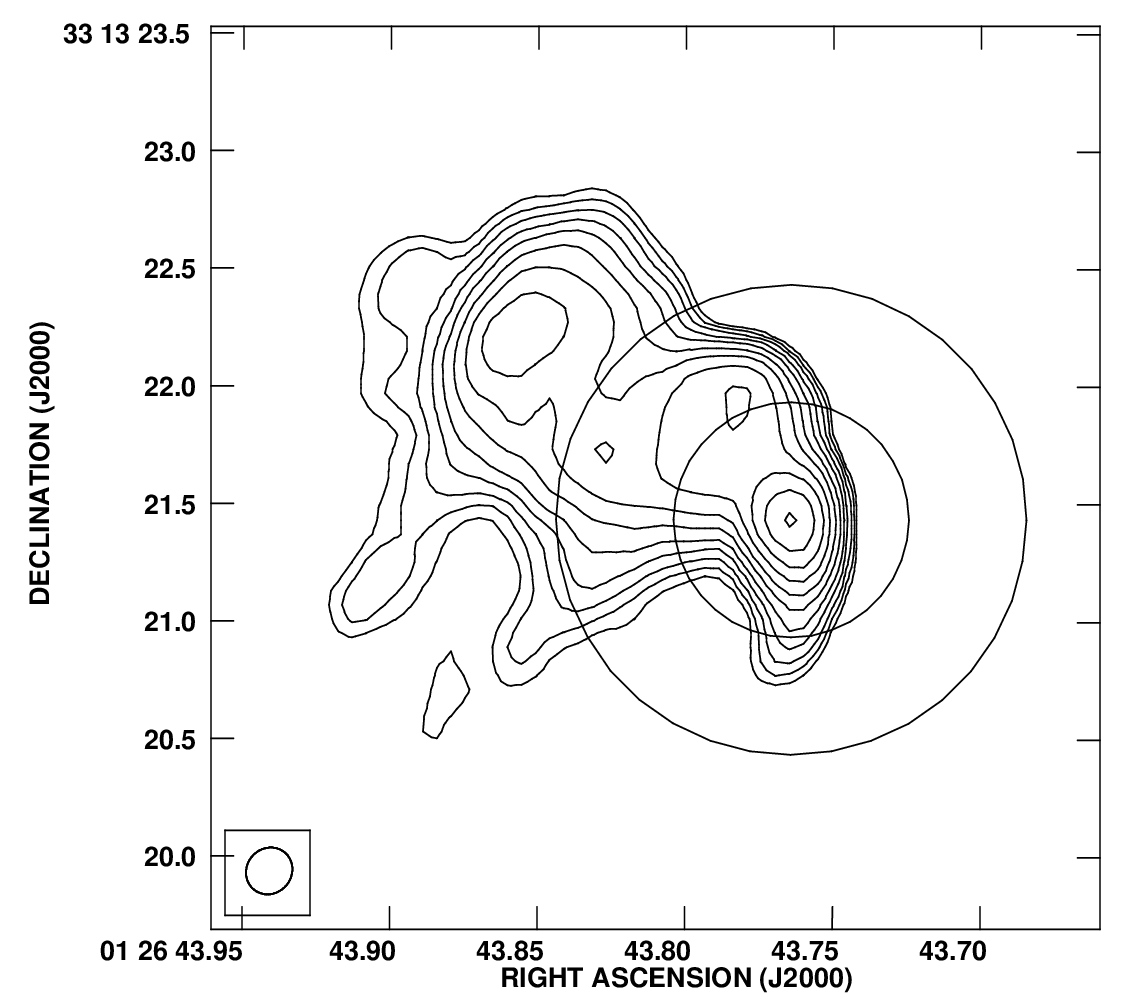}
                         \caption{3C41 north hot spot. 8.4 GHz VLA image from Kharb et al. (2008)
with CLEAN restoring beam FWHM $0.21 \times 0.19$ arcsec
(shown in the inset in the lower left).
Two circles are shown centered on the brightness peak. The larger circle has a diameter
of 2.0 arcsec, while the smaller circle has a diameter of 1.0 arcsec.
}
                  \label{fig3c41_N}
    \end{figure}

\begin{figure}
    \centering
    \includegraphics[width=\textwidth]{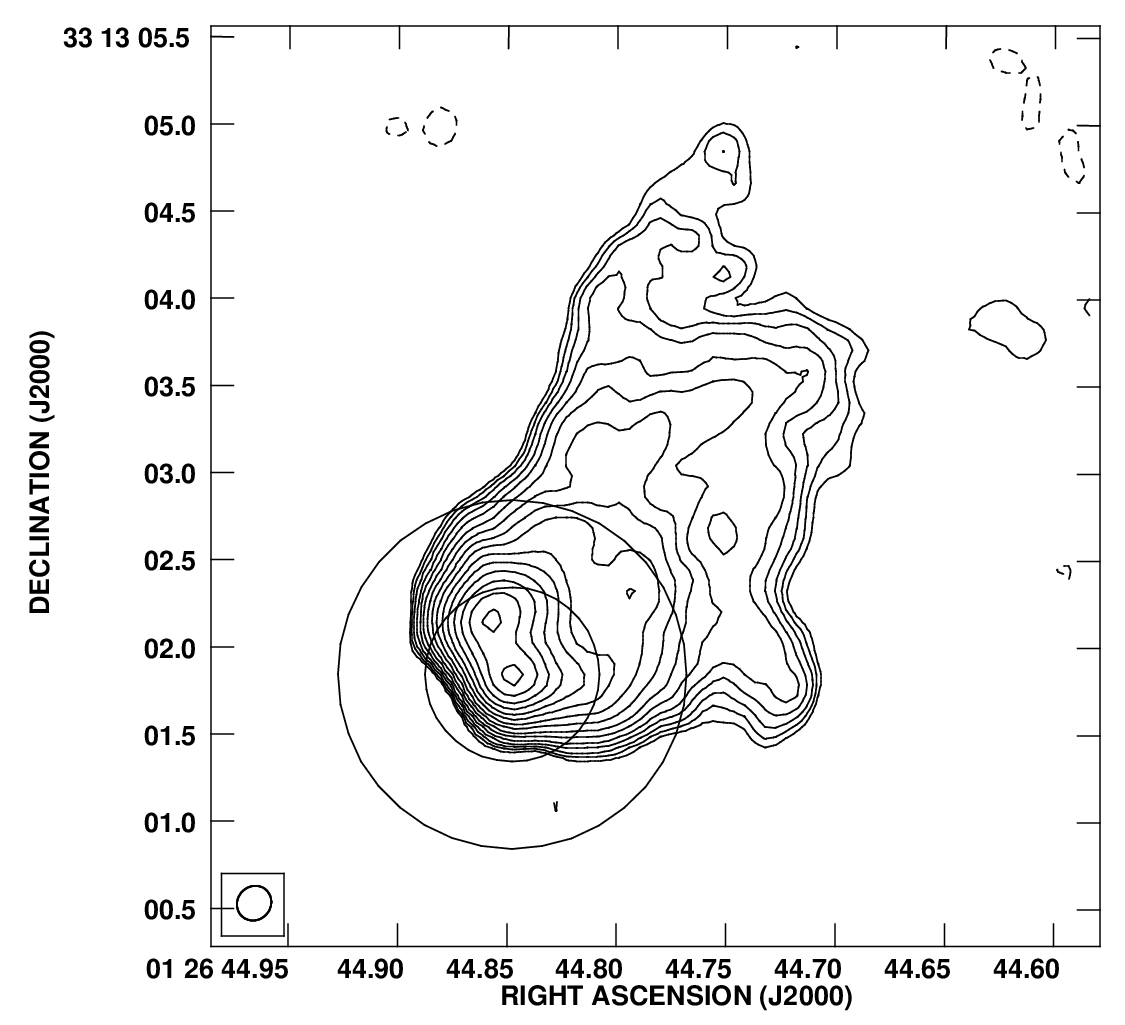}
                         \caption{3C41 south hot spot. 8.4 GHz VLA image from Kharb et al. (2008)
with CLEAN restoring beam FWHM $0.21 \times 0.19$ arcsec
(shown in the inset in the lower left).
Two circles are shown centered on the brightness peak. The larger circle has a diameter
of 2.0 arcsec, while the smaller circle has a diameter of 1.0 arcsec.
}
                  \label{fig3c41_S}
    \end{figure}

\begin{figure}
    \centering
    \includegraphics[width=\textwidth]{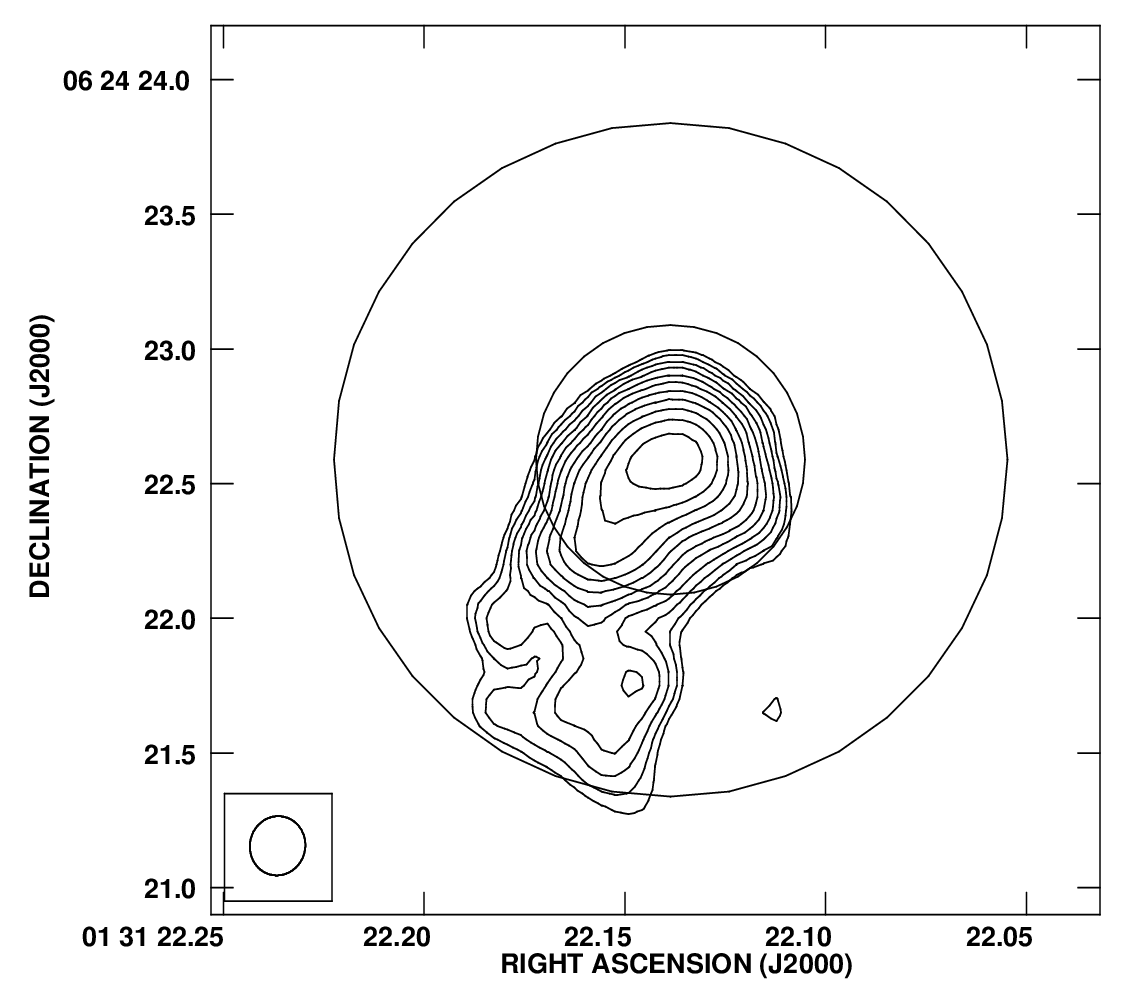}
                         \caption{3C44 north hot spot. 8.4 GHz VLA image from Kharb et al. (2008)
with CLEAN restoring beam FWHM $0.22 \times 0.21$ arcsec
(shown in the inset in the lower left).
Two circles are shown centered on the brightness peak. The larger circle has a diameter
of 2.5 arcsec, while the smaller circle has a diameter of 1.0 arcsec.
}
                  \label{fig3c44_N}
    \end{figure}

\begin{figure}
    \centering
    \includegraphics[width=\textwidth]{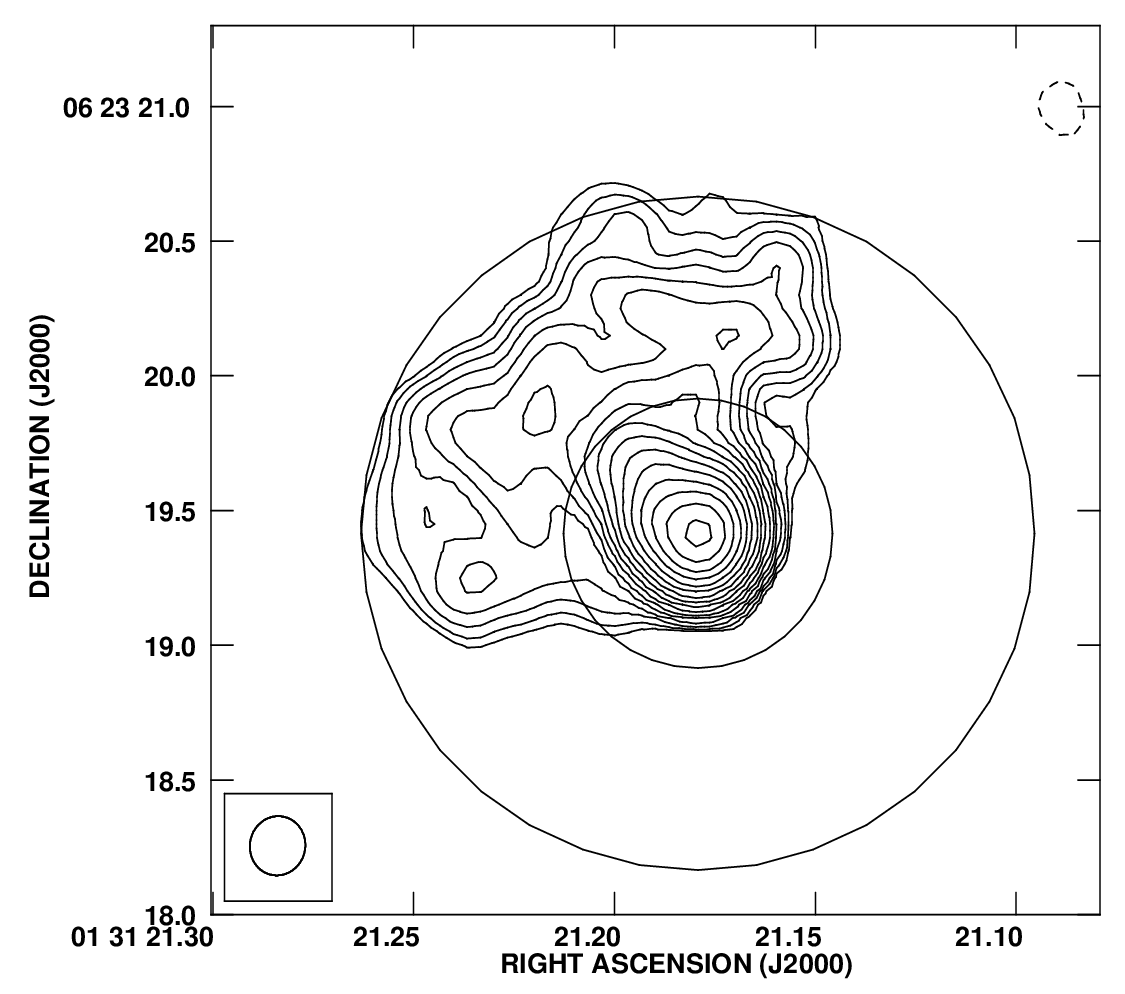}
                         \caption{3C44 south hot spot. 8.4 GHz VLA image from Kharb et al. (2008)
with CLEAN restoring beam FWHM $0.22 \times 0.21$ arcsec
(shown in the inset in the lower left).
Two circles are shown centered on the brightness peak. The larger circle has a diameter
of 2.5 arcsec, while the smaller circle has a diameter of 1.0 arcsec.
}
                  \label{fig3c44_S}
    \end{figure}

\begin{figure}
    \centering
    \includegraphics[width=\textwidth]{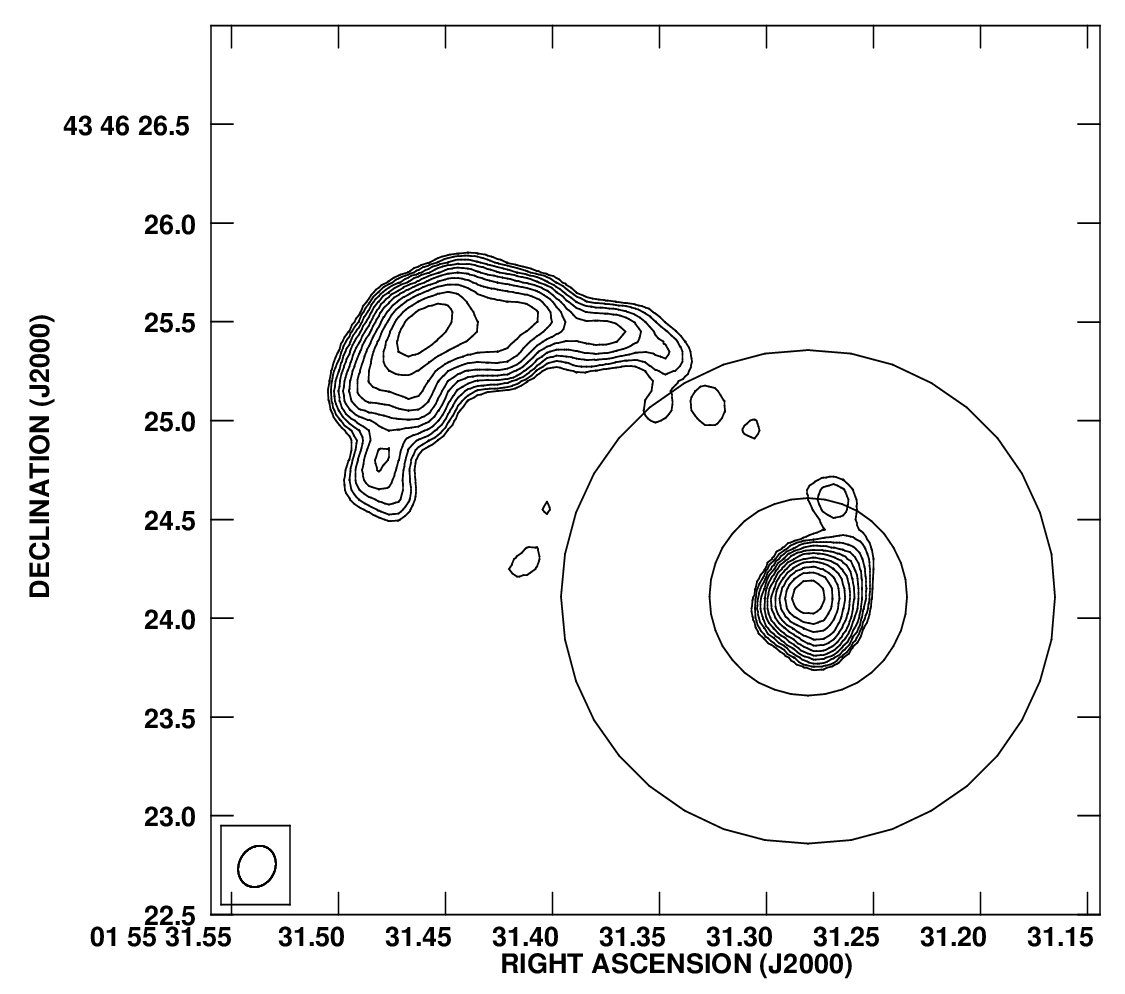}
			 \caption{3C54 north hot spot. 8.4 GHz VLA image from Kharb et al. (2008)
with CLEAN restoring beam FWHM $0.22 \times 0.18$ arcsec
(shown in the inset in the lower left).
Two circles are shown centered on the brightness peak. The larger circle has a diameter
of 2.5 arcsec, while the smaller circle has a diameter of 1.0 arcsec.
}
 		  \label{fig3c54_N}
    \end{figure}

\begin{figure}
    \centering
    \includegraphics[width=\textwidth]{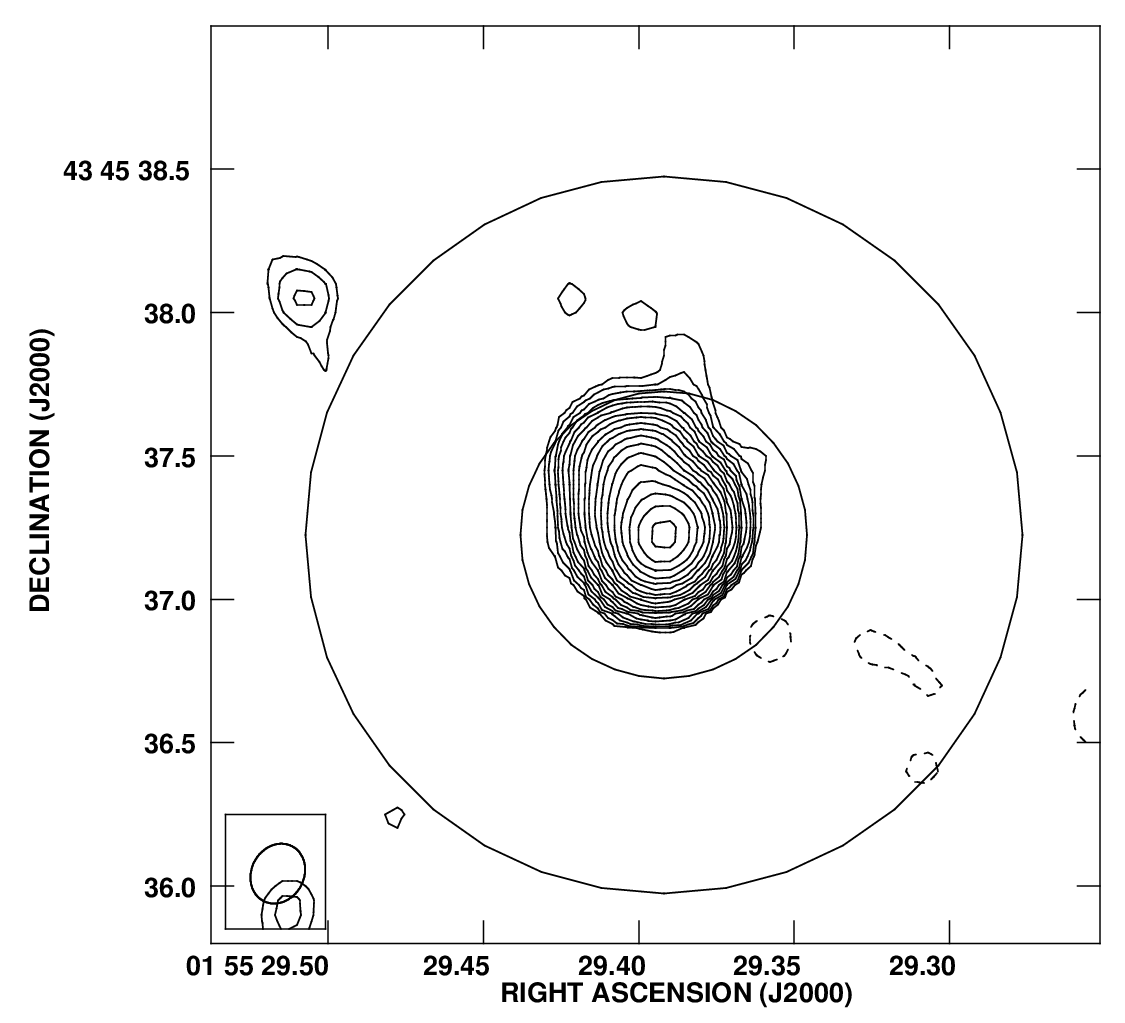}
                         \caption{3C54 south hot spot. 8.4 GHz VLA image from Kharb et al. (2008)
with CLEAN restoring beam FWHM $0.22 \times 0.18$ arcsec
(shown in the inset in the lower left).
Two circles are shown centered on the brightness peak. The larger circle has a diameter
of 2.5 arcsec, while the smaller circle has a diameter of 1.0 arcsec.
}
                  \label{fig3c54_S}
    \end{figure}

\begin{figure}
    \centering
    \includegraphics[width=\textwidth]{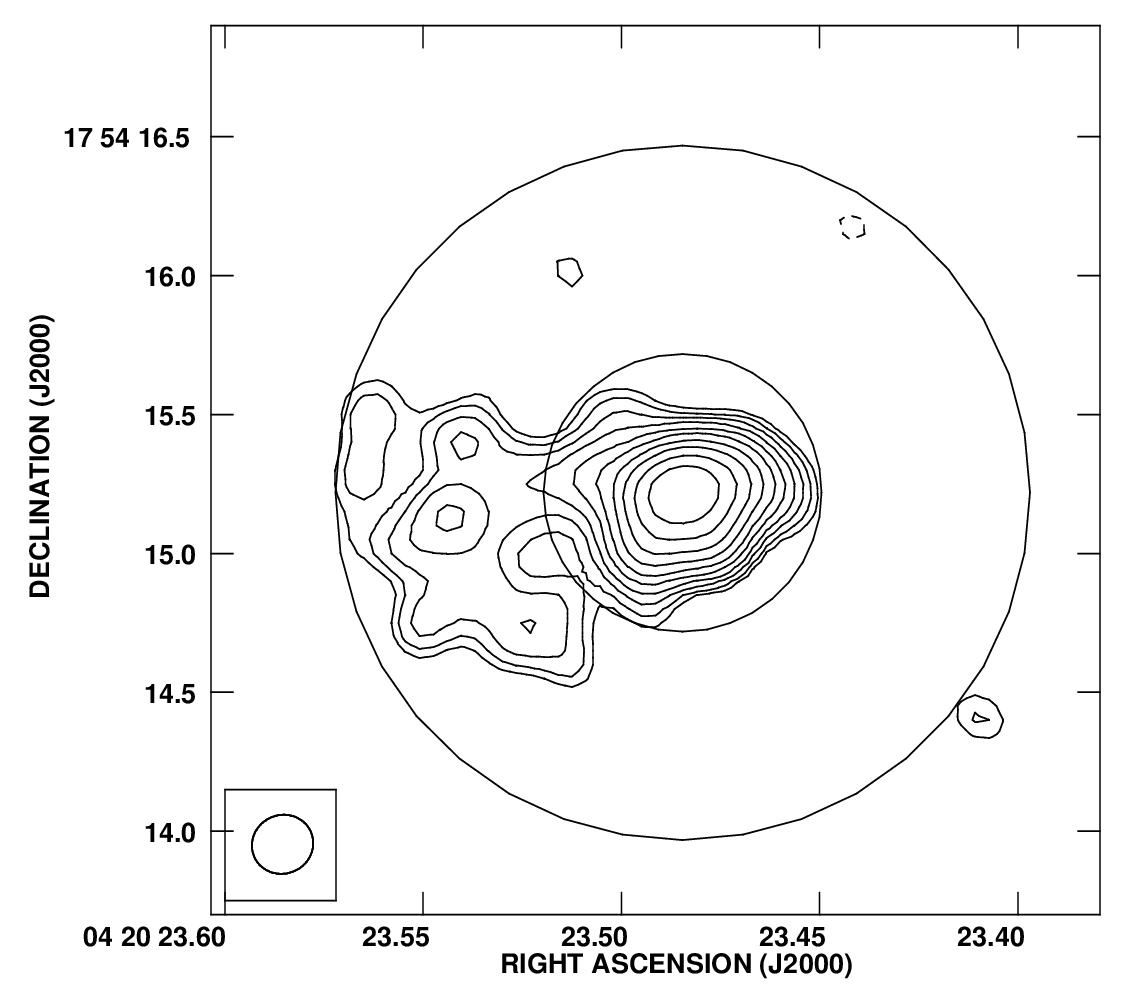}
                         \caption{3C114 north hot spot. 8.4 GHz VLA image from Kharb et al. (2008)
with CLEAN restoring beam FWHM $0.22 \times 0.21$ arcsec
(shown in the inset in the lower left).
Two circles are shown centered on the brightness peak. The larger circle has a diameter
of 2.5 arcsec, while the smaller circle has a diameter of 1.0 arcsec.
}
                  \label{fig3c114_N}
    \end{figure}

\begin{figure}
    \centering
    \includegraphics[width=\textwidth]{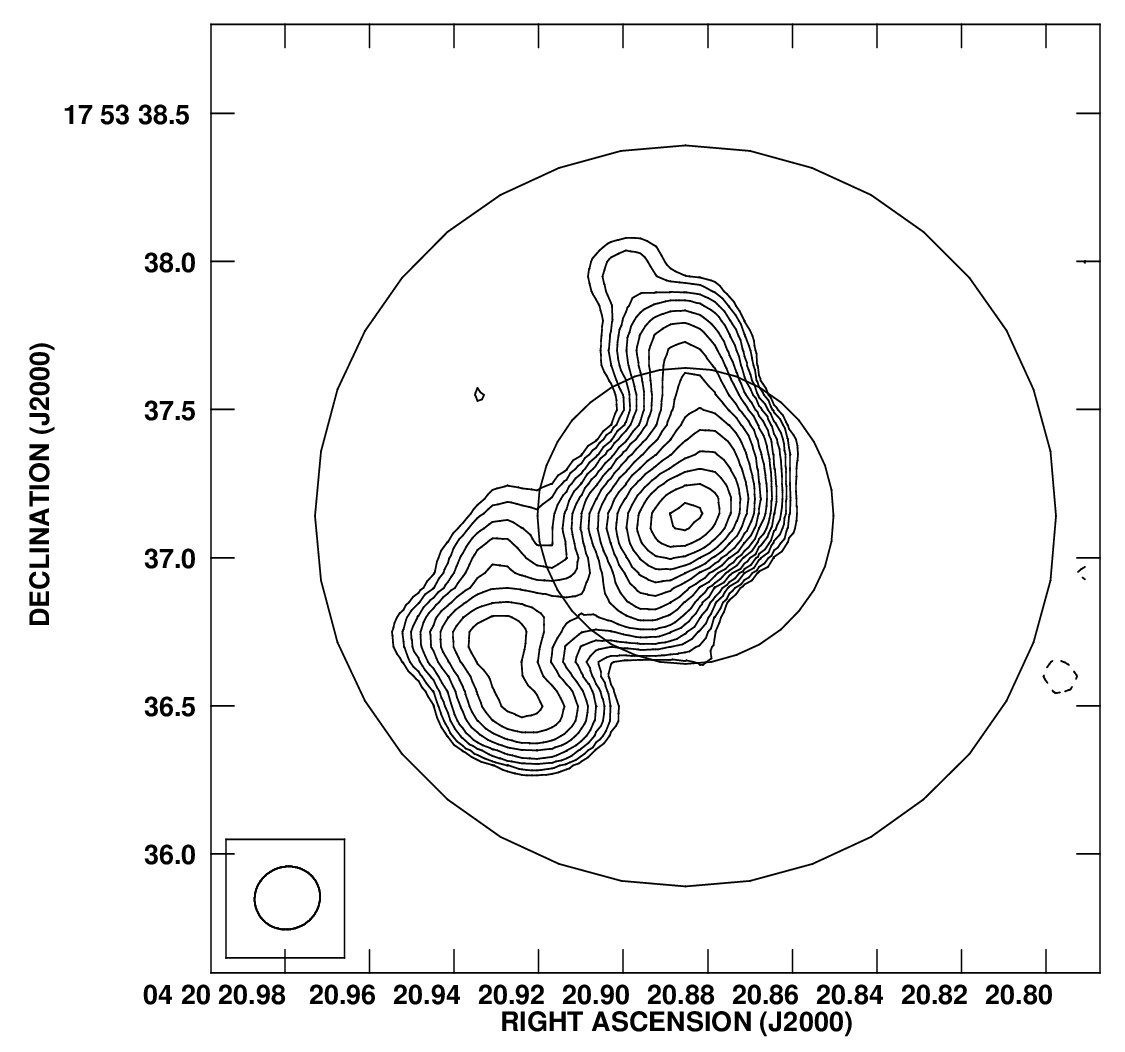}
                         \caption{3C114 south hot spot. 8.4 GHz VLA image from Kharb et al. (2008)
with CLEAN restoring beam FWHM $0.22 \times 0.21$ arcsec
(shown in the inset in the lower left).
Two circles are shown centered on the brightness peak. The larger circle has a diameter
of 2.5 arcsec, while the smaller circle has a diameter of 1.0 arcsec.
}
                  \label{fig3c114_S}
    \end{figure}

\begin{figure}
    \centering
    \includegraphics[width=\textwidth]{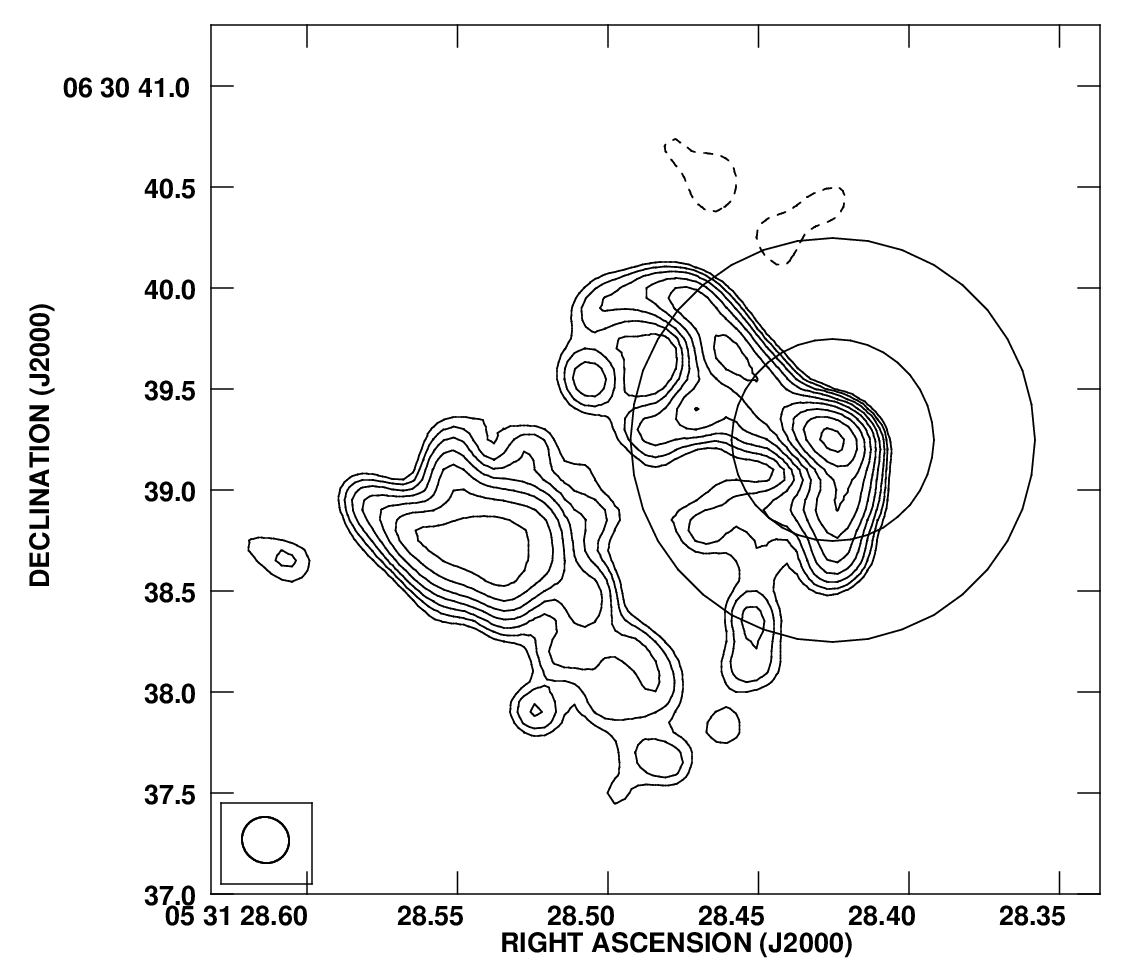}
			 \caption{3C142.1 north hot spot. 8.4 GHz VLA image from Kharb et al. (2008)
with CLEAN restoring beam FWHM $0.24 \times 0.23$ arcsec
(shown in the inset in the lower left).
Two circles are shown centered on the brightness peak. The larger circle has a diameter
of 2.0 arcsec, while the smaller circle has a diameter of 1.0 arcsec.
}
		  \label{fig3c142_N}
    \end{figure}

\begin{figure}
    \centering
    \includegraphics[width=\textwidth]{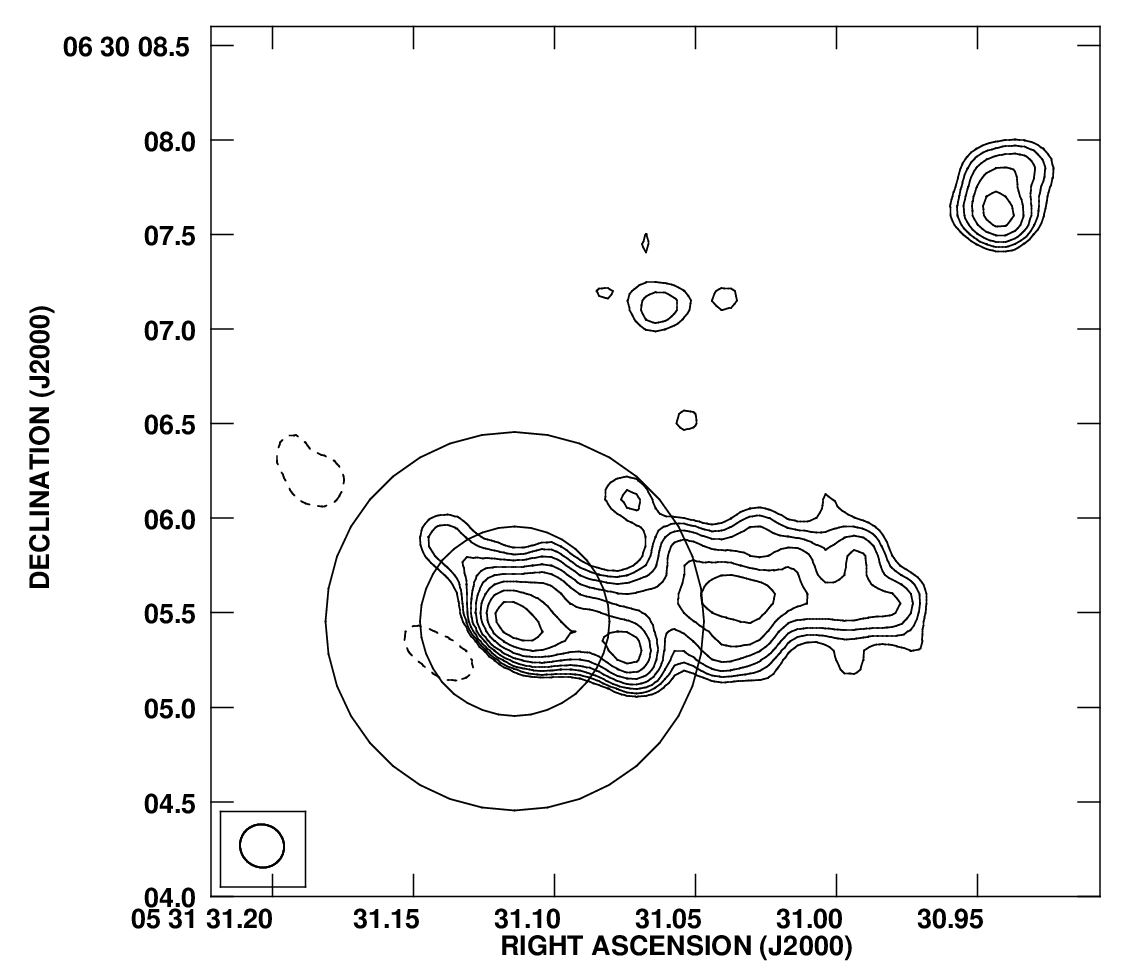}
                         \caption{3C142.1 south hot spot. 8.4 GHz VLA image from Kharb et al. (2008)
with CLEAN restoring beam FWHM $0.24 \times 0.23$ arcsec
(shown in the inset in the lower left).
Two circles are shown centered on the brightness peak. The larger circle has a diameter
of 2.0 arcsec, while the smaller circle has a diameter of 1.0 arcsec.
}
                  \label{fig3c142_S}
    \end{figure}

\begin{figure}
    \centering
    \includegraphics[width=\textwidth]{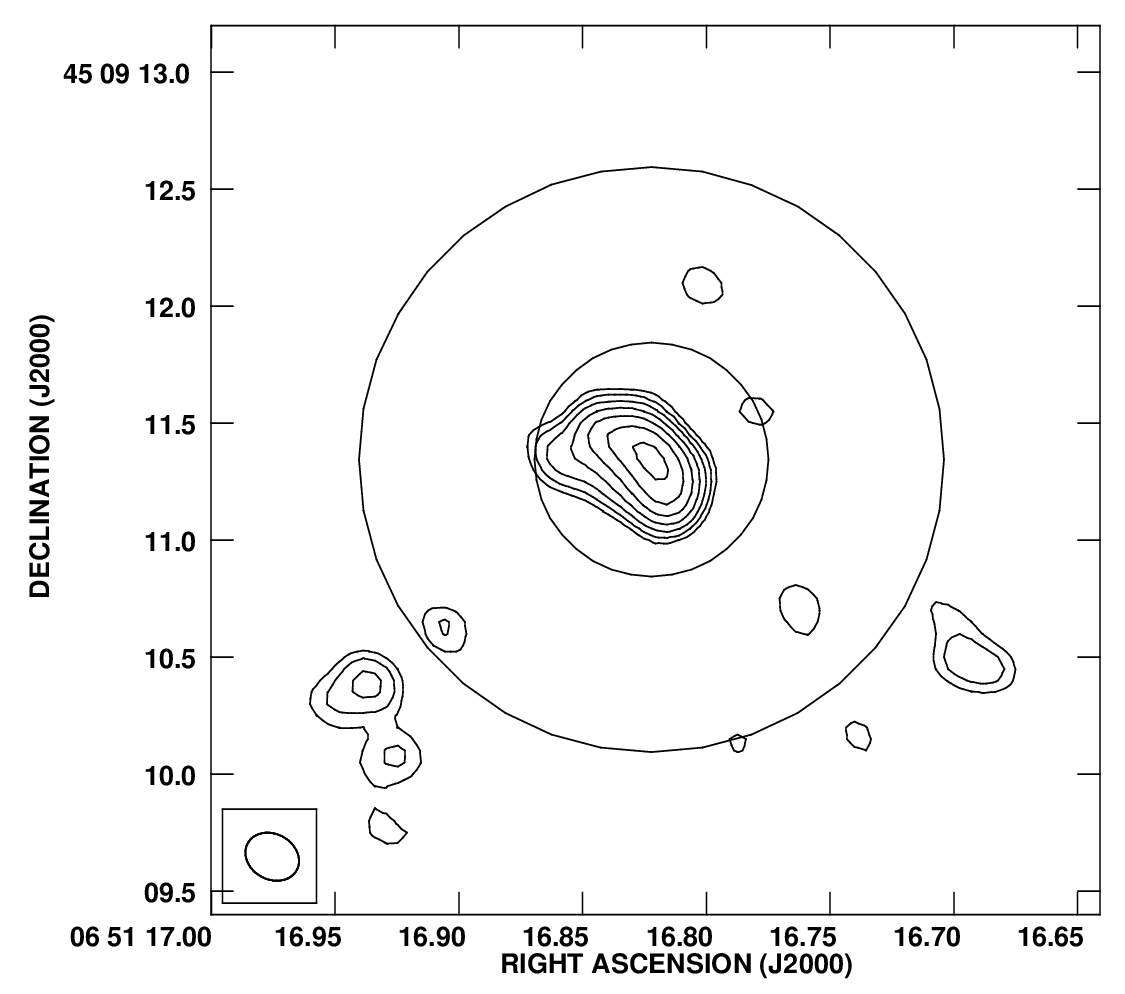}
                         \caption{3C169.1  south hot spot. 8.4 GHz VLA image from Kharb et al. (2008)
with CLEAN restoring beam FWHM $0.24 \times 0.19$ arcsec
(shown in the inset in the lower left).
Two circles are shown centered on the brightness peak. The larger circle has a diameter
of 2.5 arcsec, while the smaller circle has a diameter of 1.0 arcsec.
}
                  \label{fig3c169_S}
    \end{figure}

\clearpage

\begin{figure}
    \centering
    \includegraphics[width=\textwidth]{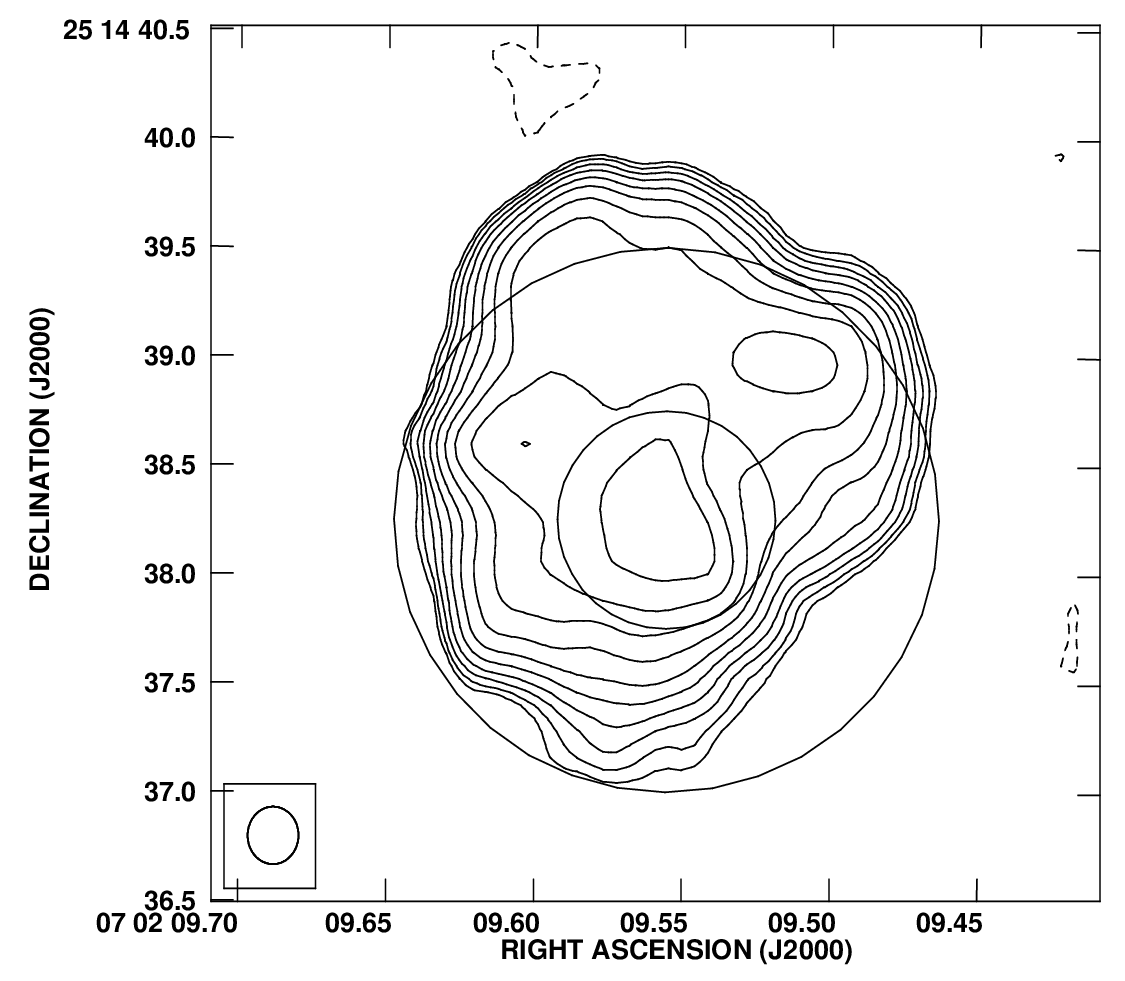}
                         \caption{3C172 north hot spot. 8.4 GHz VLA image from Kharb et al. (2008)
with CLEAN restoring beam FWHM $0.26 \times 0.23$ arcsec
(shown in the inset in the lower left).
Two circles are shown centered on the brightness peak. The larger circle has a diameter
of 2.5 arcsec, while the smaller circle has a diameter of 1.0 arcsec.
}
                  \label{fig3c172_N}
    \end{figure}

\begin{figure}
    \centering
    \includegraphics[width=\textwidth]{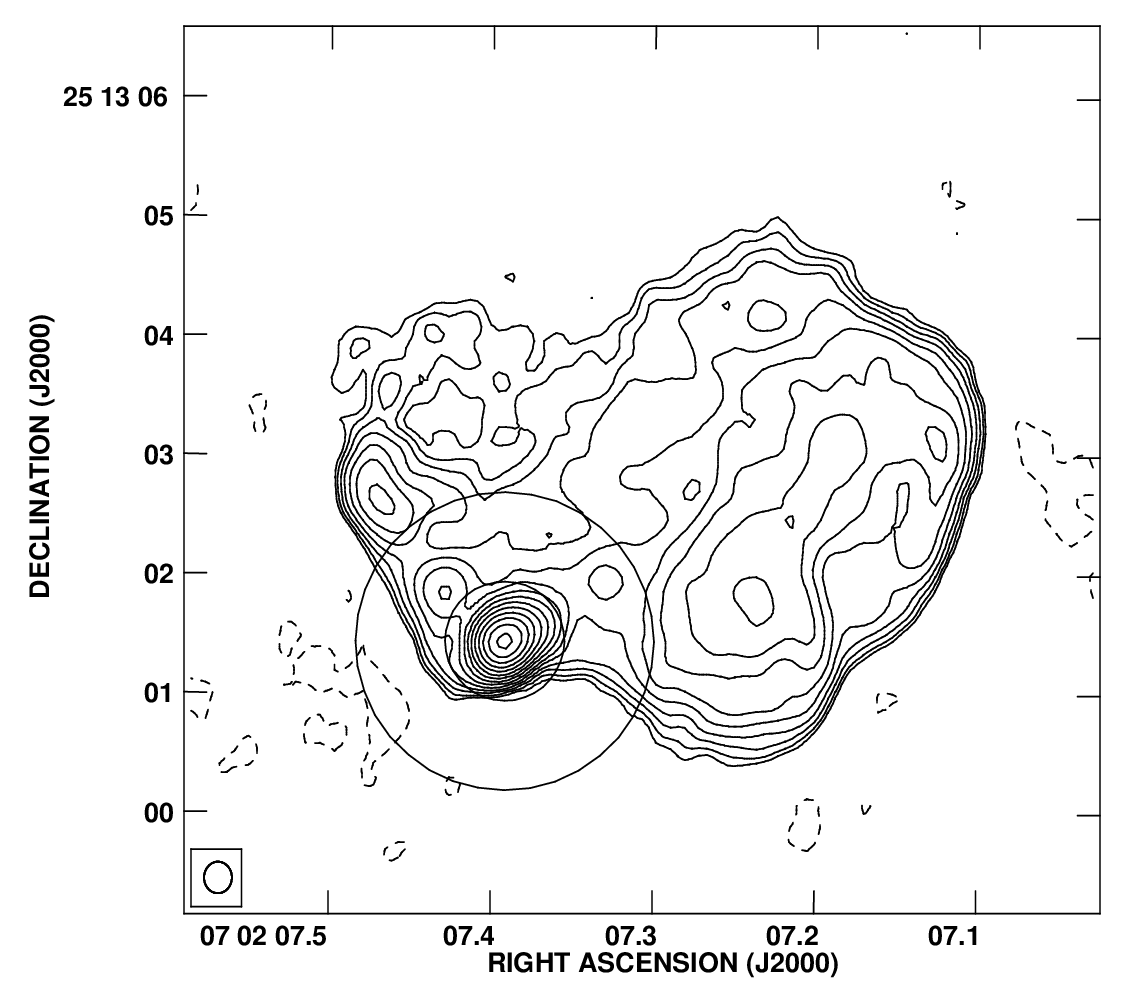}
                         \caption{3C172 south hot spot. 8.4 GHz VLA image from Kharb et al. (2008)
with CLEAN restoring beam FWHM $0.26 \times 0.23$ arcsec
(shown in the inset in the lower left).
Two circles are shown centered on the brightness peak. The larger circle has a diameter
of 2.5 arcsec, while the smaller circle has a diameter of 1.0 arcsec.
}
                  \label{fig3c172_S}
    \end{figure}

\begin{figure}
    \centering
    \includegraphics[width=\textwidth]{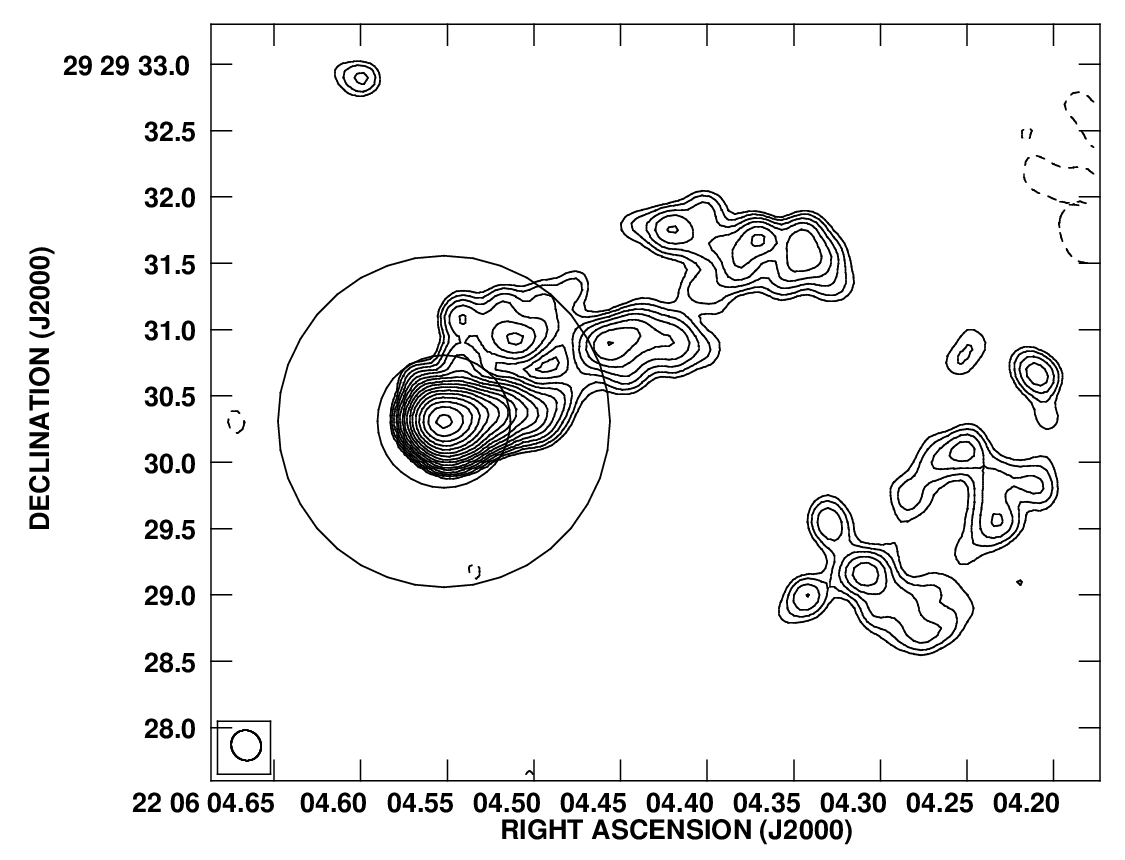}
			 \caption{3C441 north hot spot. 8.4 GHz VLA image from Kharb et al. (2008)
with CLEAN restoring beam FWHM $0.23 \times 0.22$ arcsec
(shown in the inset in the lower left).
Two circles are shown centered on the brightness peak. The larger circle has a diameter
of 2.5 arcsec, while the smaller circle has a diameter of 1.0 arcsec.
}
		  \label{fig3c441_N}
    \end{figure}

\begin{figure}
    \centering
    \includegraphics[width=\textwidth]{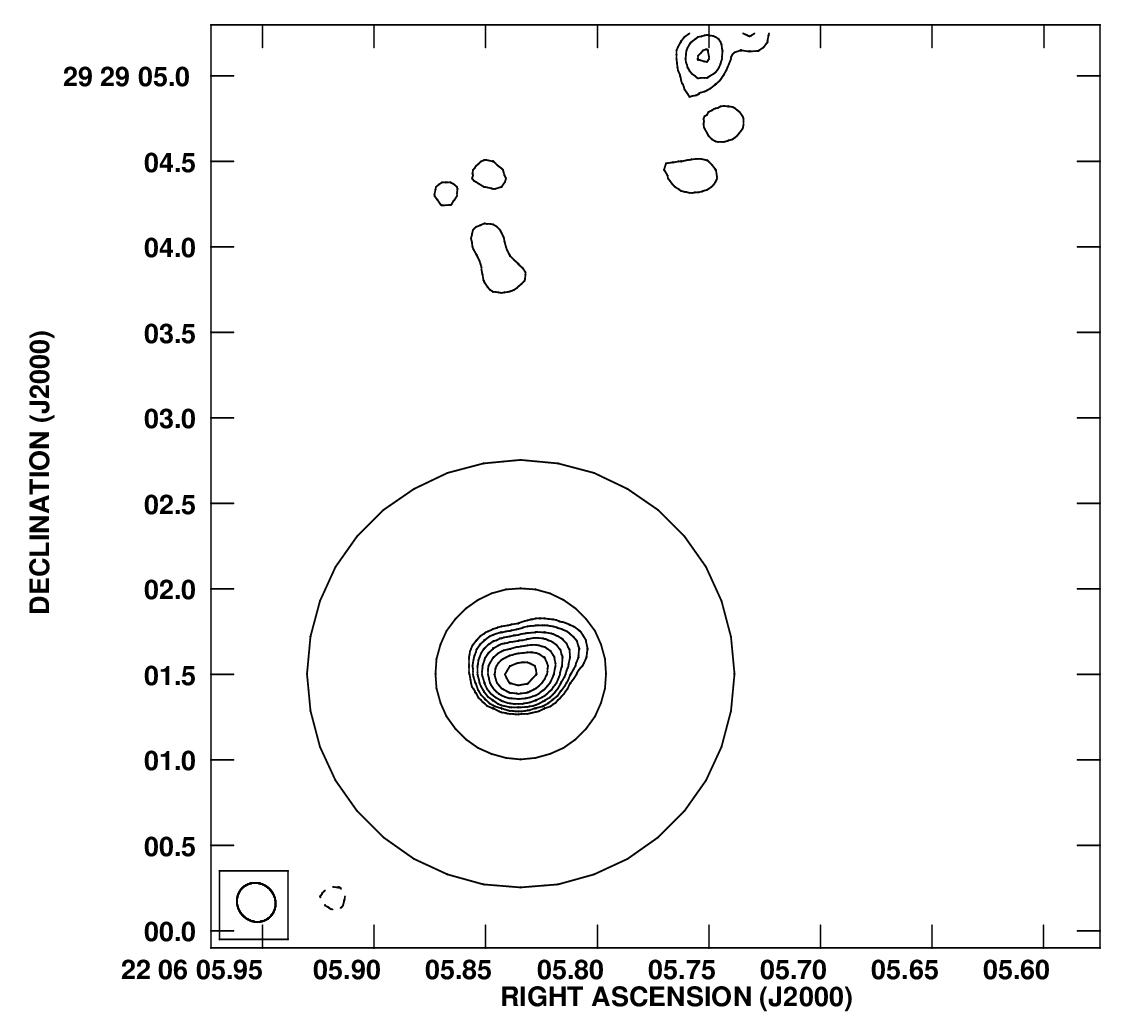}
                         \caption{3C441 south hot spot. 8.4 GHz VLA image from Kharb et al. (2008)
with CLEAN restoring beam FWHM $0.23 \times 0.22$ arcsec
(shown in the inset in the lower left).
Two circles are shown centered on the brightness peak. The larger circle has a diameter
of 2.5 arcsec, while the smaller circle has a diameter of 1.0 arcsec.
}
                  \label{fig3c441_S}
    \end{figure}

\begin{figure}
    \centering
    \includegraphics[width=\textwidth]{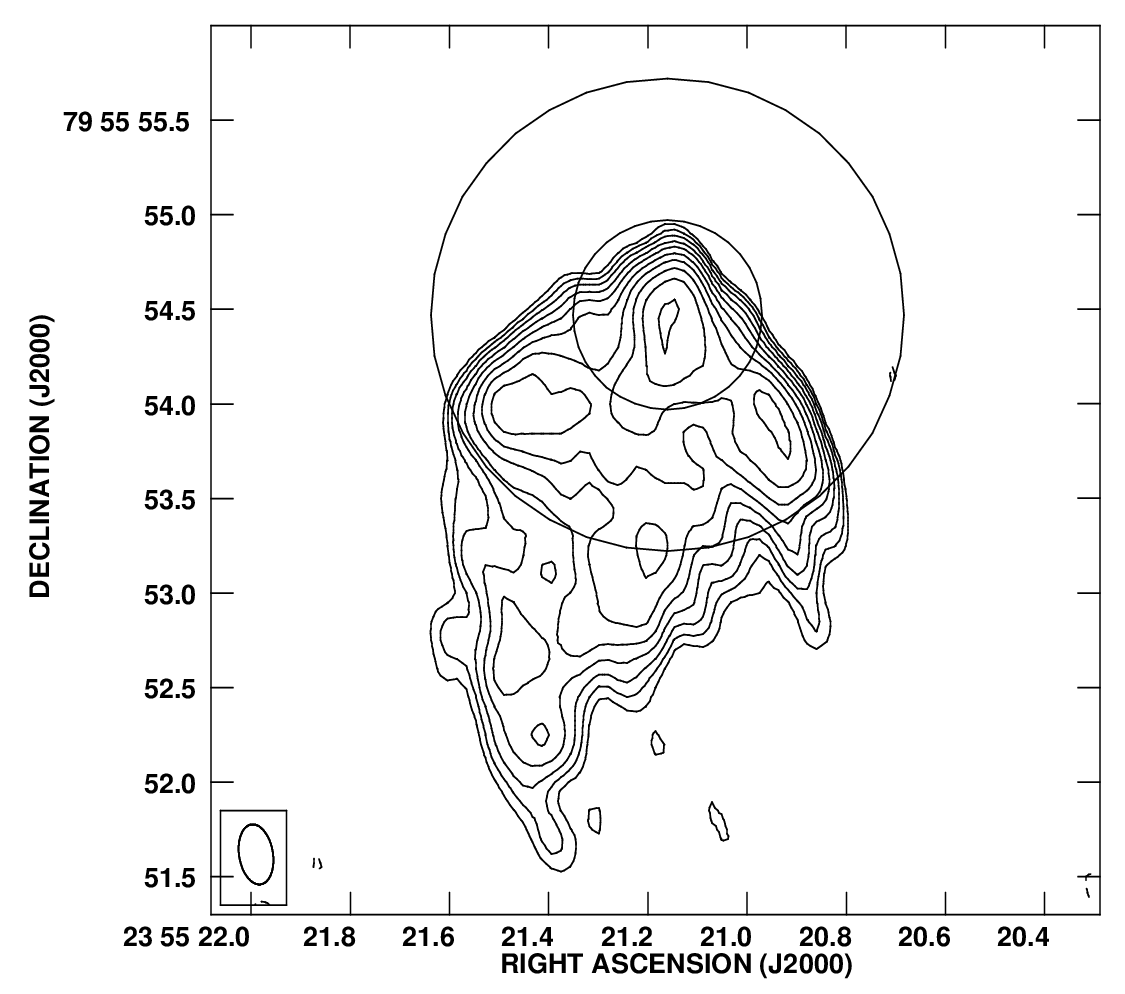} 
                         \caption{3C469.1 north hot spot. 8.4 GHz VLA image from Kharb et al. (2008)
with CLEAN restoring beam FWHM $0.32 \times 0.18$ arcsec
(shown in the inset in the lower left).
Two circles are shown centered on the brightness peak. The larger circle has a diameter
of 2.5 arcsec, while the smaller circle has a diameter of 1.0 arcsec.
}
                  \label{fig3c469_N}
    \end{figure}

\begin{figure}
    \centering
    \includegraphics[width=\textwidth]{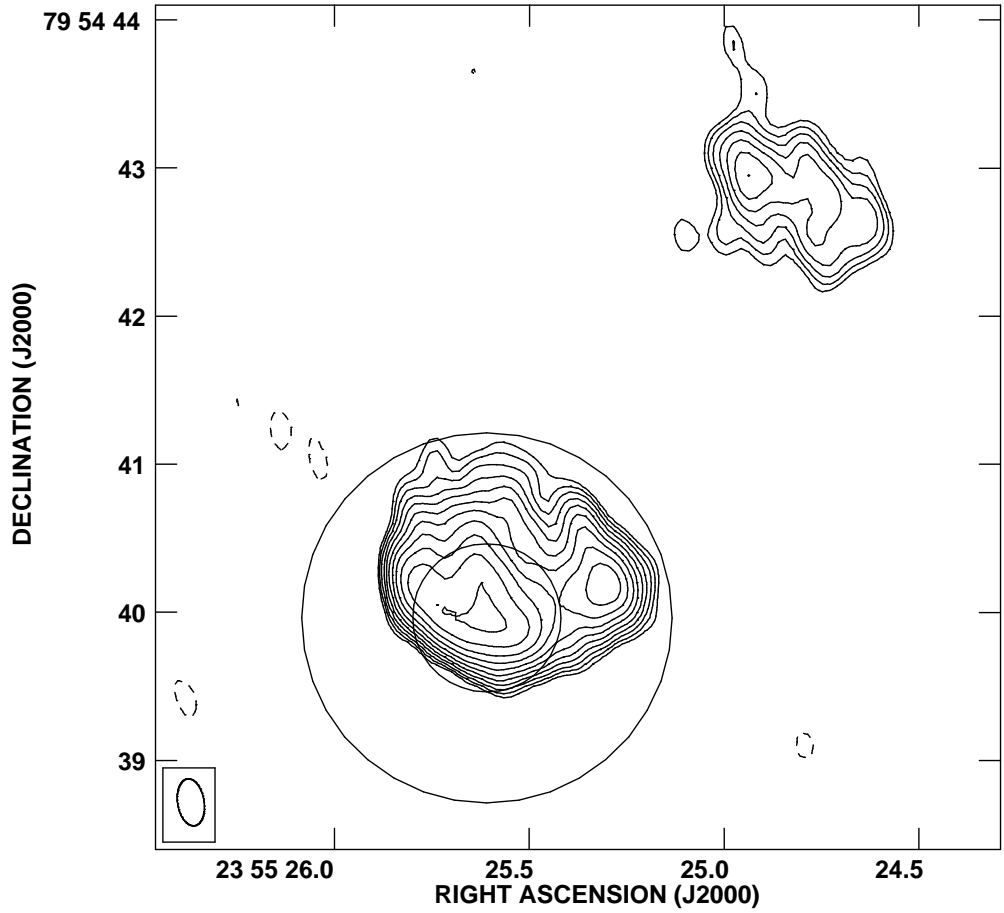} 
                         \caption{3C469.1 south hot spot. 8.4 GHz VLA image from Kharb et al. (2008)
with CLEAN restoring beam FWHM $0.32 \times 0.18$ arcsec
(shown in the inset in the lower left).
Two circles are shown centered on the brightness peak. The larger circle has a diameter
of 2.5 arcsec, while the smaller circle has a diameter of 1.0 arcsec.
}
                  \label{fig3c469_S}
    \end{figure}

\clearpage


\begin{figure}
    \centering
    \includegraphics[width=\textwidth]{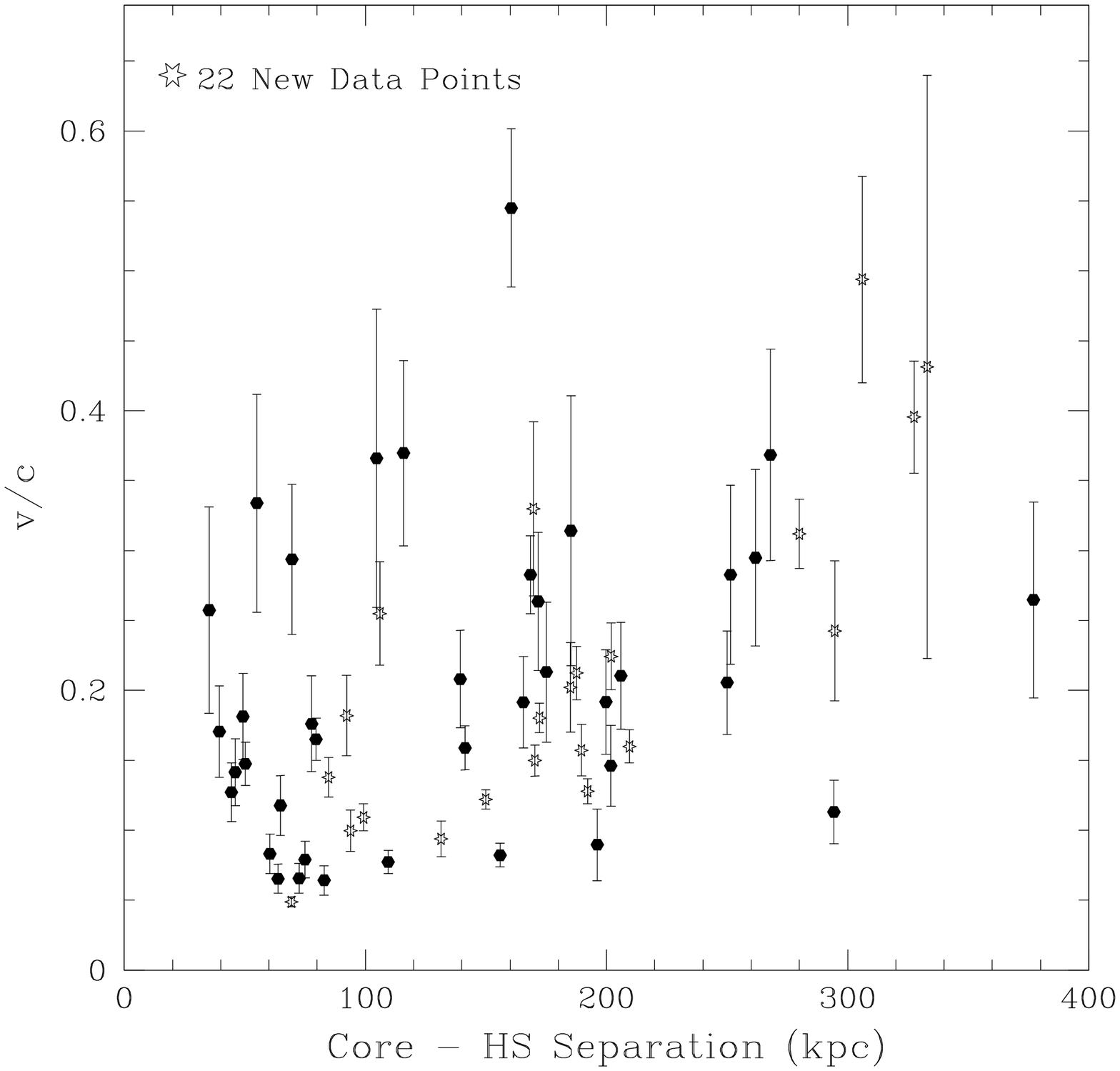}
       \caption{Source advance velocity $v/c$ as a function of core-hotspot
distance
$r$  for b=1. }
          \label{figvda}
    \end{figure}

\begin{figure}
    \centering
    \includegraphics[width=\textwidth]{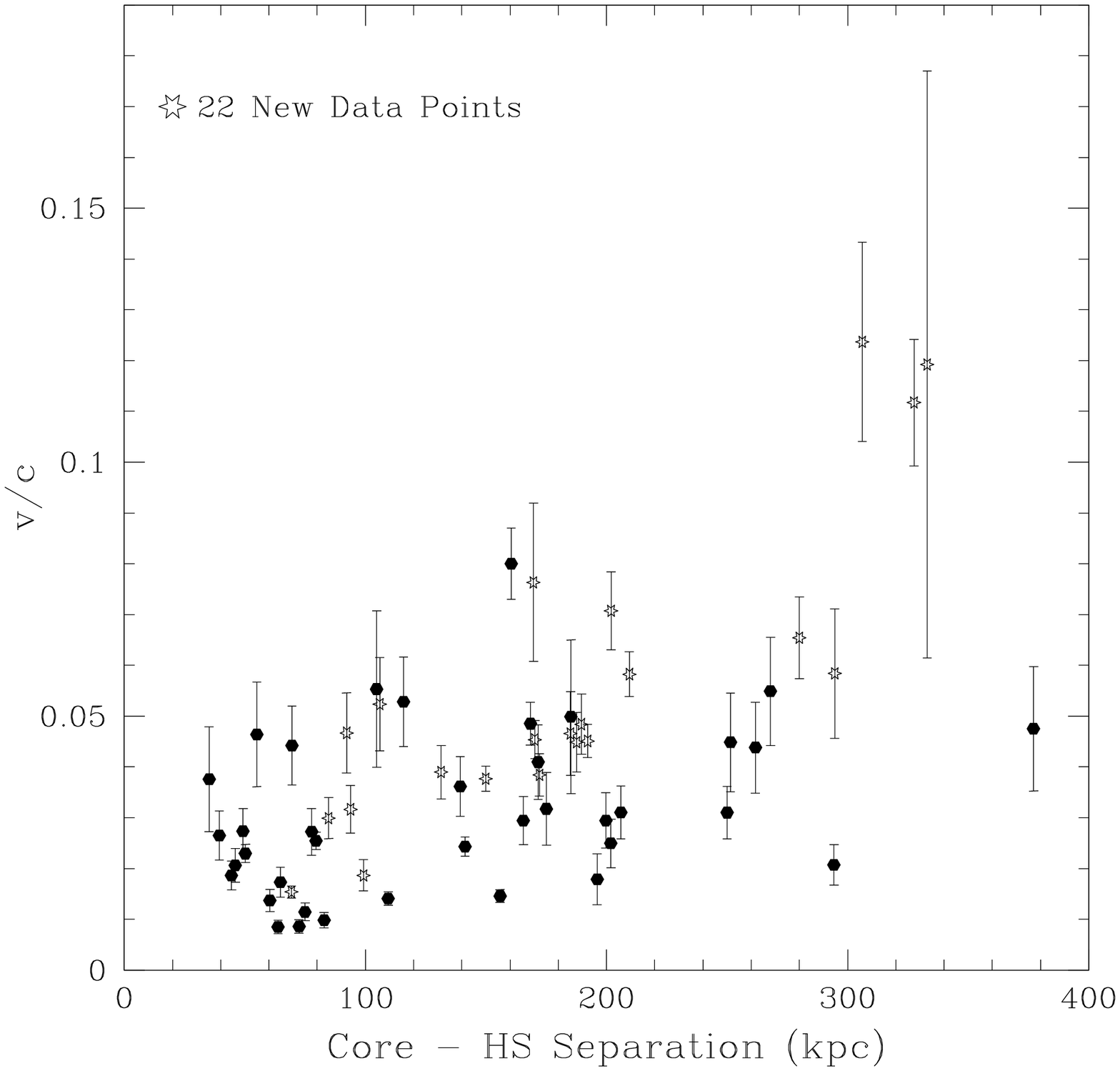}
       \caption{Source advance velocity $v/c$ as a function of core-hotspot
distance
$r$  for b=0.25. }
          \label{figvdb}
    \end{figure}


\begin{figure}
    \centering
    \includegraphics[width=\textwidth]{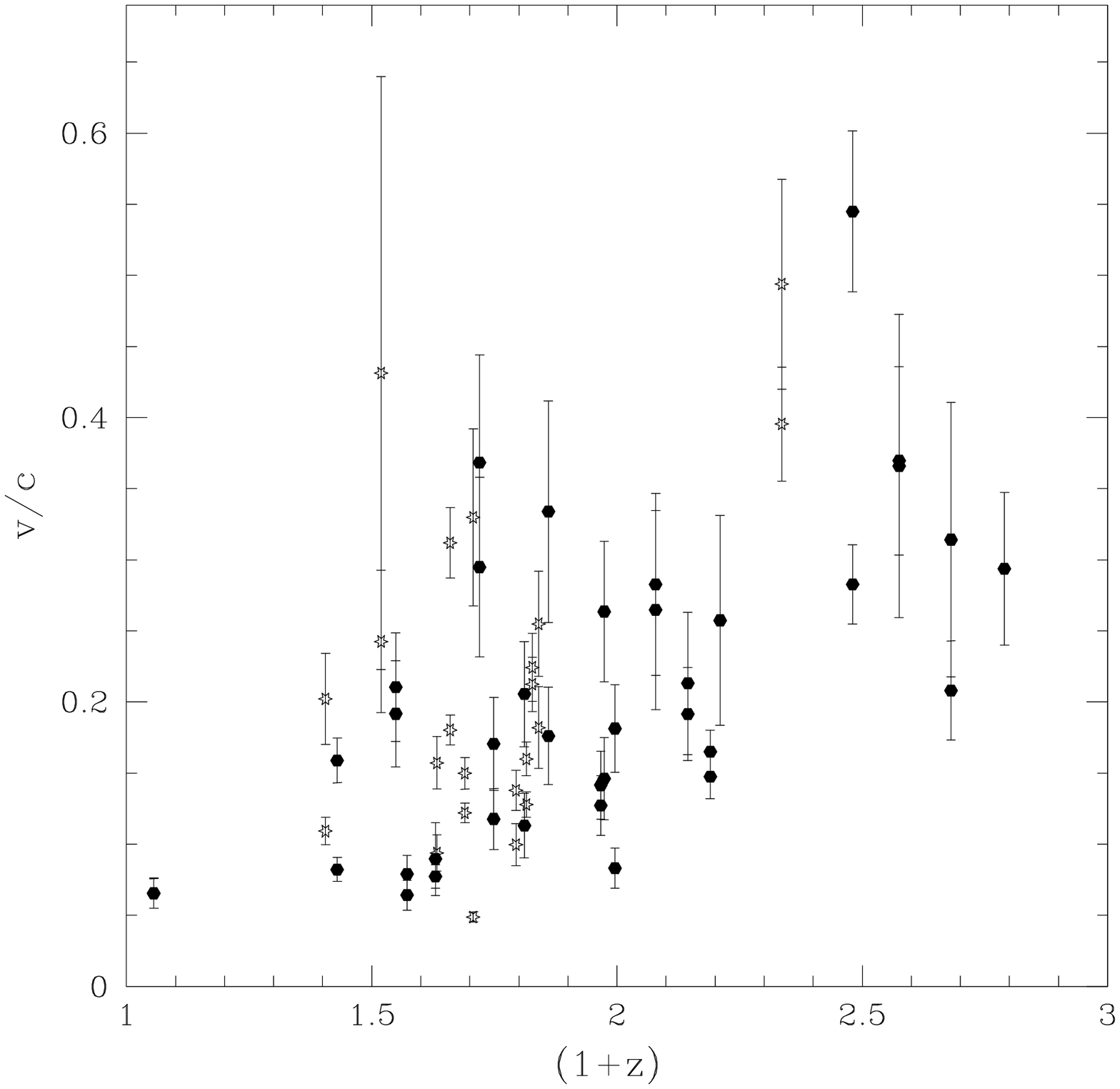}
         \caption{
Source advance velocity $v/c$ as a function of redshift (1+z) for
b=1.  }
         \label{figvza}
    \end{figure}

\begin{figure}
    \centering
    \includegraphics[width=\textwidth]{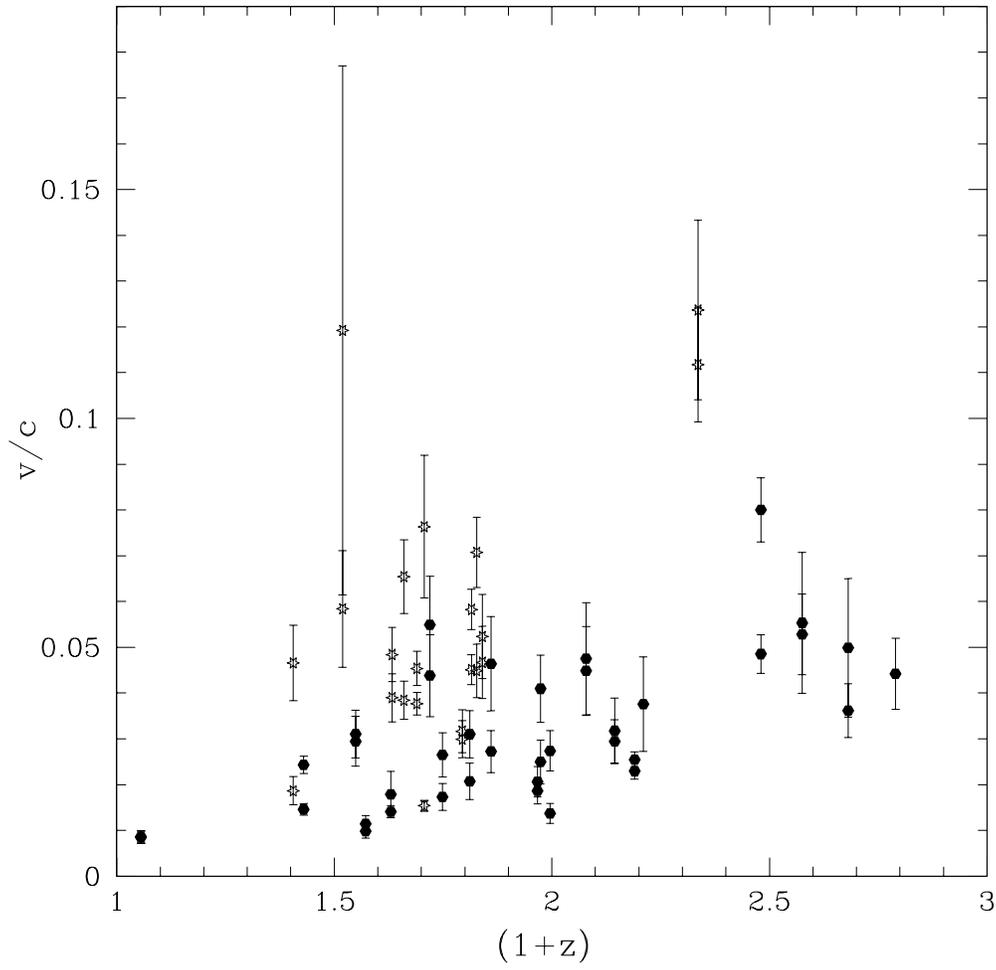}
         \caption{Source advance velocity $v/c$ as a function of redshift 
 (1+z)
  for b=0.25. }
         \label{figvzb}
    \end{figure}

\begin{figure}
    \centering
    \includegraphics[width=\textwidth]{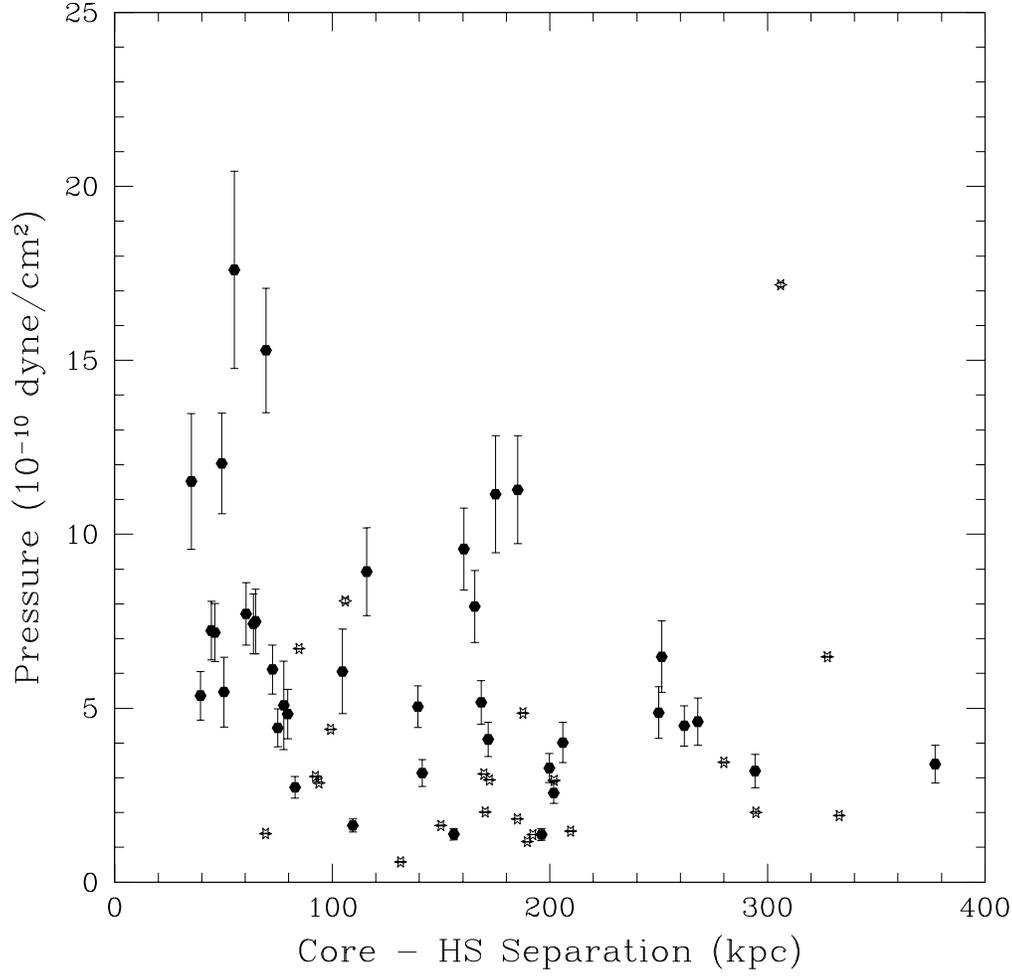}
         \caption{
Pressure in the lobe $P$  measured at 10 $h^{-1}$ kpc
from the hotspot toward the central region of the source,
as a function of core-hotspot distance $r$ for
b = 0.25. The pressure for $b=1$ is
lower by a factor of 0.218 since
$P(b=1) = 0.218 P(b=0.25)$.
  }
         \label{figpda}
    \end{figure}

\begin{figure}
    \centering
    \includegraphics[width=\textwidth]{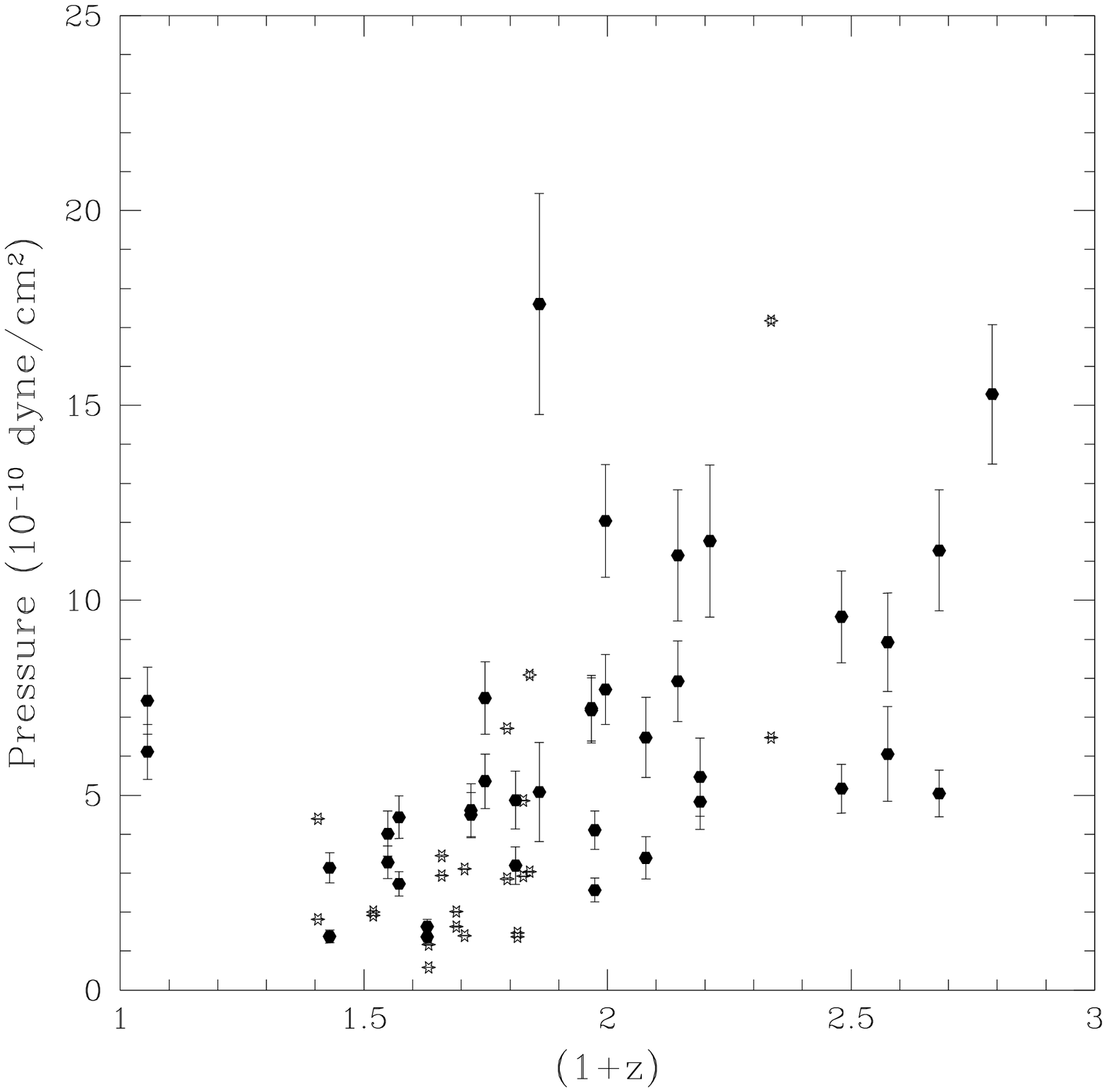}
         \caption{
Pressure in the lobe $P$  measured at 10 $h^{-1}$ kpc
from the hotspot toward the central region of the source,
as a function of $(1 + z)$  for
b = 0.25. The pressure for $b=1$ is
lower by a factor of 0.218 since
$P(b=1) = 0.218 P(b=0.25)$.
}
         \label{figpdb}
    \end{figure}

\begin{figure}
    \centering
    \includegraphics[width=\textwidth]{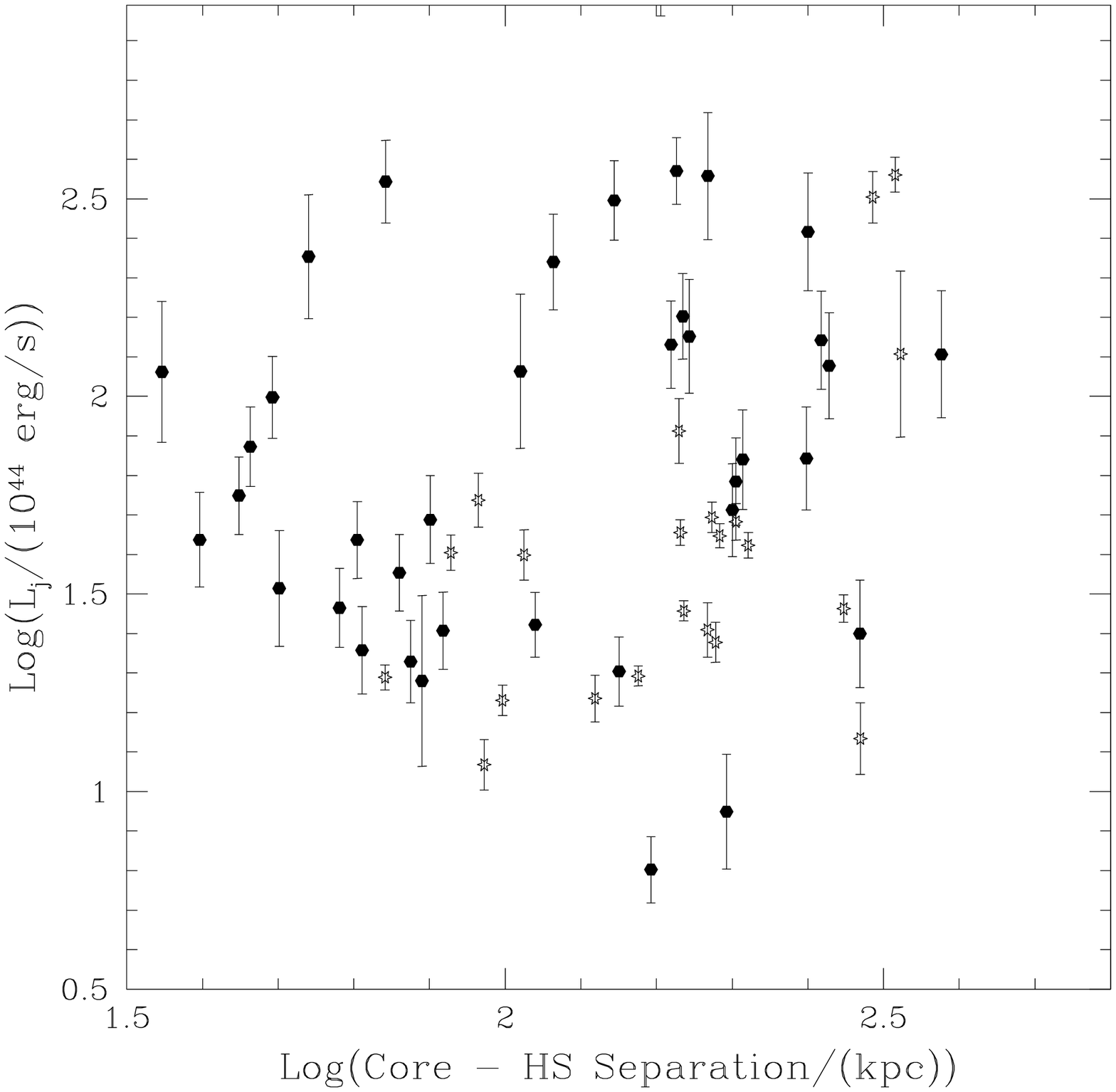}
         \caption{
Beam Power $L_j$ as a function of core-hotspot distance $r$  for
b=1.  }
          \label{figljda}
    \end{figure}

\begin{figure}
    \centering
    \includegraphics[width=\textwidth]{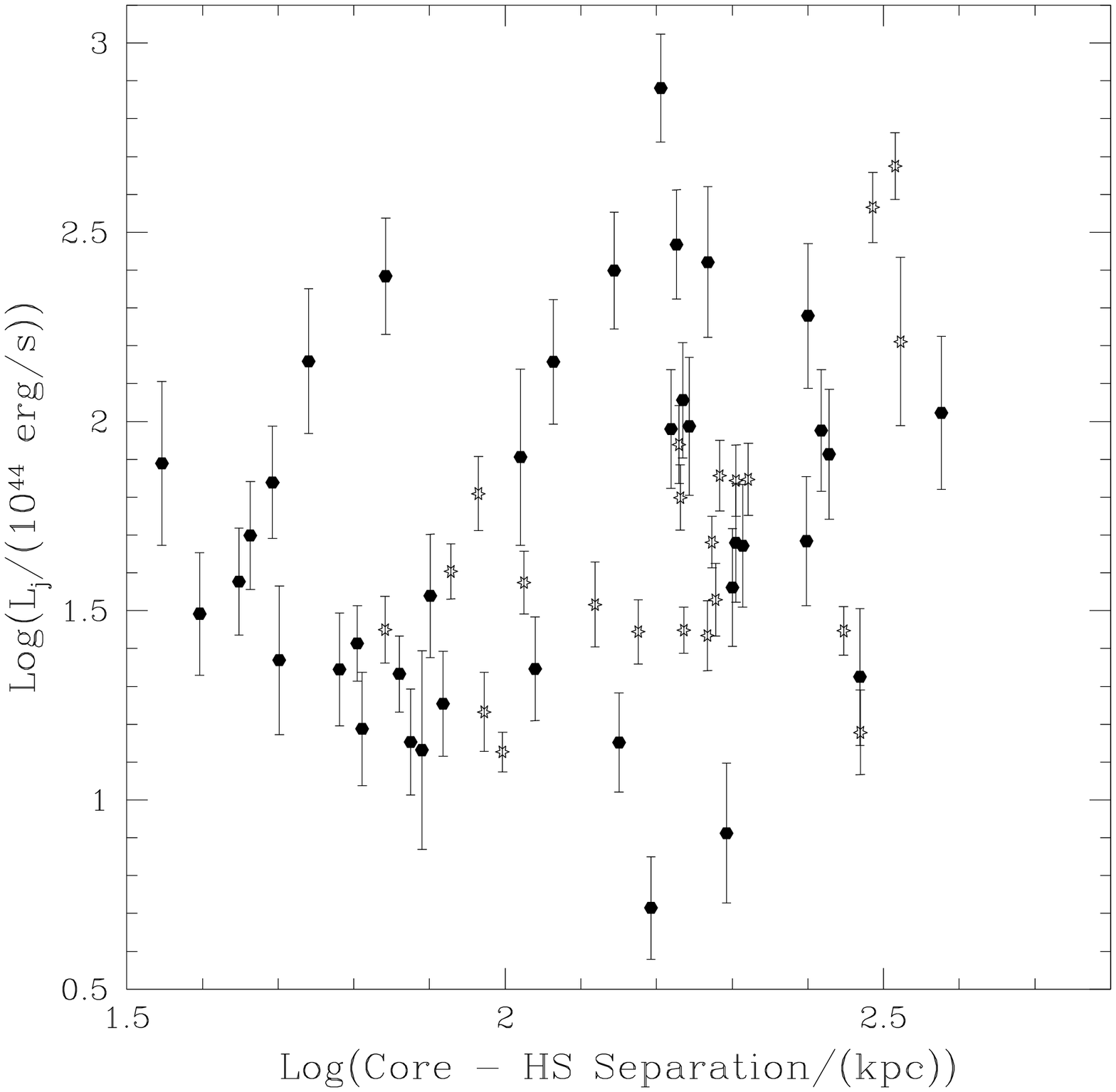}
         \caption{Beam Power $L_j$ as a function of core-hotspot distance 
 $r$
  for b=0.25. }
          \label{figljdb}
    \end{figure}

\begin{figure}
    \centering
    \includegraphics[width=\textwidth]{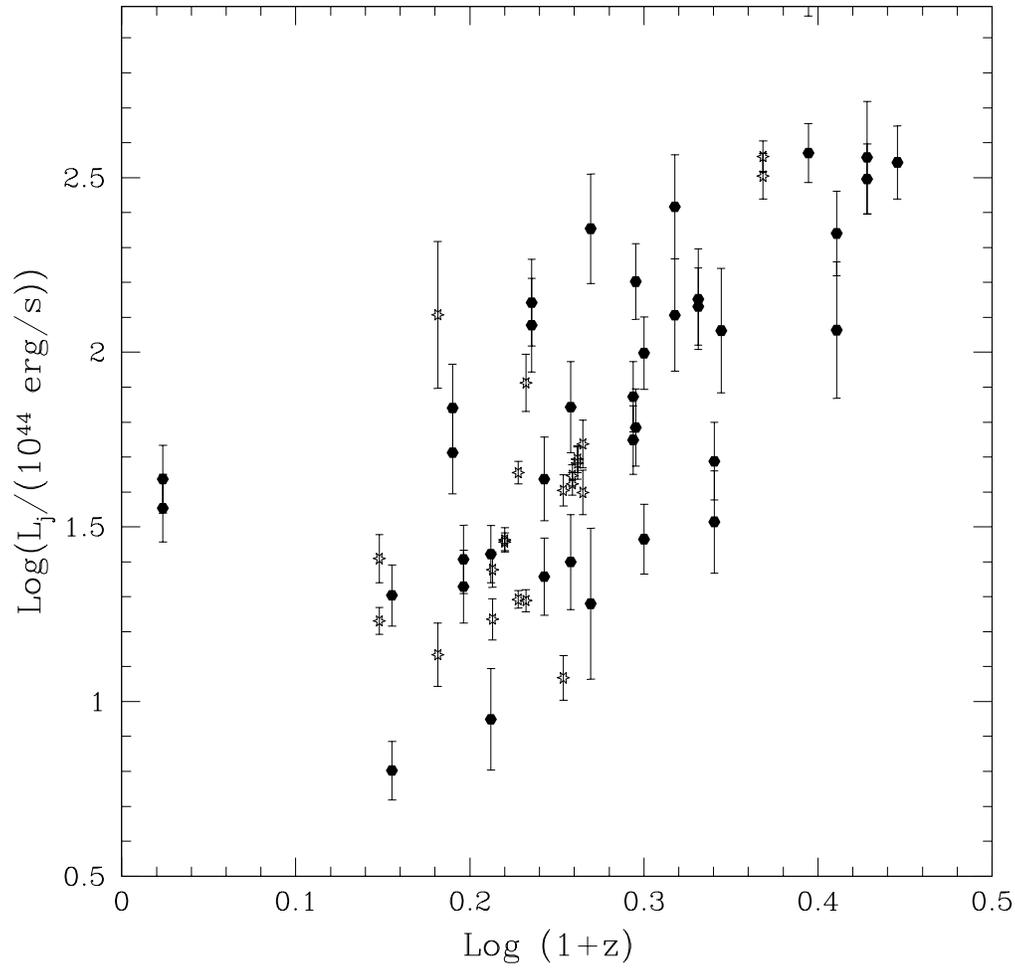}
         \caption{
Beam Power $L_j$ as a function of redshift (1+z)   for
b=1.  }
          \label{figljza}
    \end{figure}

\begin{figure}
    \centering
    \includegraphics[width=\textwidth]{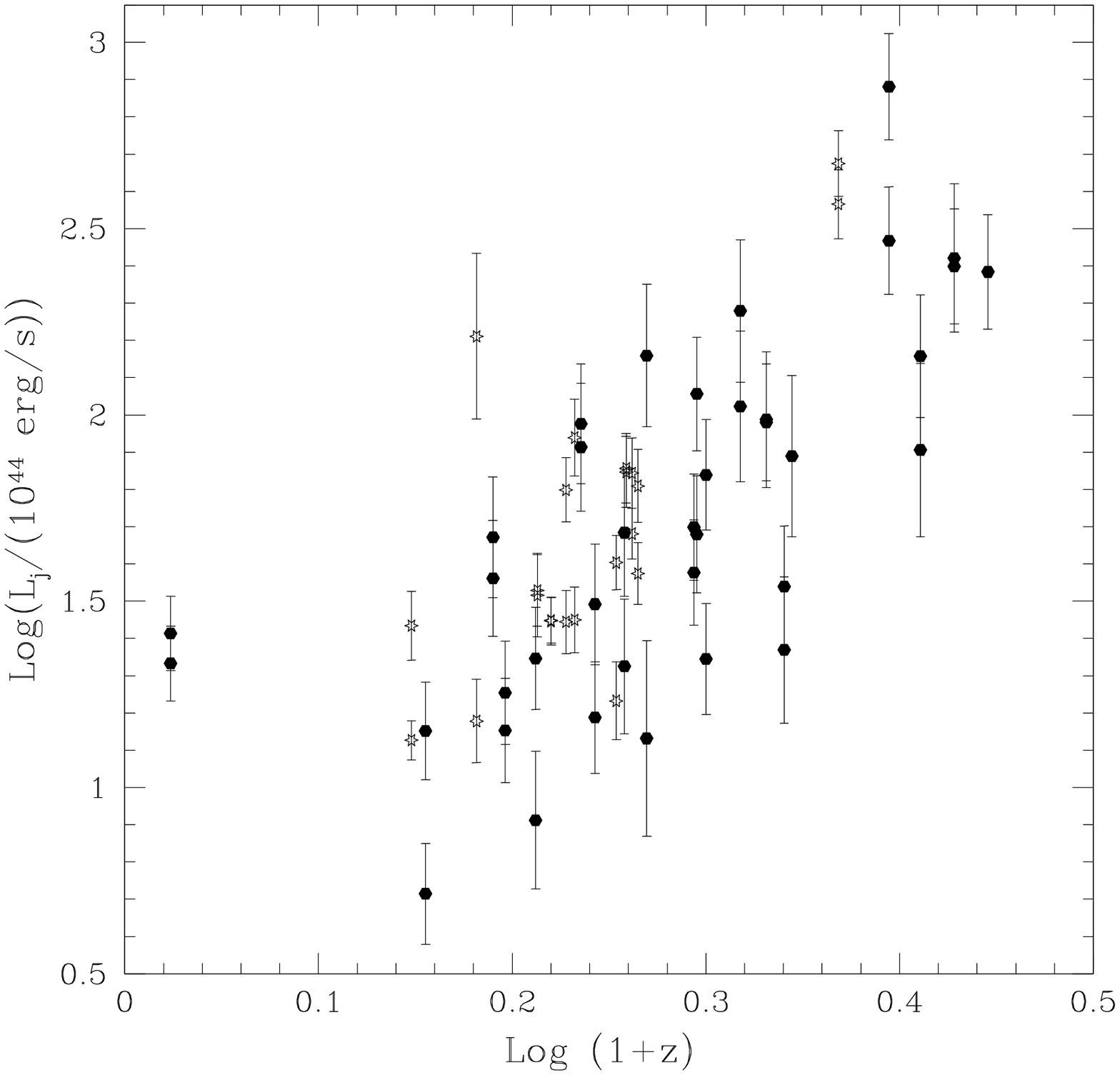}
         \caption{Beam Power $L_j$ as a function of redshift (1+z)
  for b=0.25. }
          \label{figljzb}
    \end{figure}

\clearpage

\begin{figure}
    \centering
    \includegraphics[width=\textwidth]{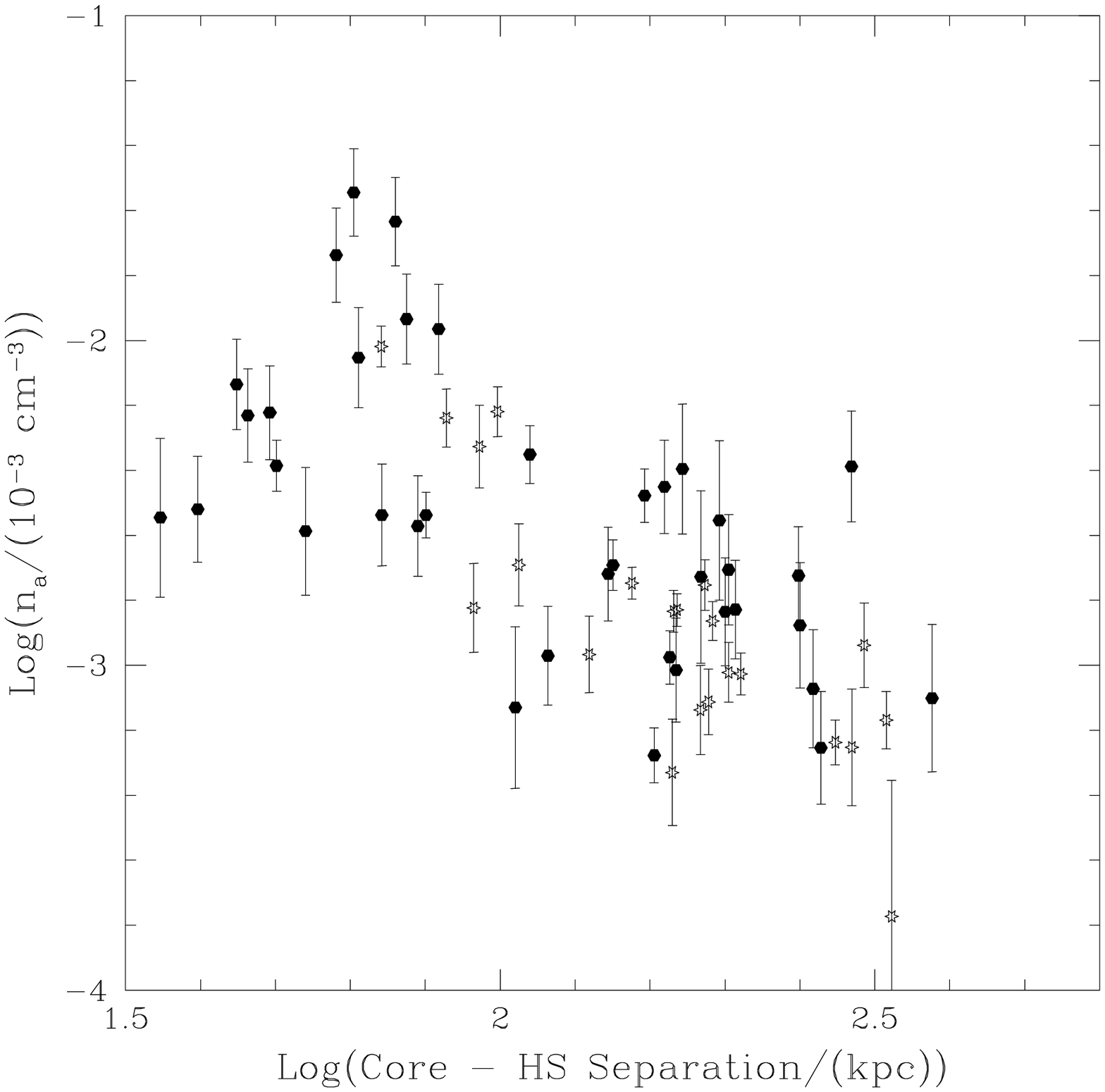}
         \caption{
Ambient density $n_a$ as a function of core-hotspot distance $r$  for
b=1.  }
          \label{fignda}
    \end{figure}

\begin{figure}
    \centering
    \includegraphics[width=\textwidth]{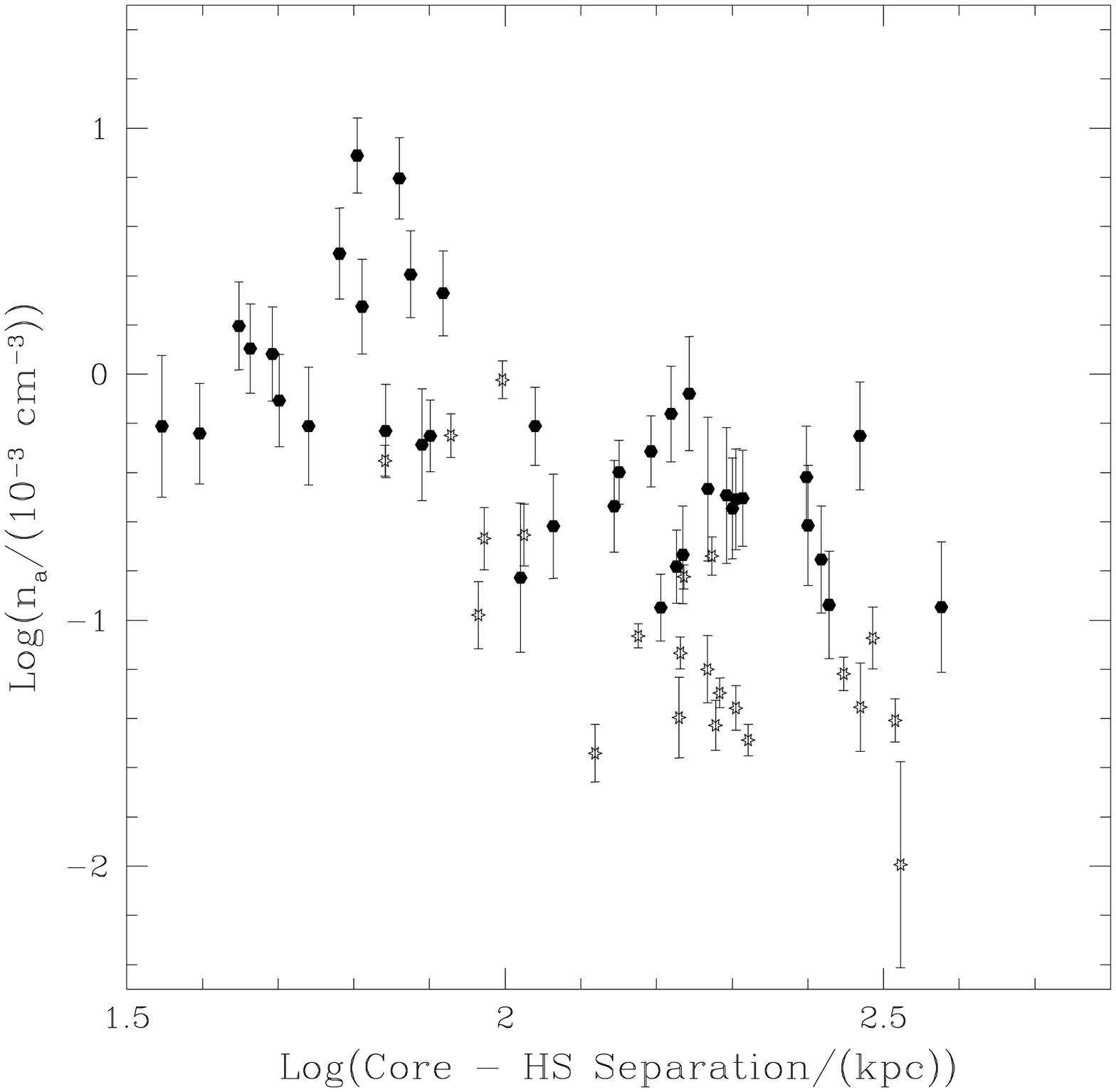}
         \caption{Ambient density $n_a$ as a function of core-hotspot distance
$r$  for
b=0.25. }
          \label{figndb}
    \end{figure}

\begin{figure}
    \centering
    \includegraphics[width=\textwidth]{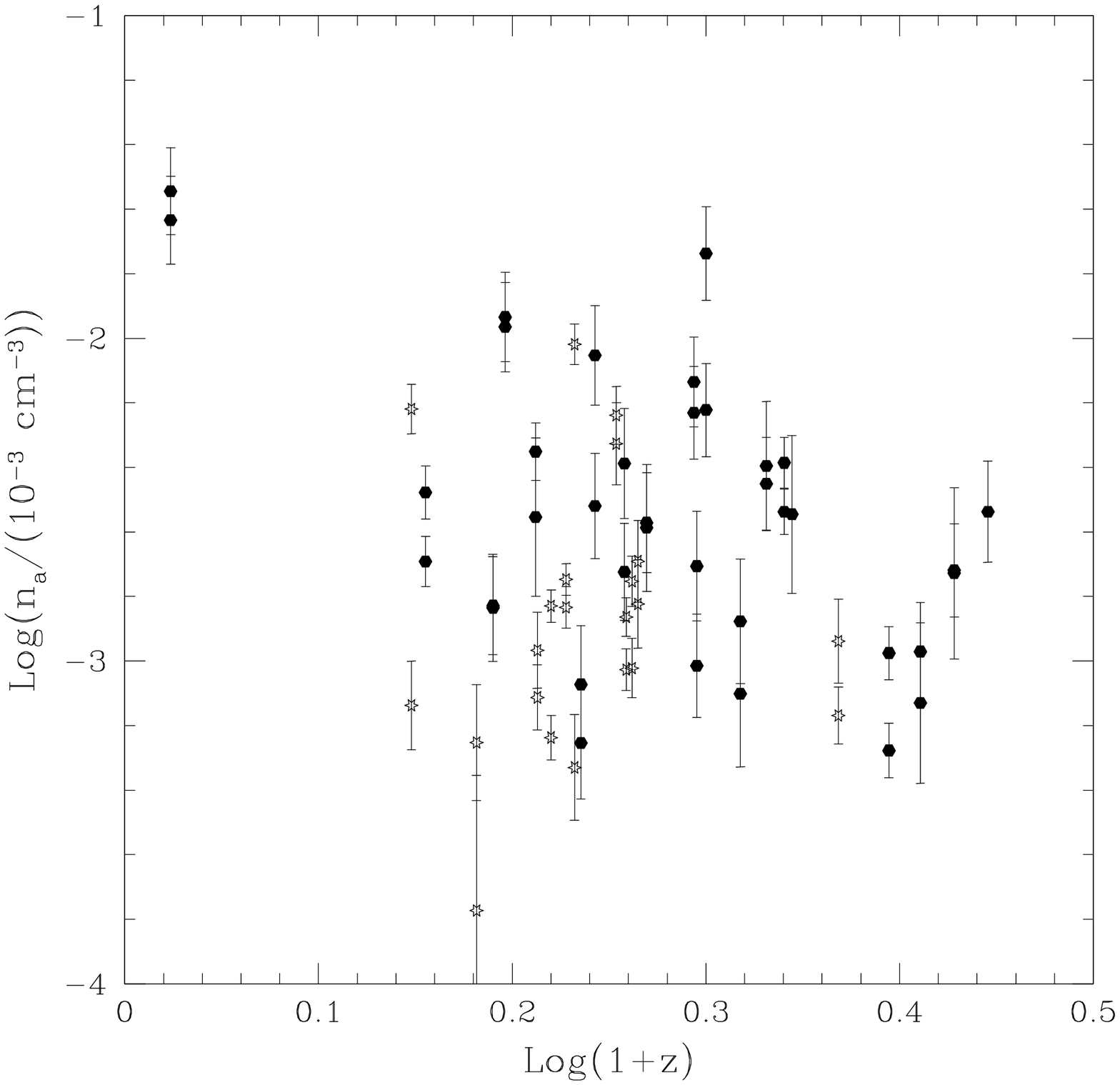}
         \caption{
Ambient density $n_a$ as a function of redshift (1+z)  for
b=1.  }
          \label{fignza}
    \end{figure}

\begin{figure}
    \centering
    \includegraphics[width=\textwidth]{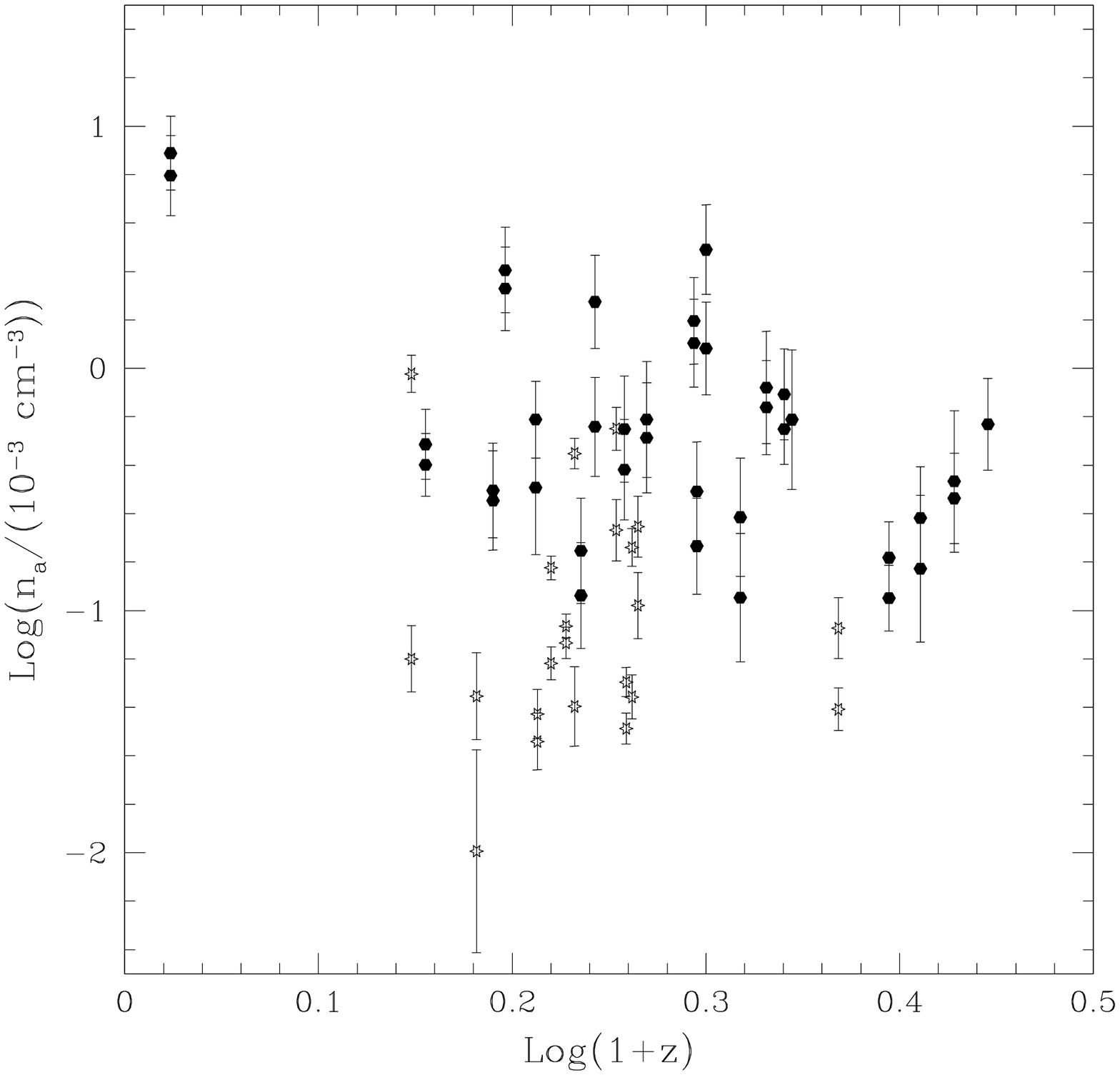}
         \caption{Ambient density $n_a$ as a function of redshift (1+z)  for
b=0.25. }
          \label{fignzb}
    \end{figure}

\begin{figure}
    \centering
    \includegraphics[width=\textwidth]{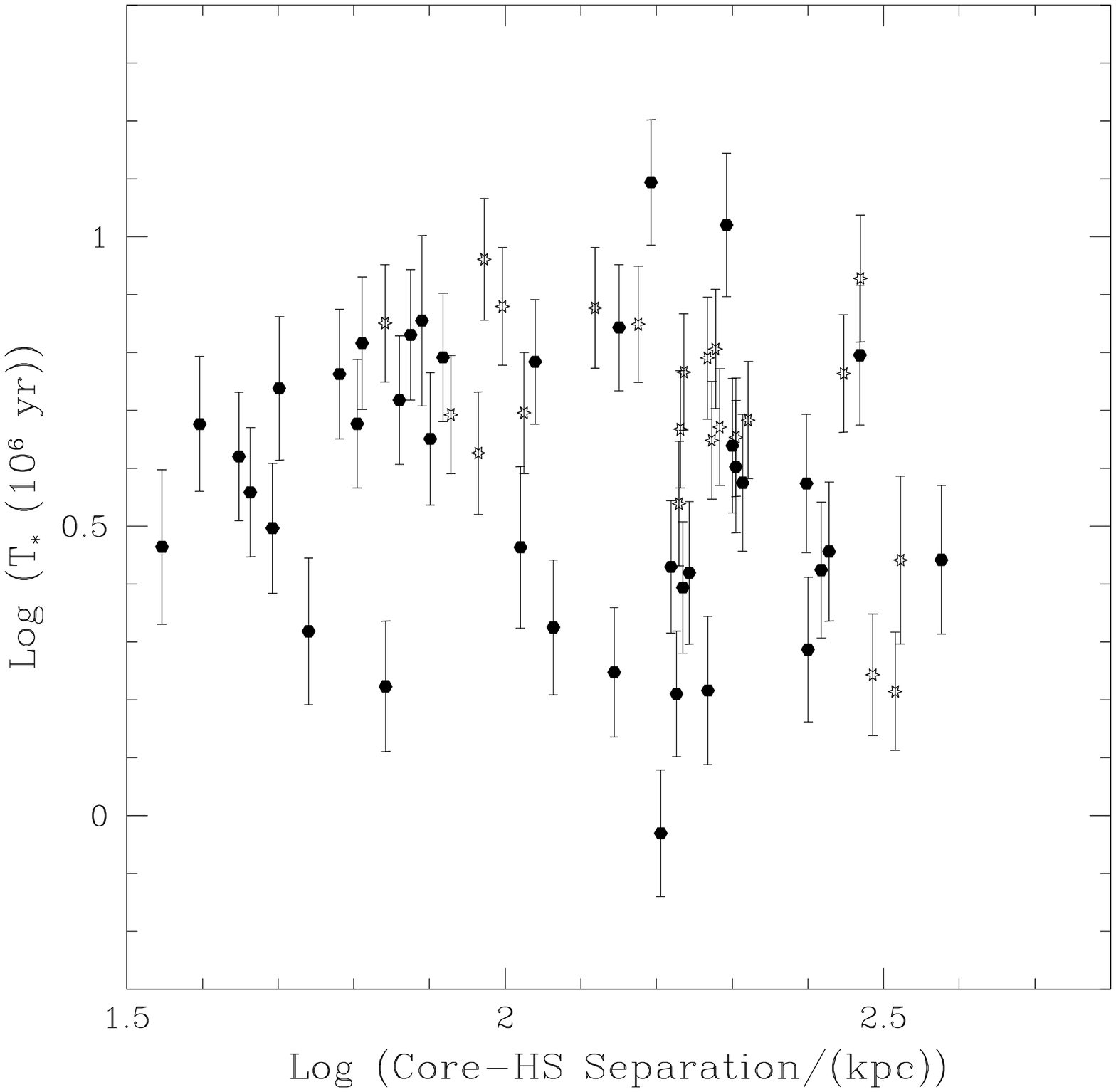}
         \caption{
Source lifetime $T_T$ as a function of core-hotspot distance $r$ for
b=1.  }
          \label{figtda}
    \end{figure}

\begin{figure}
    \centering
    \includegraphics[width=\textwidth]{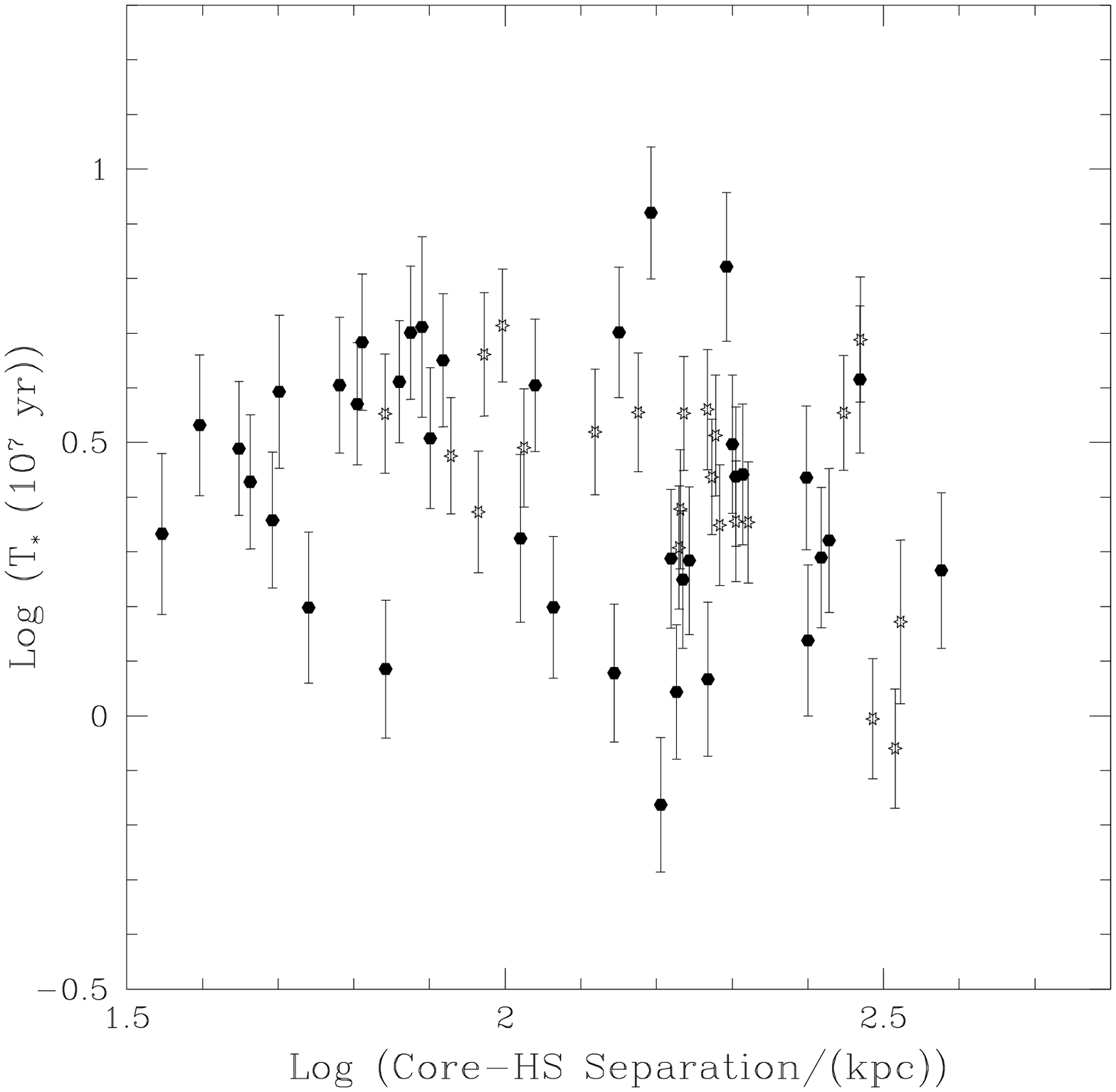}
         \caption{
Source lifetime $T_T$ as a function of core-hotspot distance $r$ for
b=0.25. }
          \label{figtdb}
    \end{figure}

\begin{figure}
    \centering
    \includegraphics[width=\textwidth]{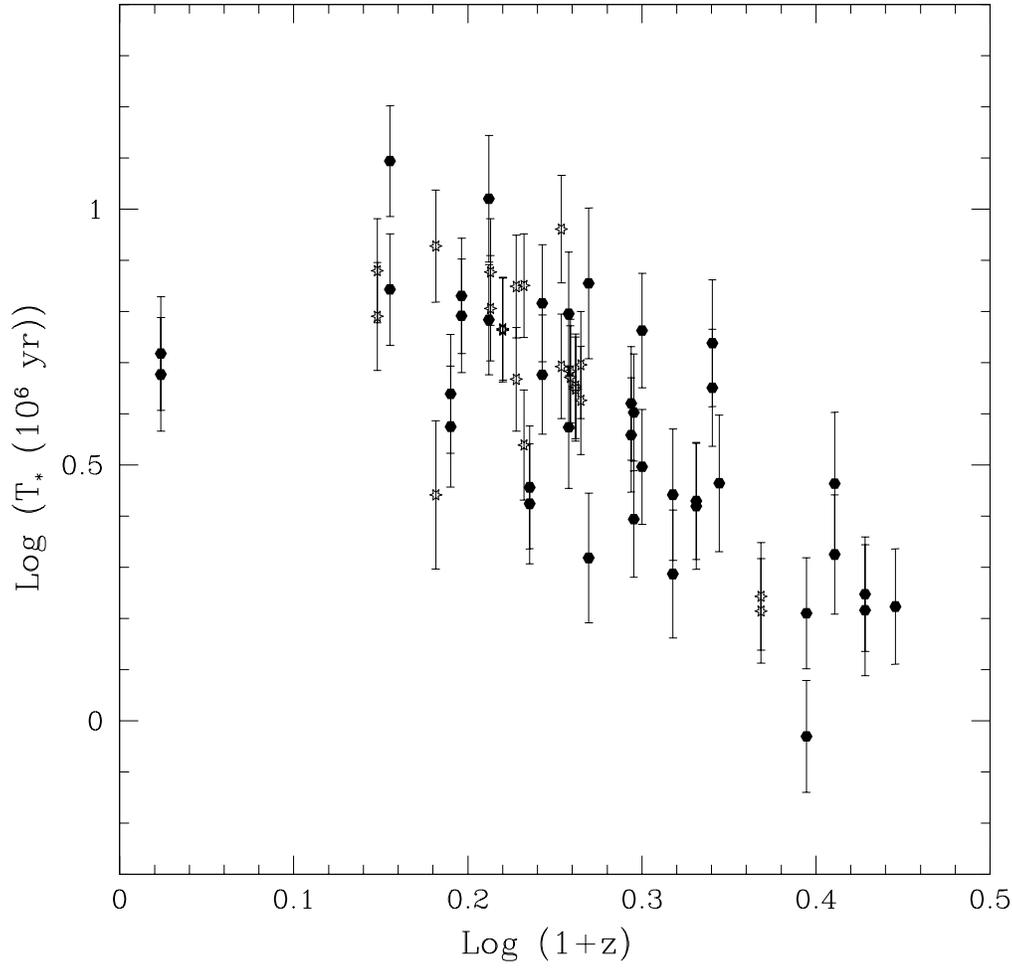}
         \caption{
Source lifetime $T_T$ as a function of redshift (1 + z)  for
b=1.  }
          \label{figtza}
    \end{figure}

\begin{figure}
    \centering
    \includegraphics[width=\textwidth]{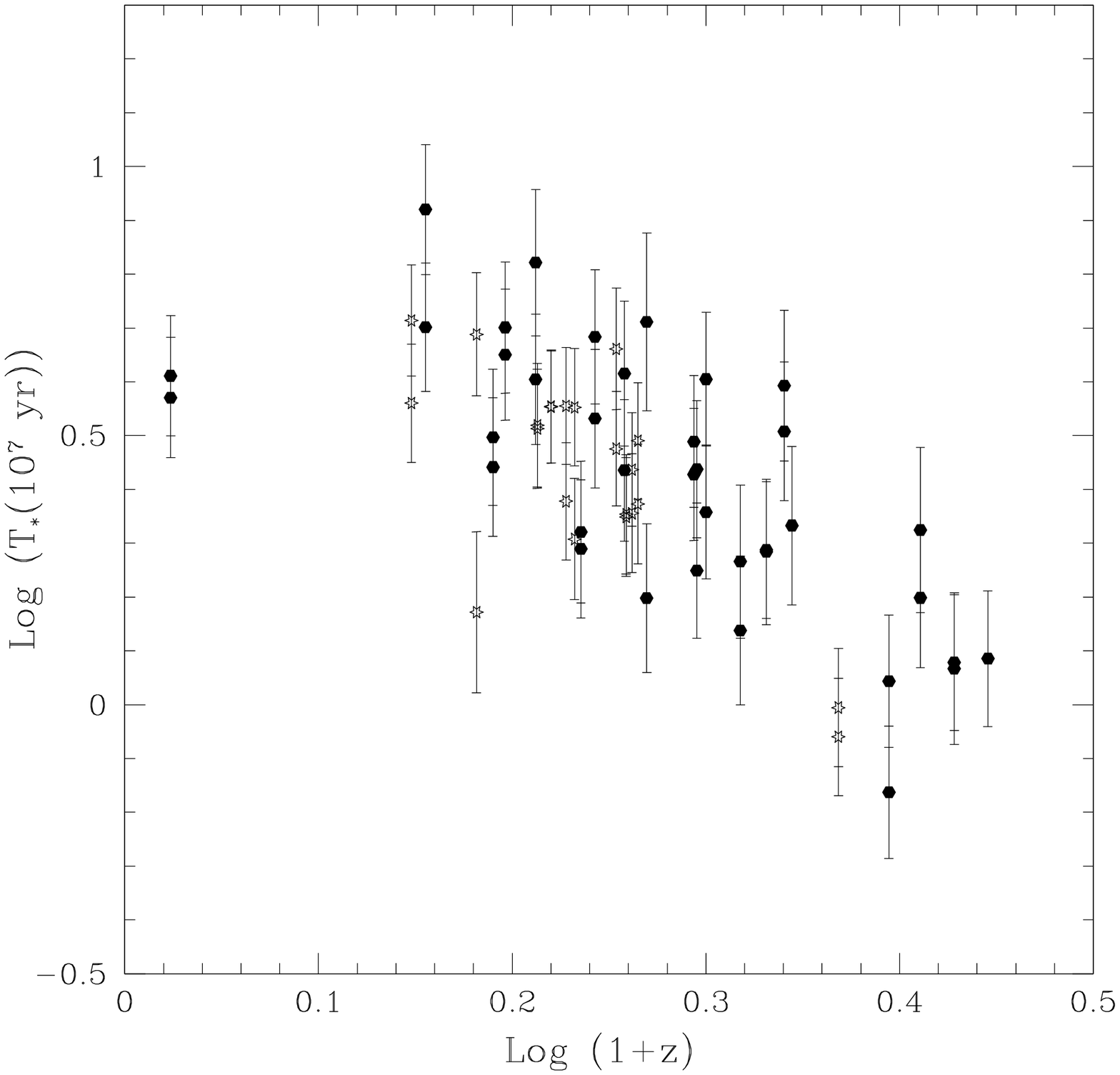}
         \caption{
Source lifetime $T_T$ as a function of redshift (1 + z) for
b=0.25. }
          \label{figtzb}
    \end{figure}

\begin{figure}
    \centering
    \includegraphics[width=\textwidth]{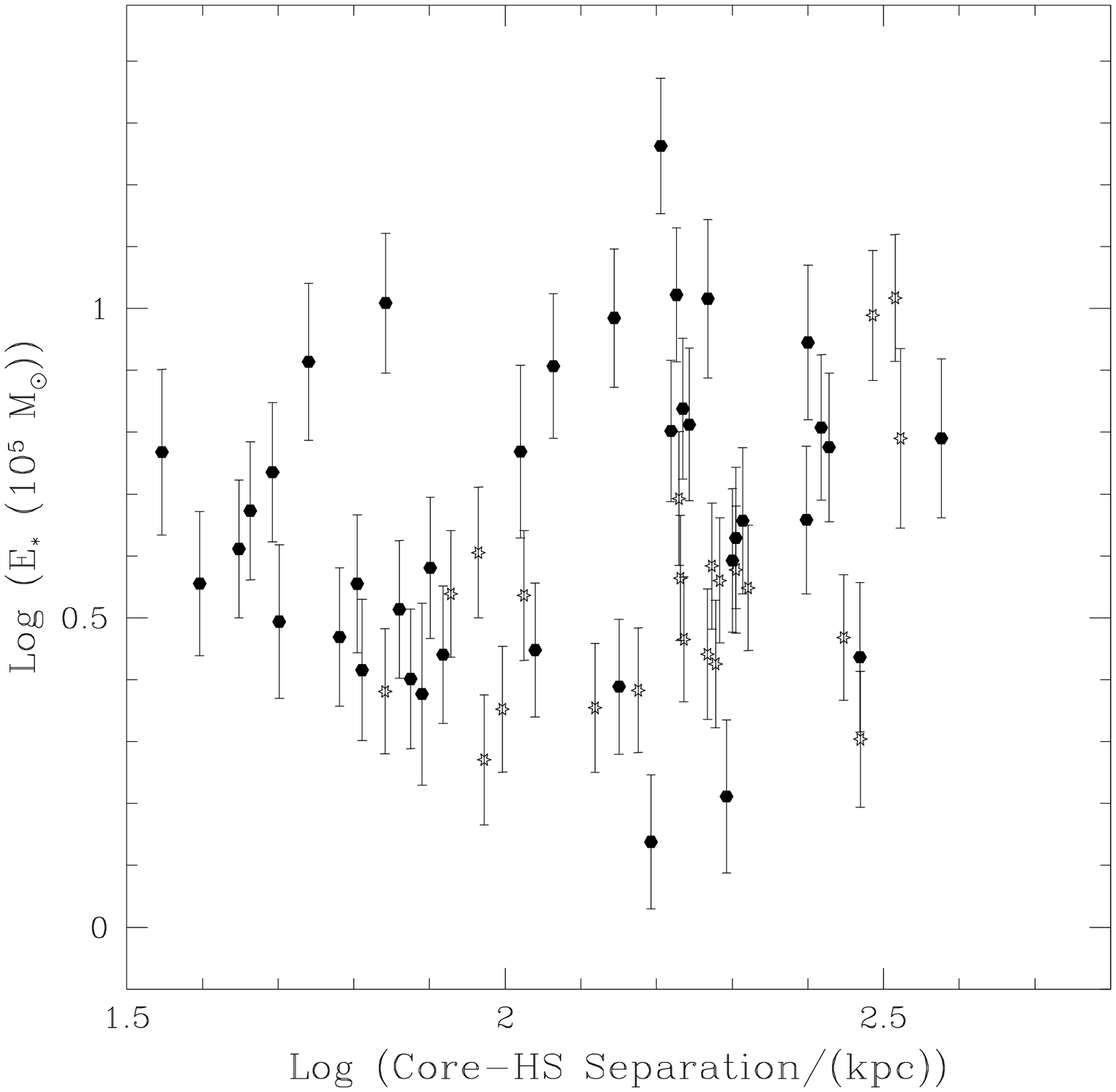}
         \caption{
Source total Energy $E_T$ as a function of core-hotspot distance $r$ for
b=1.  }
          \label{figeda}
    \end{figure}

\begin{figure}
    \centering
    \includegraphics[width=\textwidth]{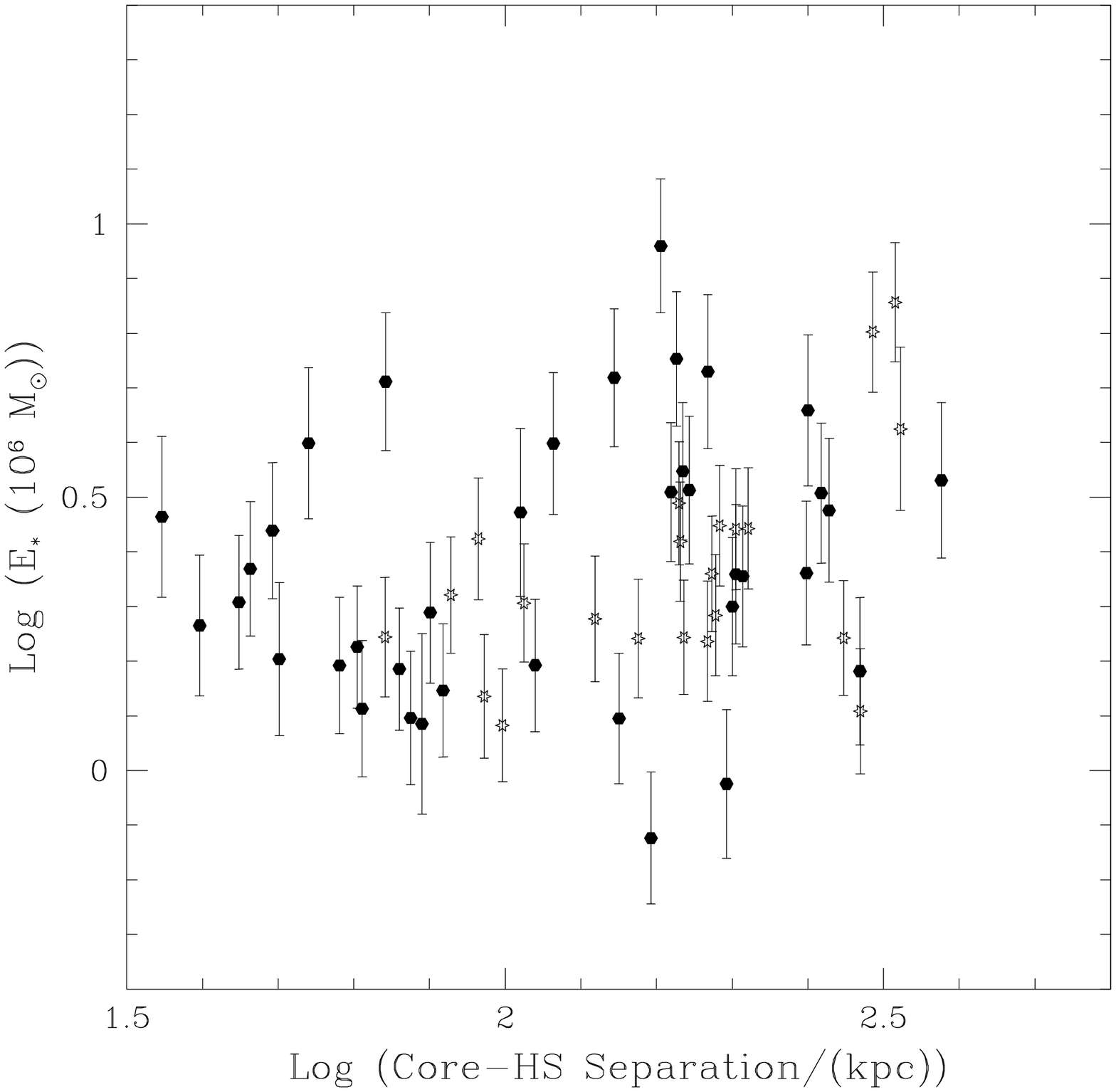}
         \caption{
Source total Energy $E_T$ as a function of core-hotspot distance $r$ for
b=0.25. }
          \label{figedb}
    \end{figure}

\begin{figure}
    \centering
    \includegraphics[width=\textwidth]{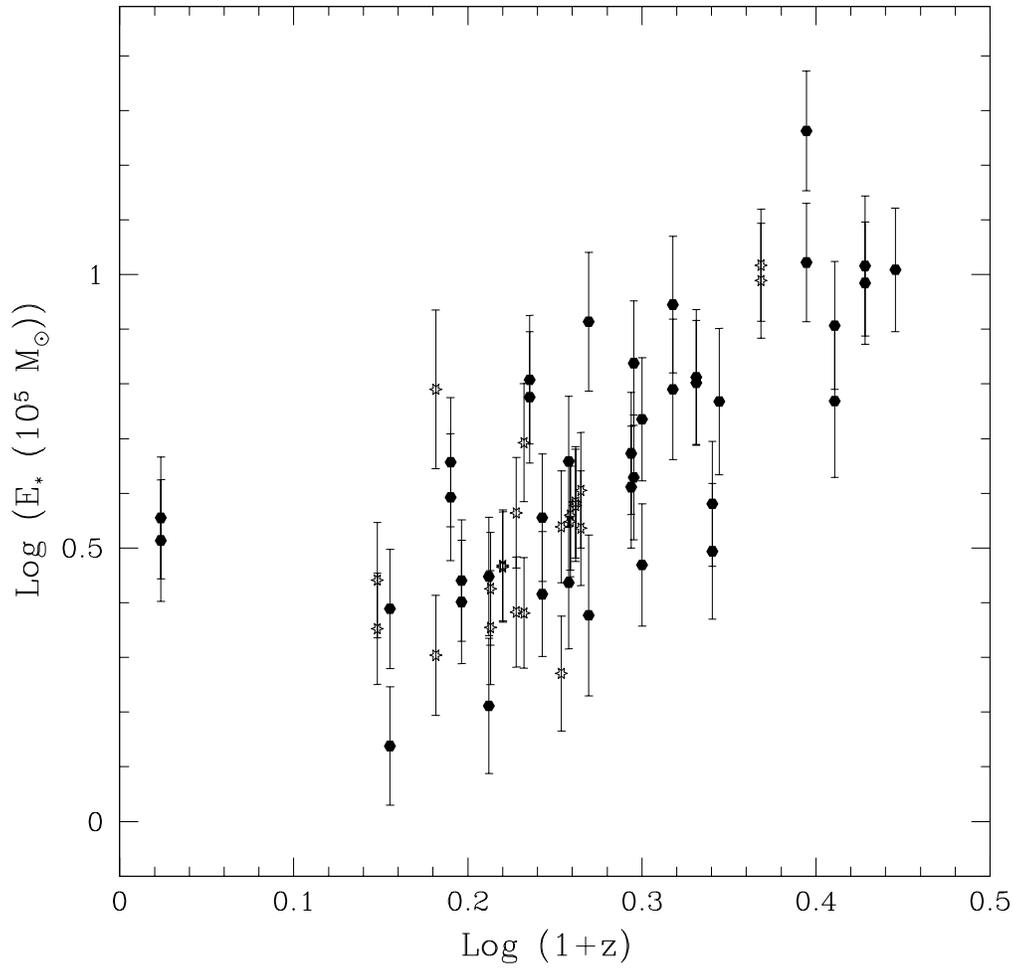}
         \caption{
Source total Energy $E_T$ as a function of redshift (1 + z)  for
b=1.  }
          \label{figeza}
    \end{figure}

\begin{figure}
    \centering
    \includegraphics[width=\textwidth]{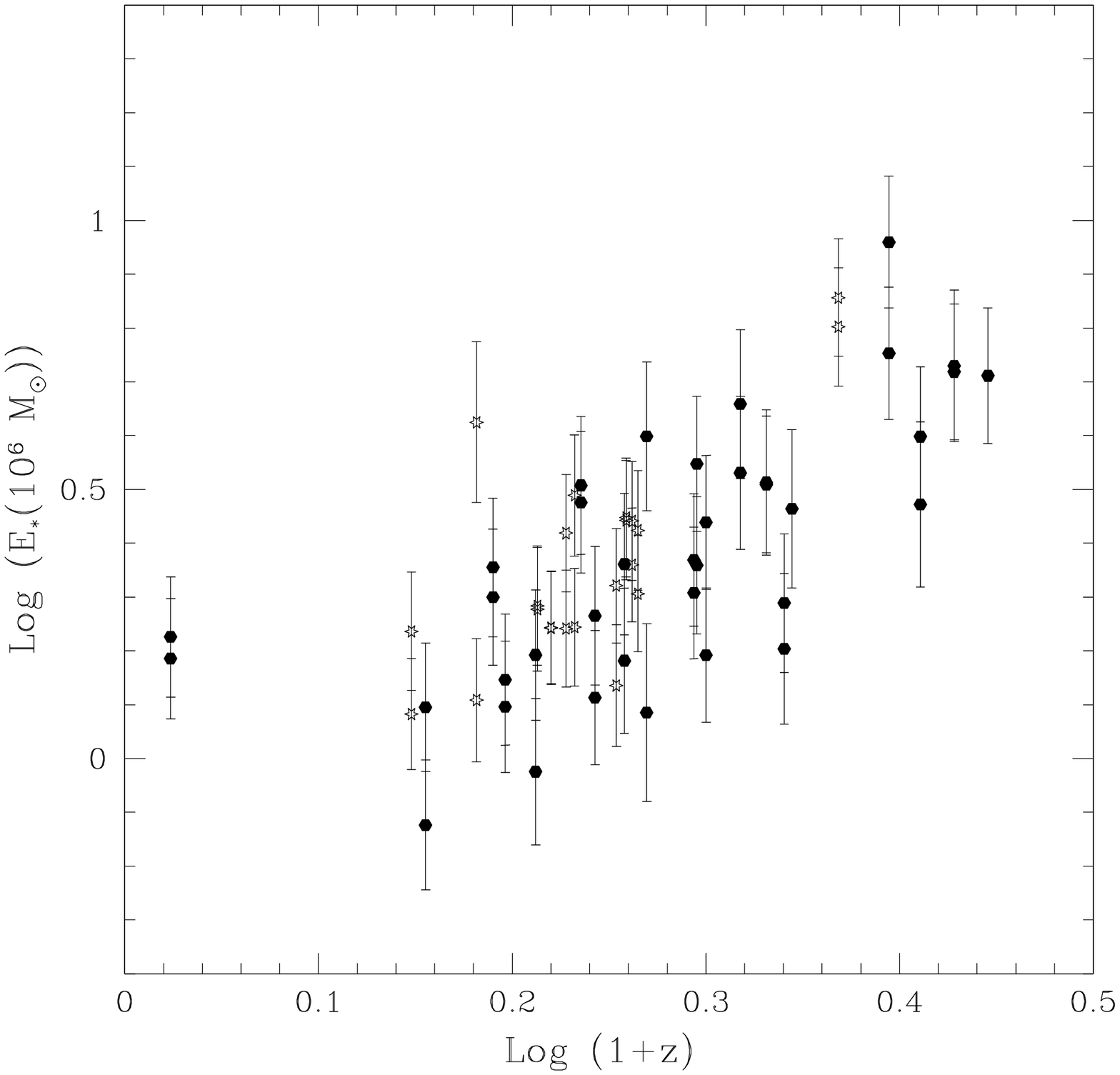}
         \caption{
Source total Energy $E_T$ as a function of redshift (1 + z) for
b=0.25. }
         \label{figezb}
    \end{figure}

\clearpage

\appendix
\section{Comparison of Methods for Determining Spectral Ages}

In sections 2.2 and 3.1 we determined the spectral
age and velocity using a two position measurement of
the spectral index gradient.
Here, we consider the velocity of the source
obtained at multiple locations on each side of each source, and
find results that are very similar and consistent with those
obtained in section 3.1.

\begin{figure}
    \centering
    \includegraphics[width=\textwidth]{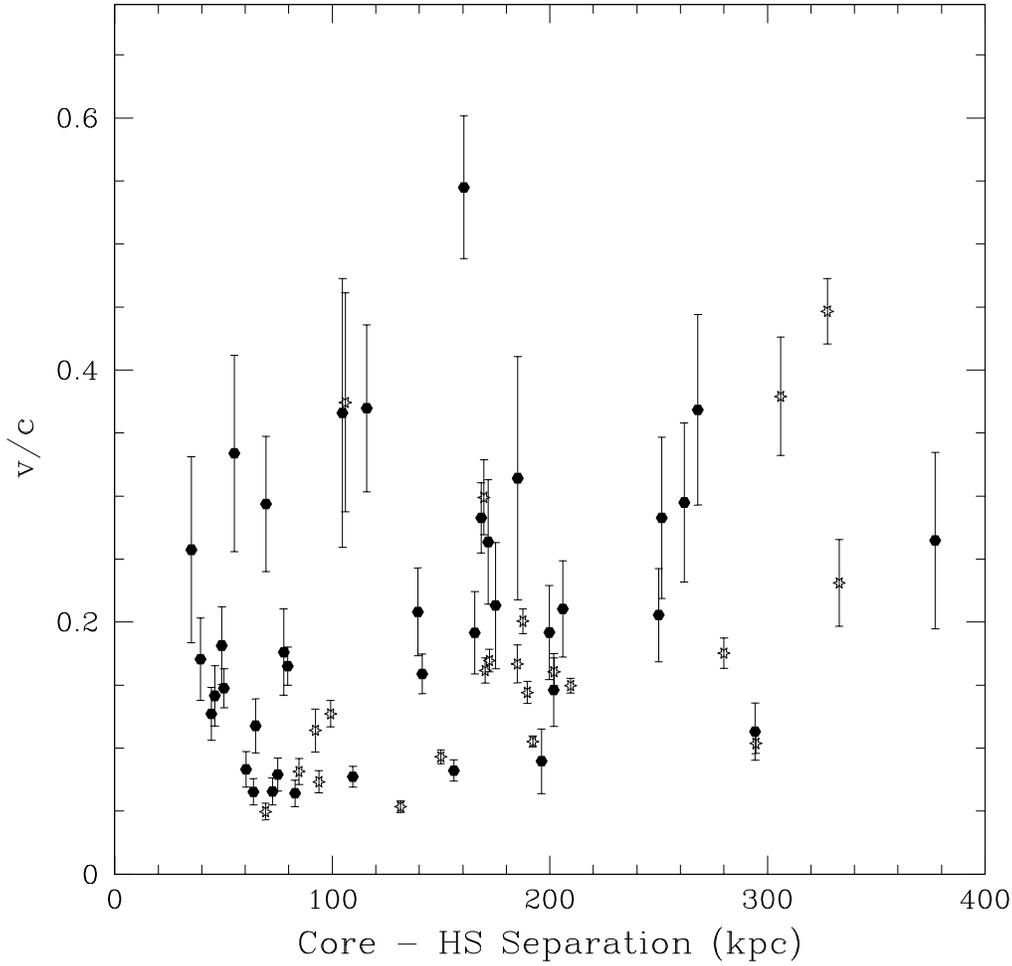}
         \caption{v/c as a function of core to hot spot distance for b=1.
For the new sample of eleven radio galaxies we use the weighted average
velocities from Table~\ref{tabposn}.
The slope is $4.0 \pm 1.0 \times 10^{-4}$
with  $\chi^2 = 929$
consistent with the results obtained using method A
(Table~\ref{tabfitsb1}).
}
         \label{figvd1}
    \end{figure}

\begin{figure}
    \centering
\includegraphics[width=\textwidth]{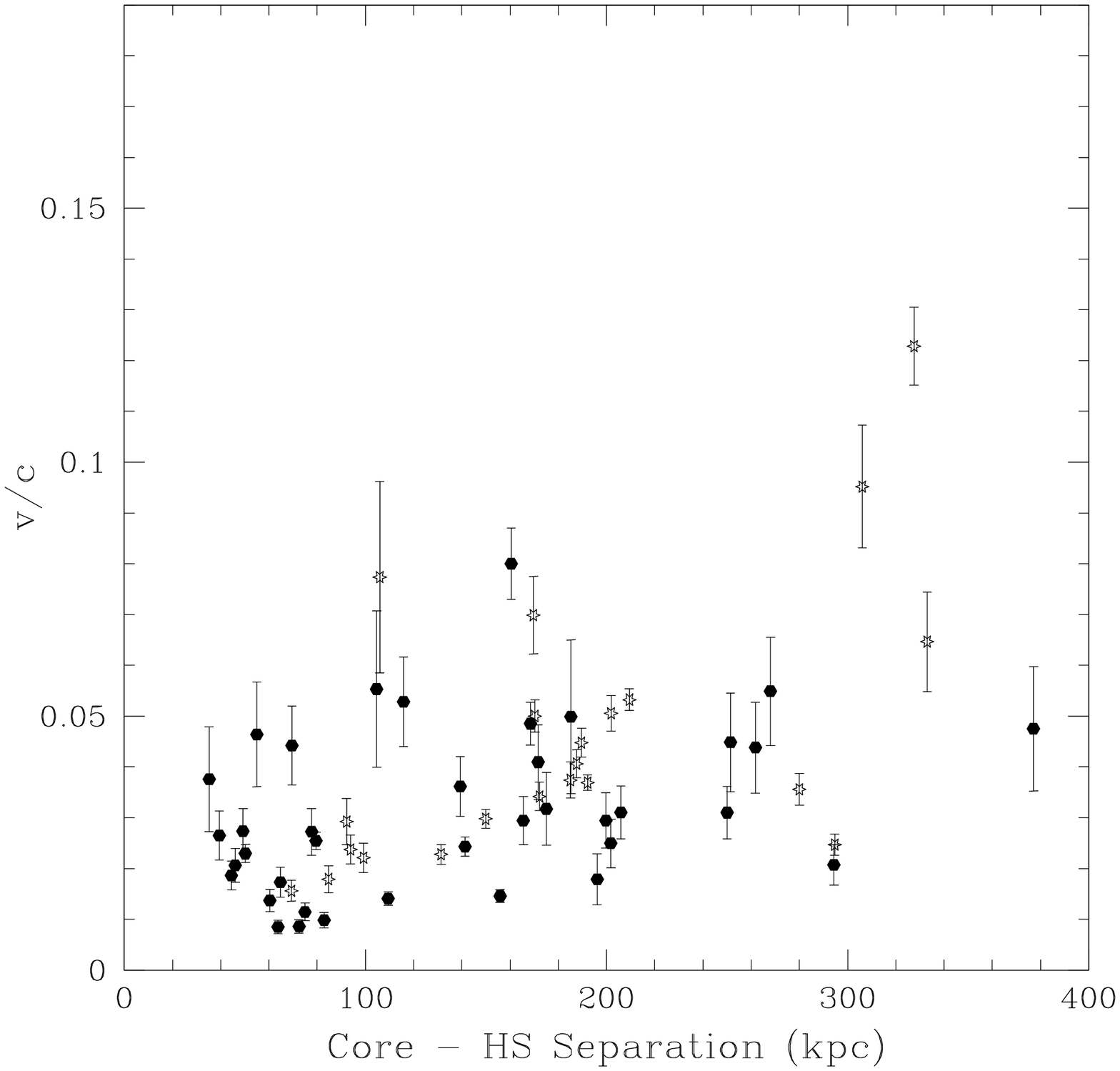}
         \caption{v/c as a function of core to hot spot distance for b=0.25.
For the new sample of eleven radio galaxies we use the weighted average
velocities from
Table~\ref{tabposn}.
The slope is $1.3 \pm 0.2 \times 10^{-4}$
with $\chi^2 = 972$
consistent with the results obtained using method A (Table~\ref{tabfits}).
}
         \label{figvd2}
    \end{figure}

\begin{figure}
    \centering
     \includegraphics[width=\textwidth]{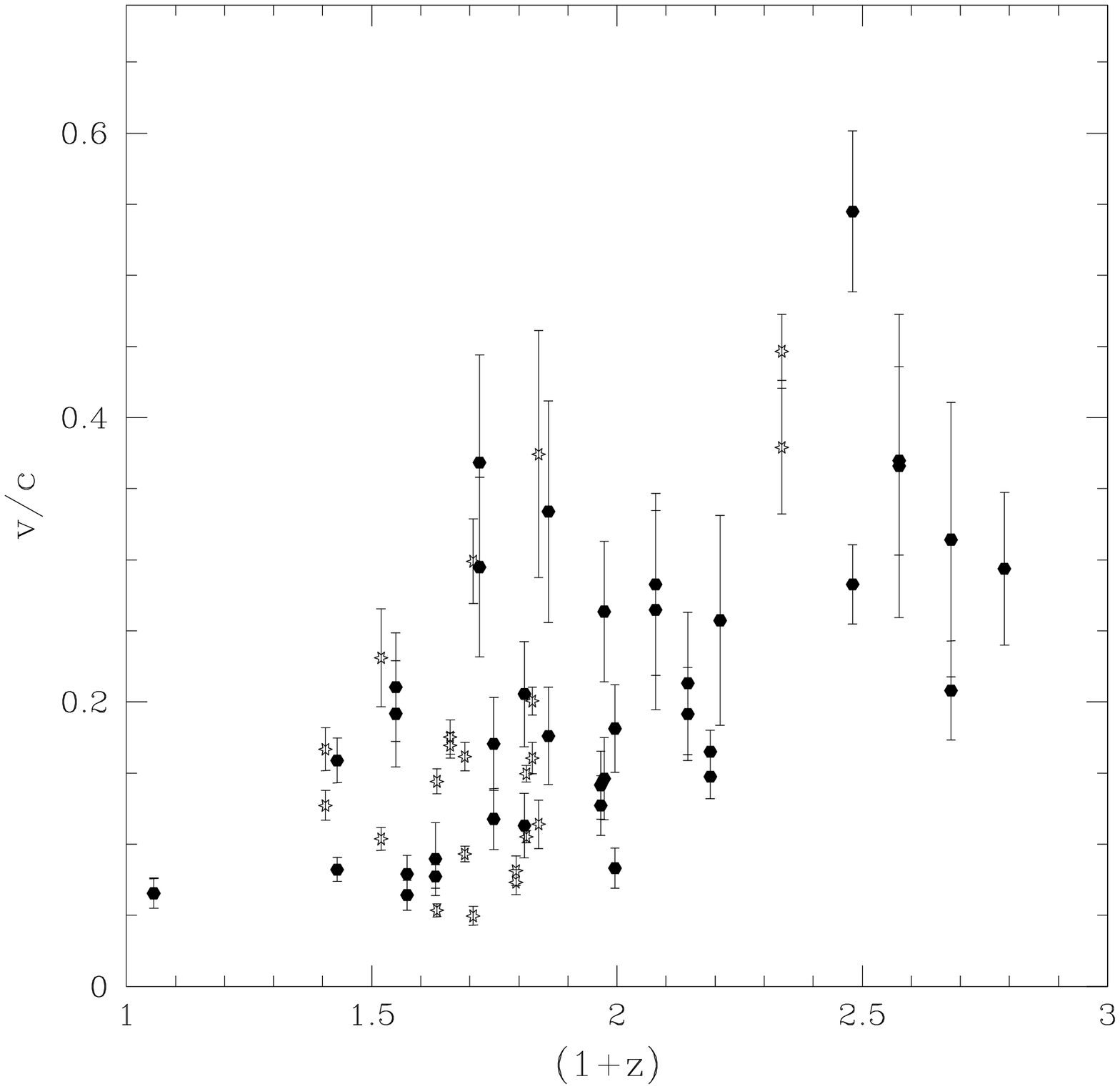}
         \caption{v/c as a function of redshift for b=1.
For the new sample of eleven radio galaxies we use the weighted average
velocities from
Table~\ref{tabposn}.
The slope is $0.11  \pm 0.03 $
with $\chi^2 = 957$
consistent with the results obtained using method A
(Table~\ref{tabfitsb1}).
}
         \label{figvz1}
    \end{figure}

\begin{figure}
    \centering
     \includegraphics[width=\textwidth]{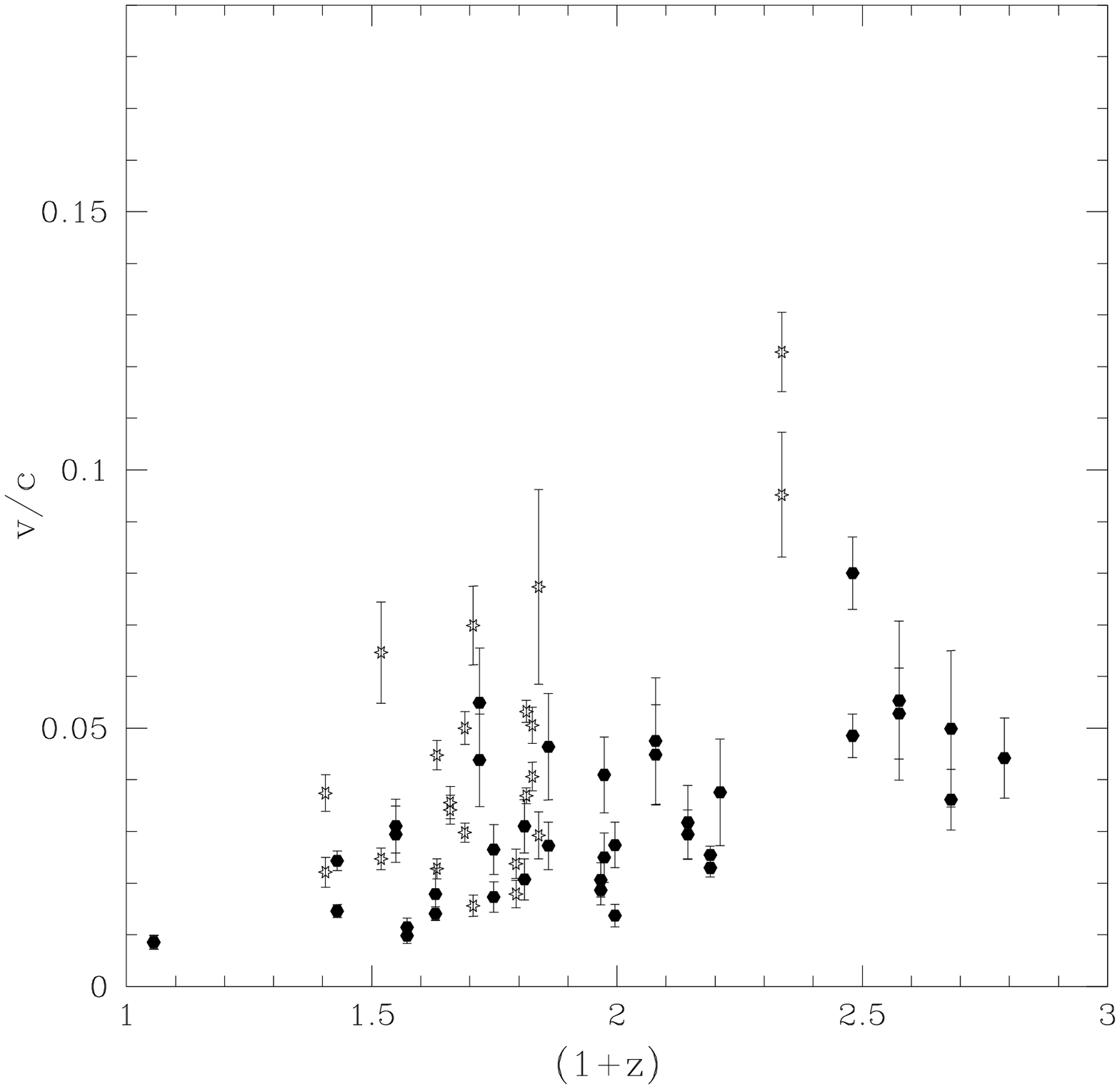}
         \caption{v/c as a function of redshift for b=0.25.
For the new sample of eleven radio galaxies we use the weighted average
velocities from
Table~\ref{tabposn}.
The slope is $0.022 \pm 0.0045$
with $\chi^2 = 1109$
consistent with the results obtained from the
two position method (Table~\ref{tabfits}).
}
         \label{figvz2}
    \end{figure}

To obtain the velocity at several different locations across the source,
we determine the spectral index at intervals of 2.5 arcsec (roughly
one CLEAN beam FWHM) from the hot spot
along the brightness ridge of the lobe. At each position
we determine the
break frequency by
comparing the observed spectral index with the calculations
of the JP model
using the
Myers and Spangler method. The one sigma range of the spectral index was
used to obtain the uncertainty of the break frequency.
The age of the electrons at that location is obtained using equation (1),
and the age and distance from the hot spot are combined to obtain
the rate of growth of the source
(the expansion velocity);
the source, angular separation of each interval from the hot spot,
break frequencies at zero redshift, and source velocities are
listed in (Table \ref{tabposn}). The average rate of growth of
each side is obtained using the weighted mean expansion velocity
for each lobe, and is also listed.
Positions near the hot spot (where there
is little spectral aging) have high and uncertain break frequencies.
The weighted average
velocities for each lobe are consistent with those we determined using the two
position method.
Figures \ref{figvd1} - \ref{figvz2} show the average velocity
as a function of core to hot
spot distance and redshift including the velocities derived from the multiple
position method. The results are
in very good agreement with our results from the two position method,
which is easily seen by comparing Figures 1, 2, 3, and 4
with Figures A.1, A.2, A.3, and A.4. The agreement between the two
methods is about 1 sigma.  If the two velocity determinations are
compared without accounting for the uncertainty of each velocity,
the scatter between the two methods is
about 24\%.

To view the comparison between the two determinations of the rate of growth
of the sources, we can consider fits to the 11 new radio galaxies only.
There are no discernible differences between fits to $v(r)$ or $v(1+z)$
for $b=1$ and $b=0.25$ using method A, which is that described in the
main text, and method B, which is described here. Generally, the
$\chi^2$ of the fits is very large, so it is appropriate to multiple
the error of the best fit parameters by $\sqrt{(\chi^2/58)}$, as described
earlier.  For $b=1$, the slope of $v_B(r) = 0.0005 \pm 0.0005$, while
$v_A(r) = 0.0009 \pm 0.0002$, both of which indicate either no or
a very weak dependence of
$v$ with $D$ and which are consistent at about one sigma;
we also find that $v_B(1+z)= 0.1 \pm 0.3$, while
$v_A(1+z)=0.1 \pm 0.3$, which are consistent
and neither of which indicates a significant trend.
Here and throughout the units of distance r are kpc and the velocity
v is relative to the speed of light.
For $b=0.25$,
the slope of $v_B(r) = 0.0001 \pm 0.0001$, while $v_A(r) = 0.0003 \pm 
0.00003$,
which are consistent at a few sigma and neither of
which indicates a strong trend; we also find
$v_B(1+z) = 0.05 \pm 0.05$ and $v_A(1+z) = 0.06 \pm 0.06$, both of
which are consistent with zero slope and with each other.

In addition, we can compare the slopes of the fits to the full samples
with the velocities of the eleven new sources determined using method A
and method B.  For $b=1$, the slope of $v_B(r) = 0.0004 \pm 0.0004$, while
$v_A(r) = 0.0001 \pm 0.0001$, both of which indicate no dependence of
$v$ with $D$; we also find that $v_B(1+z)= 0.1 \pm 0.1$, while
$v_A(1+z)=0.02 \pm 0.015$, which are consistent at the one sigma
level and neither of which indicates a significant trend. For $b=0.25$,
the slope of $v_B(r) = 0.0001 \pm 0.0001$, while $v_A(r) = 0.0007 \pm 0.0004$,
which are consistent at better than two sigma and neither of
which indicates a strong trend; we also find
$v_B(1+z) = 0.02 \pm 0.02$ and $v_A(1+z) = 0.1 \pm 0.1$, both of
which are consistent with zero slope and with each other.

Clearly, results obtained with method A and method B are very similar.

\begin{longtable}{lclrrr}
\caption{\label{tabposn} Estimates of Velocities at Multiple Positions in the
Lobes in the Eleven Radio Galaxies}\\
\hline\hline
Source & Side &    $\theta$ & $\nu_{0T}$ &      v/c (b=0.25)  & v/c (b=1) \\
         &                               &         arcsec & GHz &        \\
\hline
\endfirsthead
\caption{continued.}\\
\hline\hline
Source & Side &    $\theta$ & $\nu_{0T}$ &      v/c (b=0.25)  & v/c (b=1) \\
         &                               &         arcsec & GHz &        \\
\hline
\endhead
\hline
3C6.1   &       N       &       5       &$      57      \pm     53 
$&$     0.0629  \pm     0.0303  $&$     0.303   \pm     0.143   $\\
         &               &       7.5     &$      140     \pm     230 
 $&$     0.1510  \pm     0.1219  $&$     0.728   \pm     0.584   $\\
         &               &       10      &$      25      \pm     13 
 $&$     0.0839  \pm     0.0245  $&$     0.404   \pm     0.111   $\\
         &               &       avg     &$ 
 $&$     0.0770  \pm     0.0190  $&$     0.370   \pm     0.090   $\\
3C6.1   &       S       &       5       &$      14      \pm     4 
$&$     0.0272  \pm     0.0047  $&$     0.107   \pm     0.018   $\\
         &               &       7.5     &$      30      \pm     18 
 $&$     0.0603  \pm     0.0185  $&$     0.236   \pm     0.071   $\\
         &               &       avg     &$ 
 $&$     0.0290  \pm     0.0050  $&$     0.110   \pm     0.020   $\\
3C34    &       E       &       5       &$      66      \pm     66 
$&$     0.0388  \pm     0.0196  $&$     0.125   \pm     0.063   $\\
         &               &       7.5     &$      32      \pm     18 
 $&$     0.0405  \pm     0.0117  $&$     0.131   \pm     0.038   $\\
         &               &       10      &$      38      \pm     26 
 $&$     0.0591  \pm     0.0200  $&$     0.191   \pm     0.064   $\\
         &               &       12.5    &$      14      \pm     4 
 $&$     0.0450  \pm     0.0071  $&$     0.145   \pm     0.022   $\\
         &               &       15      &$      14      \pm     4 
 $&$     0.0532  \pm     0.0083  $&$     0.172   \pm     0.026   $\\
         &               &       17.5    &$      12      \pm     3 
 $&$     0.0572  \pm     0.0083  $&$     0.185   \pm     0.026   $\\
         &               &       20      &$      7.1     \pm     1.1 
 $&$     0.0507  \pm     0.0051  $&$     0.164   \pm     0.016   $\\
         &               &       avg     &$ 
 $&$     0.0500  \pm     0.0030  $&$     0.160   \pm     0.010   $\\
3C34    &       W       &       5       &$      18      \pm     7 
$&$     0.0202  \pm     0.0041  $&$     0.063   \pm     0.013   $\\
         &               &       7.5     &$      13      \pm     4 
 $&$     0.0261  \pm     0.0038  $&$     0.081   \pm     0.012   $\\
         &               &       10      &$      19      \pm     69 
 $&$     0.0411  \pm     0.0754  $&$     0.128   \pm     0.234   $\\
         &               &       12.5    &$      7.5     \pm     1.6 
 $&$     0.0324  \pm     0.0037  $&$     0.101   \pm     0.011   $\\
         &               &       15      &$      6.5     \pm     1.0 
 $&$     0.0363  \pm     0.0032  $&$     0.113   \pm     0.009   $\\
         &               &       avg     &$ 
 $&$     0.0298  \pm     0.0020  $&$     0.093   \pm     0.005   $\\
3C41    &       N       &       2.5     &$      29      \pm     18 
$&$     0.0165  \pm     0.0054  $&$     0.051   \pm     0.017   $\\
         &               &       5       &$      17      \pm     6 
 $&$     0.0256  \pm     0.0051  $&$     0.079   \pm     0.016   $\\
         &               &       7.5     &$      8.6     \pm     2.1 
 $&$     0.0271  \pm     0.0043  $&$     0.084   \pm     0.013   $\\
         &               &       avg     &$ 
 $&$     0.0240  \pm     0.0030  $&$     0.073   \pm     0.008   $\\
3C41    &       S       &       2.5     &$      33      \pm     22 
$&$     0.0206  \pm     0.0074  $&$     0.094   \pm     0.033   $\\
         &               &       5       &$      5.9     \pm     1.2 
 $&$     0.0175  \pm     0.0028  $&$     0.080   \pm     0.011   $\\
         &               &       avg     &$ 
 $&$     0.0180  \pm     0.0030  $&$     0.080   \pm     0.010   $\\
3C44    &       N       &       5       &$      18      \pm     8 
$&$     0.0220  \pm     0.0053  $&$     0.104   \pm     0.023   $\\
         &               &       10      &$      17      \pm     7 
 $&$     0.0433  \pm     0.0102  $&$     0.204   \pm     0.045   $\\
         &               &       15      &$      4.2     \pm     0.9 
 $&$     0.0337  \pm     0.0048  $&$     0.159   \pm     0.017   $\\
         &               &       20      &$      16      \pm     8 
 $&$     0.0836  \pm     0.0218  $&$     0.394   \pm     0.096   $\\
         &               &       25      &$      6.5     \pm     1.1 
 $&$     0.0666  \pm     0.0089  $&$     0.314   \pm     0.030   $\\
         &               &       avg     &$ 
 $&$     0.0360  \pm     0.0030  $&$     0.170   \pm     0.010   $\\
3C44    &       S       &       5       &$      28      \pm     15 
$&$     0.0277  \pm     0.0080  $&$     0.131   \pm     0.036   $\\
         &               &       10      &$      7.6     \pm     1.8 
 $&$     0.0290  \pm     0.0044  $&$     0.137   \pm     0.017   $\\
         &               &       15      &$      5.2     \pm     0.6 
 $&$     0.0358  \pm     0.0041  $&$     0.170   \pm     0.011   $\\
         &               &       20      &$      3.4     \pm     0.3 
 $&$     0.0390  \pm     0.0042  $&$     0.185   \pm     0.010   $\\
         &               &       avg     &$ 
 $&$     0.0340  \pm     0.0030  $&$     0.170   \pm     0.010   $\\
3C54    &       N       &       2.5     &$      140     \pm     220 
$&$     0.0391  \pm     0.0317  $&$     0.124   \pm     0.100    $\\
         &               &       5       &$      16      \pm     7 
 $&$     0.0267  \pm     0.0058  $&$     0.085   \pm     0.018   $\\
         &               &       7.5     &$      15      \pm     6 
 $&$     0.0394  \pm     0.0084  $&$     0.125   \pm     0.027   $\\
         &               &       10      &$      23      \pm     12 
 $&$     0.0646  \pm     0.0167  $&$     0.205   \pm     0.053   $\\
         &               &       12.5    &$      21      \pm     10 
 $&$     0.0760  \pm     0.0185  $&$     0.241   \pm     0.059   $\\
         &               &       15      &$      15      \pm     5 
 $&$     0.0775  \pm     0.0151  $&$     0.245   \pm     0.047   $\\
         &               &       17.5    &$      15      \pm     6 
 $&$     0.0919  \pm     0.0182  $&$     0.291   \pm     0.057   $\\
         &               &       20      &$      7.4     \pm     1.7 
 $&$     0.0729  \pm     0.0098  $&$     0.231   \pm     0.031   $\\
         &               &       22.5    &$      6.0     \pm     1.2 
 $&$     0.0734  \pm     0.0091  $&$     0.233   \pm     0.028   $\\
         &               &       avg     &$ 
 $&$     0.0510  \pm     0.0030  $&$     0.160   \pm     0.010   $\\
3C54    &       S       &       2.5     &$      34      \pm     23 
$&$     0.0232  \pm     0.0085  $&$     0.110   \pm     0.039   $\\
         &               &       5       &$      18      \pm     8 
 $&$     0.0344  \pm     0.0086  $&$     0.163   \pm     0.038   $\\
         &               &       7.5     &$      9.8     \pm     2.6 
 $&$     0.0376  \pm     0.0067  $&$     0.178   \pm     0.027   $\\
         &               &       10      &$      7.6     \pm     1.4 
 $&$     0.0442  \pm     0.0066  $&$     0.209   \pm     0.024   $\\
         &               &       12.5    &$      4.9     \pm     0.6 
 $&$     0.0442  \pm     0.0056  $&$     0.209   \pm     0.018   $\\
         &               &       15      &$      4.0     \pm     0.4 
 $&$     0.0480  \pm     0.0060  $&$     0.228   \pm     0.019   $\\
         &               &       avg     &$ 
 $&$     0.0410  \pm     0.0030  $&$     0.200   \pm     0.010   $\\
3C114   &       N       &       2.5     &$      210     \pm     370 
$&$     0.0463  \pm     0.0406  $&$     0.132   \pm     0.116 $\\
         &               &       5       &$      210     \pm     368 
 $&$     0.0926  \pm     0.0812  $&$     0.264   \pm     0.231   $\\
         &               &       7.5     &$      35      \pm     22 
 $&$     0.0570  \pm     0.0178  $&$     0.162   \pm     0.051   $\\
         &               &       10      &$      7.6     \pm     1.4 
 $&$     0.0353  \pm     0.0039  $&$     0.100   \pm     0.011   $\\
         &               &       12.5    &$      5.4     \pm     0.9 
 $&$     0.0370  \pm     0.0036  $&$     0.105   \pm     0.010   $\\
         &               &       15      &$      2.5     \pm     0.3 
 $&$     0.0304  \pm     0.0027  $&$     0.087   \pm     0.007   $\\
         &               &       17.5    &$      2.7     \pm     0.5 
 $&$     0.0370  \pm     0.0038  $&$     0.105   \pm     0.011   $\\
         &               &       20      &$      3.8     \pm     0.4 
 $&$     0.0498  \pm     0.0037  $&$     0.142   \pm     0.010   $\\
         &               &       avg     &$ 
 $&$     0.0370  \pm     0.0010  $&$     0.100   \pm     0.004   $\\
3C114   &       S       &       2.5     &$      86      \pm     110 
$&$     0.0296  \pm     0.0190  $&$     0.083   \pm     0.053 $\\
         &               &       5       &$      24      \pm     12 
 $&$     0.0315  \pm     0.0081  $&$     0.088   \pm     0.023   $\\
         &               &       7.5     &$      14      \pm     6 
 $&$     0.0364  \pm     0.0072  $&$     0.102   \pm     0.020   $\\
         &               &       10      &$      11      \pm     3 
 $&$     0.0423  \pm     0.0061  $&$     0.118   \pm     0.017   $\\
         &               &       12.5    &$      21      \pm     10 
 $&$     0.0740  \pm     0.0179  $&$     0.207   \pm     0.050   $\\
         &               &       15      &$      6.6     \pm     1.5 
 $&$     0.0493  \pm     0.0056  $&$     0.138   \pm     0.016   $\\
         &               &       17.5    &$      6.2     \pm     1.1 
 $&$     0.0557  \pm     0.0050  $&$     0.156   \pm     0.014   $\\
         &               &       20      &$      6.9     \pm     1.3 
 $&$     0.0672  \pm     0.0064  $&$     0.188   \pm     0.018   $\\
         &               &       22.5    &$      4.9     \pm     0.6 
 $&$     0.0634  \pm     0.0040  $&$     0.178   \pm     0.011   $\\
         &               &       avg     &$ 
 $&$     0.0530  \pm     0.0020  $&$     0.150   \pm     0.006   $\\
3C142.1 
&       N       &       2.5     &$      380     \pm     900     $&$ 
0.0306  \pm     0.0360  $&$     0.185   \pm
0.216   $\\
         &               &       5       &$      140     \pm     190 
 $&$     0.0368  \pm     0.0263  $&$     0.222   \pm     0.156   $\\
         &               &       7.5     &$      96      \pm     96 
 $&$     0.0460  \pm     0.0239  $&$     0.278   \pm     0.139   $\\
         &               &       10      &$      18      \pm     7 
 $&$     0.0267  \pm     0.0062  $&$     0.162   \pm     0.030   $\\
         &               &       12.5    &$      6.5     \pm     1.1 
 $&$     0.0200  \pm     0.0034  $&$     0.121   \pm     0.011   $\\
         &               &       avg     &$ 
 $&$     0.0220  \pm     0.0030  $&$     0.130   \pm     0.010   $\\
3C142.1 
&       S       &       2.5     &$      14000   \pm     330000  $&$ 
0.1279  \pm     1.5354  $&$     0.568   \pm
6.814   $\\
         &               &       5       &$      14000   \pm     330000 
 $&$     0.2559  \pm     3.0707  $&$     1.136   \pm     13.628  $\\
         &               &       7.5     &$      61      \pm     57 
 $&$     0.0256  \pm     0.0121  $&$     0.114   \pm     0.053   $\\
         &               &       10      &$      38      \pm     24 
 $&$     0.0269  \pm     0.0088  $&$     0.120   \pm     0.038   $\\
         &               &       12.5    &$      29      \pm     16 
 $&$     0.0291  \pm     0.0083  $&$     0.129   \pm     0.036   $\\
         &               &       15      &$      35      \pm     21 
 $&$     0.0384  \pm     0.0120  $&$     0.170   \pm     0.051   $\\
         &               &       17.5    &$      22      \pm     11 
 $&$     0.0358  \pm     0.0091  $&$     0.159   \pm     0.039   $\\
         &               &       20      &$      38      \pm     24 
 $&$     0.0539  \pm     0.0176  $&$     0.239   \pm     0.076   $\\
         &               &       22.5    &$      18      \pm     8 
 $&$     0.0411  \pm     0.0095  $&$     0.183   \pm     0.040   $\\
         &               &       25      &$      26      \pm     14 
 $&$     0.0556  \pm     0.0153  $&$     0.247   \pm     0.065   $\\
         &               &       27.5    &$      35      \pm     21 
 $&$     0.0704  \pm     0.0220  $&$     0.312   \pm     0.094   $\\
         &               &       30      &$      17      \pm     7 
 $&$     0.0539  \pm     0.0122  $&$     0.239   \pm     0.051   $\\
         &               &       avg     &$ 
 $&$     0.0370  \pm     0.0030  $&$     0.170   \pm     0.010   $\\
3C169.1 
&       N       &       2.5     &$      540     \pm     1900    $&$ 
0.0460  \pm     0.0827  $&$     0.108   \pm 0.194
$\\
         &               &       5       &$      240     \pm     570 
 $&$     0.0613  \pm     0.0735  $&$     0.144   \pm     0.173   $\\
         &               &       7.5     &$      34      \pm     22 
 $&$     0.0345  \pm     0.0113  $&$     0.081   \pm     0.026   $\\
         &               &       10      &$      26      \pm     15 
 $&$     0.0409  \pm     0.0119  $&$     0.096   \pm     0.028   $\\
         &               &       12.5    &$      4.9     \pm     0.8 
 $&$     0.0219  \pm     0.0020  $&$     0.051   \pm     0.005   $\\
         &               &       avg     &$ 
 $&$     0.0023  \pm     0.0020  $&$     0.053   \pm     0.004   $\\
3C169.1 
&       S       &       2.5     &$      13000   \pm     300000  $&$ 
0.2368  \pm     2.6043  $&$     0.766   \pm
8.424   $\\
         &               &       5       &$      130     \pm     200 
 $&$     0.0474  \pm     0.0380  $&$     0.153   \pm     0.123   $\\
         &               &       7.5     &$      49      \pm     48 
 $&$     0.0431  \pm     0.0210  $&$     0.139   \pm     0.068   $\\
         &               &       10      &$      52      \pm     52 
 $&$     0.0592  \pm     0.0298  $&$     0.191   \pm     0.096   $\\
         &               &       12.5    &$      30      \pm     20 
 $&$     0.0564  \pm     0.0190  $&$     0.182   \pm     0.061   $\\
         &               &       15      &$      14      \pm     6 
 $&$     0.0466  \pm     0.0095  $&$     0.151   \pm     0.030   $\\
         &               &       17.5    &$      15      \pm     6 
 $&$     0.0552  \pm     0.0115  $&$     0.179   \pm     0.037   $\\
         &               &       20      &$      11      \pm     3 
 $&$     0.0541  \pm     0.0083  $&$     0.175   \pm     0.026   $\\
         &               &       22.5    &$      5.0     \pm     0.8 
 $&$     0.0410  \pm     0.0039  $&$     0.133   \pm     0.012   $\\
         &               &       25      &$      4.4     \pm     1.6 
 $&$     0.0431  \pm     0.0082  $&$     0.139   \pm     0.026   $\\
         &               &       avg     &$ 
 $&$     0.0450  \pm     0.0030  $&$     0.140   \pm     0.010   $\\
3C172   &       N       &       2.5     &$      60      \pm     48 
$&$     0.0121  \pm     0.0050  $&$     0.050   \pm     0.020   $\\
         &               &       5       &$      52      \pm     46 
 $&$     0.0227  \pm     0.0101  $&$     0.093   \pm     0.041   $\\
         &               &       7.5     &$      13      \pm     4 
 $&$     0.0167  \pm     0.0030  $&$     0.069   \pm     0.011   $\\
         &               &       10      &$      130     \pm     240 
 $&$     0.0725  \pm     0.0656  $&$     0.297   \pm     0.268   $\\
         &               &       12.5    &$      18      \pm     7 
 $&$     0.0336  \pm     0.0069  $&$     0.138   \pm     0.026   $\\
         &               &       15      &$      39      \pm     30 
 $&$     0.0588  \pm     0.0228  $&$     0.241   \pm     0.092   $\\
         &               &       17.5    &$      39      \pm     30 
 $&$     0.0686  \pm     0.0266  $&$     0.281   \pm     0.107   $\\
         &               &       20      &$      60      \pm     48 
 $&$     0.0967  \pm     0.0396  $&$     0.396   \pm     0.160   $\\
         &               &       22.5    &$      32      \pm     25 
 $&$     0.0796  \pm     0.0318  $&$     0.326   \pm     0.128   $\\
         &               &       25      &$      79      \pm     86 
 $&$     0.1394  \pm     0.0761  $&$     0.572   \pm     0.309   $\\
         &               &       27.5    &$      32      \pm     22 
 $&$     0.0973  \pm     0.0343  $&$     0.399   \pm     0.138   $\\
         &               &       30      &$      12      \pm     34 
 $&$     0.0659  \pm     0.0107  $&$     0.270   \pm     0.039   $\\
         &               &       32.5    &$      5.4     \pm     1.1 
 $&$     0.0471  \pm     0.0063  $&$     0.193   \pm     0.022   $\\
         &               &       avg     &$ 
 $&$     0.0250  \pm     0.0020  $&$     0.100   \pm     0.008   $\\
3C172   &       S       &       2.5     &$      440     \pm     1500 
$&$        0.0312  \pm     0.0539  $&$     0.112   \pm     0.193
$\\
         &               &       5       &$      13000   \pm     270000 
 $&$     0.3431  \pm     3.4312  $&$     1.227   \pm     12.273  $\\
         &               &       7.5     &$      13000   \pm     270000 
 $&$     0.5147  \pm     5.1469  $&$     1.841   \pm     18.410
$\\
         &               &       10      &$      13000   \pm     270000 
 $&$     0.6862  \pm     6.8625  $&$     2.455   \pm     24.547
$\\
         &               &       12.5    &$      110     \pm     180 
 $&$     0.0780  \pm     0.0640  $&$     0.279   \pm     0.229   $\\
         &               &       15      &$      13000   \pm     270000 
 $&$     1.0294  \pm     10.2937 $&$     3.682   \pm     36.820
$\\
         &               &       17.5    &$      210     \pm     520 
 $&$     0.1501  \pm     0.1879  $&$     0.537   \pm     0.672   $\\
         &               &       20      &$      13000   \pm     270000 
 $&$     1.3725  \pm     13.7250 $&$     4.909   \pm     49.093
$\\
         &               &       22.5    &$      110     \pm     180 
 $&$     0.1404  \pm     0.1152  $&$     0.502   \pm     0.412   $\\
         &               &       25      &$      210     \pm     520 
 $&$     0.2145  \pm     0.2685  $&$     0.767   \pm     0.960   $\\
         &               &       27.5    &$      15      \pm     5 
 $&$     0.0629  \pm     0.0113  $&$     0.225   \pm     0.040   $\\
         &               &       32.5    &$      13      \pm     9 
 $&$     0.0708  \pm     0.0230  $&$     0.253   \pm     0.082   $\\
         &               &       avg     &$ 
 $&$     0.0650  \pm     0.0100  $&$     0.230   \pm     0.030   $\\
3C441   &       N       &       2.5     &$      26      \pm     18 
$&$     0.0128  \pm     0.0043  $&$     0.041   \pm     0.014   $\\
         &               &       5       &$      11      \pm     3 
 $&$     0.0165  \pm     0.0024  $&$     0.052   \pm     0.007   $\\
         &               &       avg     &$ 
 $&$     0.0156  \pm     0.0020  $&$     0.050   \pm     0.007   $\\
3C441   &       S       &       2.5     &$      1500 \pm        9900 
$&$        0.1076  \pm     0.3589  $&$     0.464   \pm
1.546   $\\
         &               &       5       &$      86      \pm     100 
 $&$     0.0517  \pm     0.0314  $&$     0.223   \pm     0.134   $\\
         &               &       7.5     &$      150     \pm     190 
 $&$     0.1020  \pm     0.0652  $&$     0.439   \pm     0.279   $\\
         &               &       10      &$      34      \pm     23 
 $&$     0.0646  \pm     0.0235  $&$     0.278   \pm     0.099   $\\
         &               &       12.5    &$      150     \pm     190 
 $&$     0.1699  \pm     0.1086  $&$     0.732   \pm     0.465   $\\
         &               &       15      &$      34      \pm     24 
 $&$     0.0969  \pm     0.0352  $&$     0.417   \pm     0.148   $\\
         &               &       17.5    &$      10      \pm     3 
 $&$     0.0628  \pm     0.0100  $&$     0.271   \pm     0.038   $\\
         &               &       20      &$      16      \pm     5 
 $&$     0.0891  \pm     0.0177  $&$     0.384   \pm     0.070   $\\
         &               &       avg     &$ 
 $&$     0.0699  \pm     0.0080  $&$     0.300   \pm     0.030   $\\
3C469.1 
&       N       &       2.5     &$      150     \pm     160     $&$ 
0.1145  \pm     0.0617  $&$     0.456   \pm
0.245   $\\
         &               &       5       &$      840     \pm     3800 
 $&$     0.5444  \pm     1.2266  $&$     2.173   \pm     4.896   $\\
         &               &       7.5     &$      9.1     \pm     1.9 
 $&$     0.0848  \pm     0.0132  $&$     0.339   \pm     0.051   $\\
         &               &       10      &$      18      \pm     6 
 $&$     0.1584  \pm     0.0341  $&$     0.632   \pm     0.134   $\\
         &               &       avg     &$ 
 $&$     0.0950  \pm     0.0120  $&$     0.380   \pm     0.050   $\\
3C469.1 
&       S       &       2.5     &$      130     \pm     200     $&$ 
0.1050  \pm     0.0793  $&$     0.383   \pm
0.288   $\\
         &               &       5       &$      130     \pm     210 
 $&$     0.2100  \pm     0.1689  $&$     0.766   \pm     0.615   $\\
         &               &       7.5     &$      13      \pm     2 
 $&$     0.0985  \pm     0.0123  $&$     0.359   \pm     0.042   $\\
         &               &       10      &$      9.3     \pm     2.0 
 $&$     0.1106  \pm     0.0148  $&$     0.403   \pm     0.051   $\\
         &               &       12.5    &$      9.3     \pm     2.0 
 $&$     0.1382  \pm     0.0185  $&$     0.504   \pm     0.063   $\\
         &               &       15      &$      11      \pm     3 
 $&$     0.1800  \pm     0.0297  $&$     0.656   \pm     0.104   $\\
         &               &       17.5    &$      8.4     \pm     1.7 
 $&$     0.1838  \pm     0.0239  $&$     0.670   \pm     0.081   $\\
          &              &       avg      & 
 &$
0.1200   \pm    0.0080  $&$     0.450   \pm     0.030   $\\
\end{longtable}

Col 1. Source Name. Col 2. Side of the source on which the spectral aging
was determined. Col 3.
The distance from the hotspot. The estimated error is 0.5 arcsec.
Col 4. The estimated break frequency.
Col 5. The estimated hot spot advance speed assuming the magnetic field is
equal
to 0.25 the averaged equipartition value
given in Table~\ref{tabaging}.
Col 6. The estimated hot spot advance speed assuming the magnetic field is
equal to
the averaged equipartition value given in Table~\ref{tabaging}.
The last row for each side gives the weighted mean velocity.

\clearpage

  \end{document}